\documentclass[reprint,prx,aps,amsmath,amssymb,floatfix,superscriptaddress,longbibliography]{revtex4-2}

\usepackage{mathtools,amsmath}
\usepackage{graphicx} 
\usepackage[breaklinks=true,colorlinks,citecolor=teal,linkcolor=teal,urlcolor=teal]{hyperref}
\usepackage{physics}
\usepackage{siunitx}
\usepackage{bbold}
\usepackage{soul}
\usepackage{bm}
\usepackage{multirow}
\usepackage[normalem]{ulem}
\usepackage{times}
\usepackage{enumerate}  
\usepackage{float}
\usepackage{amssymb}
\usepackage{xcolor} 
\usepackage{booktabs}

\begin{document}

\title{Symmetrically Threaded Superconducting Quantum Interference Devices As Next Generation Kerr-cat Qubits}
\author{Bibek Bhandari}
\email{These authors contributed equally; Correspondence should be addressed to: bbhandari@chapman.edu}
\affiliation{Institute for Quantum Studies, Chapman University, Orange, CA 92866, USA}

\author{Irwin Huang}
\email{These authors contributed equally; Correspondence should be addressed to: bbhandari@chapman.edu}
\affiliation{Department of Physics and Astronomy, University of Rochester, Rochester, NY 14627, USA}

\author{Ahmed Hajr}
\affiliation{Quantum Nanoelectronics Laboratory, Department of Physics, University of California at Berkeley, Berkeley, CA 94720, USA}
\affiliation{Graduate Group in Applied Science and Technology, University of California at Berkeley, Berkeley, CA 94720, USA }

\author{Kagan Yanik}
\affiliation{Department of Physics and Astronomy, University of Rochester, Rochester, NY 14627, USA}

\author{Bingcheng Qing}
\affiliation{Quantum Nanoelectronics Laboratory, Department of Physics, University of California at Berkeley, Berkeley, CA 94720, USA}

\author{Ke Wang}
\affiliation{Quantum Nanoelectronics Laboratory, Department of Physics, University of California at Berkeley, Berkeley, CA 94720, USA}

\author{David I Santiago}
\affiliation{Quantum Nanoelectronics Laboratory, Department of Physics, University of California at Berkeley, Berkeley, CA 94720, USA}
\affiliation{Computational Research Division, Lawrence Berkeley National Laboratory, Berkeley, Berkeley, CA 94720, USA}

\author{Justin Dressel}
\affiliation{Schmid College of Science and Technology, Chapman University, Orange, CA 92866, USA}
\affiliation{Institute for Quantum Studies, Chapman University, Orange, CA 92866, USA}

\author{Irfan Siddiqi}
\affiliation{Quantum Nanoelectronics Laboratory, Department of Physics, University of California at Berkeley, Berkeley, CA 94720, USA}
\affiliation{Computational Research Division, Lawrence Berkeley National Laboratory, Berkeley, Berkeley, CA 94720, USA}
\author{Andrew N Jordan}
\affiliation{The Kennedy Chair in Physics, Chapman University, Orange, CA 92866, USA}
\affiliation{Schmid College of Science and Technology, Chapman University, Orange, CA 92866, USA}
\affiliation{Institute for Quantum Studies, Chapman University, Orange, CA 92866, USA}
\affiliation{Department of Physics and Astronomy, University of Rochester, Rochester, NY 14627, USA}

          
\begin{abstract}
Kerr-cat qubits are bosonic qubits offering autonomous bit-flip protection, traditionally studied using driven Superconducting Nonlinear Asymmetric Inductive eLement (SNAIL) oscillators. Here, we theoretically explore an alternative circuit for Kerr-cat qubits based on symmetrically threaded Superconducting Quantum Interference Devices (SQUID). The Symmetrically Threaded SQUIDs (STS) architecture employs a simplified flux-pumped design that suppresses two-photon dissipation, a dominant loss mechanism in high-Kerr regimes, by engineering the drive Hamiltonian’s flux operator to generate only even-order harmonics.  By fulfilling two critical criteria for practical Kerr-cat qubit operation, the STS emerges as an ideal platform: (1) a static Hamiltonian with diluted Kerr nonlinearity (achieved via the STS's middle branch) and (2) a drive Hamiltonian restricted to even harmonics, which ensures robust two-photon driving with reduced dissipation. For weak Kerr nonlinearity, we find that the coherent state lifetime ($T_\alpha$) is similar between STS and SNAIL circuits. However, STS Kerr-cat qubits exhibit enhanced resistance to higher-order photon dissipation, enabling significantly extended $T_\alpha$ even with stronger Kerr nonlinearities ($\sim 10$ MHz). In contrast to SNAIL, STS Kerr-cat qubits display a $T_\alpha$ dip under weak two-photon driving for high Kerr coefficient. We demonstrate that this dip can be suppressed by applying drive-dependent detuning, enabling Kerr-cat qubit operation with only eight Josephson junctions (of energies 80 GHz); fewer junctions suffice for higher junction energies. We further validate the robustness of the STS design by studying the impact of strong flux driving and asymmetric Josephson junctions on $T_\alpha$. With the proposed design and considering a cat size of 10 photons, we predict $T_\alpha$ of the order of tens of milliseconds, even in the presence of multi-photon heating and dephasing effects. The robustness of the STS Kerr-cat qubit makes it a promising component for fault-tolerant quantum processors.

\end{abstract}

\maketitle


\section{Introduction}
\label{sec:intro}

\begin{figure}[hbt!]
\includegraphics[width=\columnwidth]{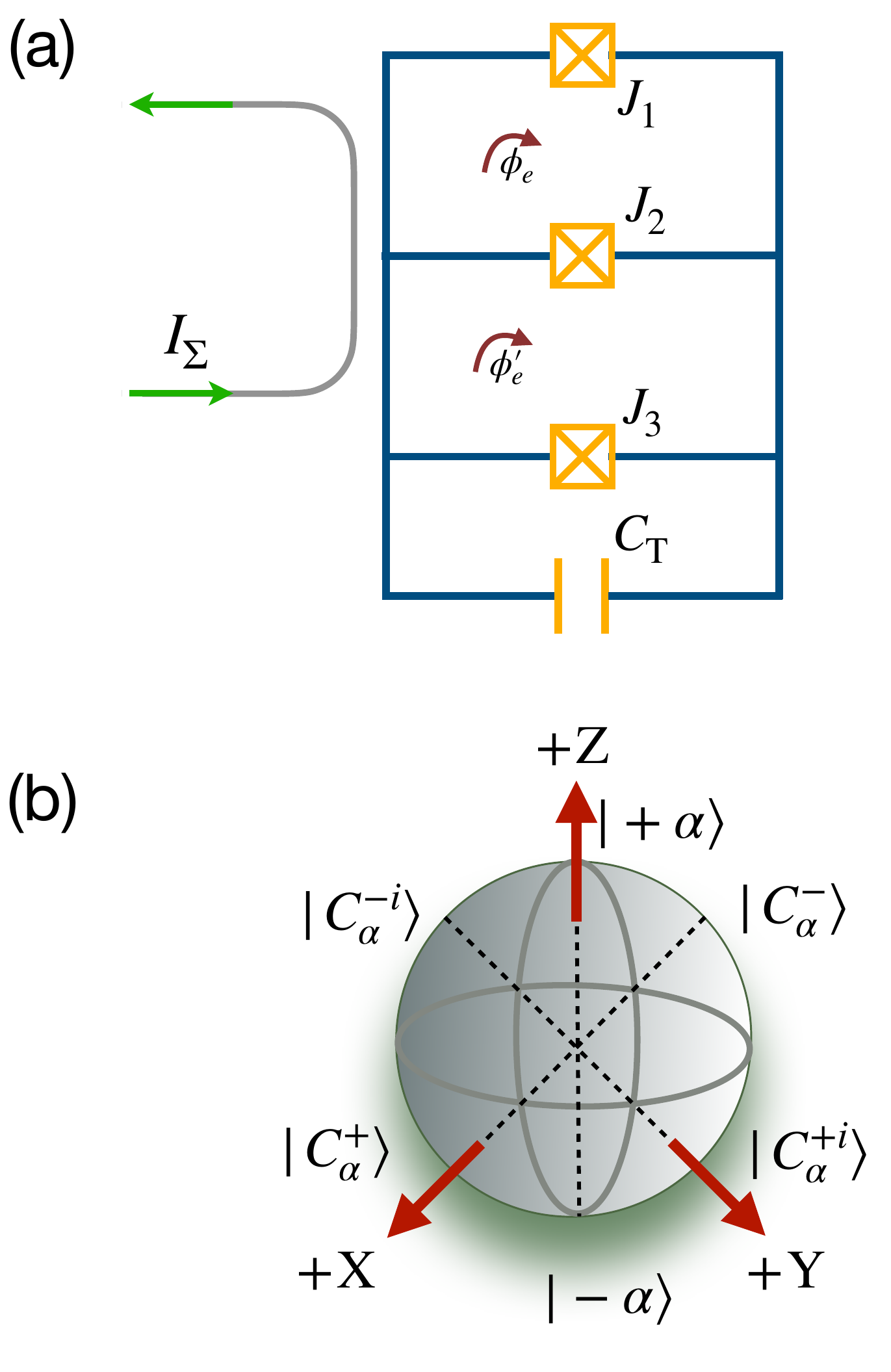}
 \caption{(a) STS design for Kerr-cat qubit. The junctions ${\rm J}_{1}$ and ${\rm J}_{3}$ compose the ``SQUID branch" whereas the junction ${\rm J}_{2}$ gives the ``transmon branch". The loop between ${\rm J}_{1}$ and ${\rm J}_{2}$ is threaded by an external flux $\phi_e$ whereas the loop between ${\rm J}_{2}$ and ${\rm J}_{3}$ is threaded by an external flux $\phi_e^\prime$. We consider $\phi_e^\prime = \phi_e = -\pi/2 + \delta\phi \cos(\omega_{\rm d}t)$. $C_{\rm T}$ is the shunt capacitor associated with the transmon branch. (b) Bloch sphere representation of the Kerr-cat qubit states.}
\label{fig:sketch_design}
\end{figure}
Creating artificial atoms using superconducting circuits is one of the most promising approaches to developing qubits for fault-tolerant quantum computation \cite{scqubit_nakamura,noise_koch,scqubit_fluxonium,grimm2020,noise_nguyen}. However, many of the superconducting qubit architectures suffer from low coherence time, limiting the gates and readout fidelities \cite{noise_anton,noise_ithier,noise_koch,noise_long,briseno2025dynamical,noise_nguyen,noise_yoshihara,scqubit_fluxonium,grimm2020}. Here we investigate a noise-biased Kerr-cat qubit \cite{puri2017,goto2016universal, puri2019, grimm2020, frattini2022,jaya2023,ahmed,iyama2024observation,chavez2023,gautier2022,kwon2022,puri2020,venkatraman2022,You2019,ruiz2023,kerr_andersen,kerr_miano2022,kerr_puri2021pr,kerr_he,kerr_mirrahimi,qing2024benchmarking}
which has a long coherence time, observed close to a millisecond experimentally \cite{frattini2022,jaya2023,ahmed,qing2024benchmarking}. The static effective Hamiltonian describing a Kerr nonlinear oscillator subject to a two-photon drive (commonly referred to as a squeeze drive) consists of a double-well in the phase space representation. The tunneling between the two wells destructively interferes under certain conditions and gives rise to noise-biased qubits which are strongly protected against bit-flip channels \cite{thesis_jaya} (see Fig.~\ref{fig:sketch_design} (b)). The static effective Hamiltonian, which governs the dynamics, obtained using fast time-periodic driving, cannot be obtained using a static Hamiltonian. One can make a direct analogy to the case of a driven classical oscillator, namely the Kapitza pendulum, where the oscillator can be dynamically stabilized in an otherwise unstable point \cite{kapitza1965dynamic}.

There are two main approaches to obtaining a cat qubit: 1) a quantum harmonic oscillator in the presence of engineered two-photon driven dissipation \cite{Dissipative_cat_intro,lescanne2020exponential, Dissipative_cat_intro3, Dissipative_cat_intro4,gautier2022,diss_gautier,diss_guillaud,diss_putterman,diss_touzard} and 2) two-photon driving of a Kerr non-linear oscillator with a finite but small Kerr coefficient \cite{puri2017,goto2016universal,grimm2020,frattini2022, jaya2023,ahmed}. The second approach, i.e., a Kerr-cat qubit, allows for simpler high-fidelity non-demolition readout and faster gate operations. Kerr-cat qubits have been extensively studied, both theoretically \cite{venkatraman2022,jaya2023,puri2017,puri2020,aoki2024,kang2023effective,suzuki2023} and experimentally \cite{grimm2020,frattini2022,ahmed,qing2024benchmarking}, in a setup based on a Superconducting Nonlinear Asymmetric Inductive eLement (SNAIL) oscillator, which consists of a loop of multiple large Josephson junctions and a smaller Josephson junction, threaded by a DC magnetic flux. In this article, we present an alternate design based on symmetrically threaded Superconducting Quantum Interference Devices (SQUID) as shown in Fig.~\ref{fig:sketch_design}(a).

The flux-driven Symmetrically Threaded SQUIDs (STS) design differs fundamentally from the charge-driven SNAIL architecture \cite{grimm2020,frattini2022} in two critical ways. First, the STS exhibits a drive-dependent detuning term in its static effective Hamiltonian \cite{misc_siddiqi_amp}, a feature absent in the SNAIL. Second, the STS distinguishes between even and odd harmonics by coupling them to different orders of the zero-point phase spread ($\varphi_{\rm zps}$), determined by the symmetry of the SQUID junctions: even harmonics arise from symmetric junction parameters, while odd harmonics originate from asymmetries. Crucially, we find that the dominant multi-photon heating effects, known to degrade the coherence time $T_{\alpha}$	
in SNAILs at large Kerr coefficients, scale with junction asymmetry in the STS. By optimizing junction symmetry, these heating processes can be suppressed, enabling robust $T_\alpha$ even in the high-Kerr regime. Additionally, the STS design enables 2-photon driving at second order in $\varphi_{\rm zps}$, whereas the SNAIL requires third-order terms in non-linearity for the same effect. This lower-order coupling results in enhanced two-photon driving strength for a similar modulation of the two designs. We demonstrate that the STS preserves the Kerr-cat energy spectrum and resists multiphoton heating even with Kerr coefficients exceeding $10~{\rm MHz}$. These properties allow STS-based Kerr-cat qubits to achieve faster quantum gates and high-fidelity readout without sacrificing coherence, offering a scalable pathway toward high-performance quantum operations.

Heating effects caused by single- and multi-photon excitations in qubits significantly degrade their coherence and operational lifetime. These effects persist even at zero temperature in strongly driven quantum systems, as demonstrated in Ref.~\cite{venkatraman2022}. A promising strategy to suppress such decoherence lies in leveraging the degenerate energy spectrum of the Kerr-cat Hamiltonian. This work explores two complementary approaches to amplify spectral degeneracy in the superconducting STS Kerr-cat qubit, thereby mitigating heating-induced losses. The first approach involves enhancing the two-photon drive strength, a method extensively studied in Refs.~\cite{frattini2022,jaya2023,ahmed}. While a stronger two-photon drive enhances noise bias (leading to longer bit-flip lifetimes), it typically comes at the expense of a reduced phase-flip lifetime. The second approach, which overcomes this trade-off, involves tuning the system’s detuning to match non-negative even integer multiples of the Kerr nonlinearity, as discussed in Refs.\cite{frattini2022,ahmed,jaya2023,qing2024benchmarking}. In Ref.~\cite{qing2024benchmarking}, it was shown that such detuning increases the bit-flip lifetime without compromising the phase-flip lifetime. Moreover, we find that combining both strategies, simultaneously increasing the two-photon drive and applying detuning, further enhances the noise bias by accelerating the growth of spectral degeneracies.

In the following section, we introduce the theoretical model for the Kerr-cat Hamiltonian. Building on this framework, Sec.~\ref{sec:ham} investigates an alternate circuit based on driven STS for the Kerr-cat qubit, followed by a derivation of its static effective Hamiltonian. In Sec.~\ref{sec:single_squid}, we compare the performance of the STS design with the single SQUID design, establishing the STS as a better candidate for Kerr-cat qubit implementation in flux-pumped superconducting circuit platform. In Sec.~\ref{sec:master_equation}, we investigate the static effective master equation up to the leading order in system-environment coupling and fourth order in zero point phase spread, with a discussion on the effects of asymmetry in the Josephson junction. Leveraging the derived master equations, in Secs.~\ref{sec:lifetime_resonant} and \ref{sec:lifetime_detuned}, we study $T_\alpha$ of the STS Kerr-cat qubit as a function of two-photon drive strength and detuning.  Under zero detuning, adiabatic ramping of the two-photon drive causes the first two Fock states to converge to cat-like superpositions, a behavior that breaks down for detuned systems. Sec.~\ref{sec:qbt_ini} investigates initialization protocols tailored for detuned Kerr-cat qubits to address this limitation. Finally, Sec.~\ref{sec:conc} summarizes our key findings and conclusions.

\section{Kerr-cat Hamiltonian}
\label{sec:model}
We analyze two distinct implementations of Kerr-cat qubits: the detuned and resonant configurations. The detuned Kerr-cat qubit is governed by the Hamiltonian~\cite{milburn1, milburn2,puri2017,goto2016universal}
\begin{equation}
\hat{H}_{\rm DKC} = \Delta \hat{a}^\dagger \hat{a}+\epsilon_2(\hat{a}^{\dagger 2}+\hat{a}^2)-K\hat{a}^{\dagger 2}\hat{a}^2,
\label{eq:Kerrcat}
\end{equation}
where $\Delta$ is the detuning term, $K$ is the Kerr nonlinearity, and $\epsilon_2$ is the two-photon drive strength that excites and annihilates photons in pairs. The two-photon drive, typically realized in SNAIL oscillators via a charge drive at twice the oscillator frequency \cite{grimm2020,frattini2022,jaya2023,ahmed}, excites and annihilates photon pairs. For $\Delta=0$, the system reduces to the resonant Kerr-cat qubit described by
\begin{equation}
\hat{H}_{\rm RKC}=\epsilon_2(\hat{a}^{\dagger 2}+\hat{a}^2)-K\hat{a}^{\dagger 2}\hat{a}^2.
\label{eq:res_Kerrcat}
\end{equation}
The coherent states $|\pm \alpha\rangle$ with $\alpha = \sqrt{\epsilon_2/K}$ are the eigenstates of $\hat H_{\rm RKC}$. The cat states are the even/odd parity states formed by the superposition of coherent states
\begin{equation}
\ket{{C}_\alpha^{\pm}} = \frac{1}{\sqrt{2}}\frac{1}{\sqrt{1\pm e^{-2\alpha^2}}}\Big(|\alpha\rangle \pm |-\alpha\rangle\Big),
\end{equation}
which approximate the eigenstates of the resonant Kerr-cat qubit for  $|\alpha|^2 \gg 1$ ($e^{-2\alpha^2}\rightarrow  0$). In the effective low-dimensional subspace, the eigenstates of ${\bf Z}$, ${\bf Z} |\pm Z\rangle= \pm |\pm Z\rangle$ are shown as bases in the Bloch sphere representation (see Fig.~\ref{fig:sketch_design}(b)) where
\begin{equation}
|\pm Z\rangle = \frac{1}{\sqrt{2}}\left(\ket{{ C}_\alpha^{+}}\pm \ket{{ C}_{\alpha}^{-}} \right) \approx |\pm \alpha\rangle.
\label{eq:Z_bloch}
\end{equation}

The remaining Pauli operators are defined analogously \cite{thesis_frattini}. Interestingly, coherent states act as eigenstates for both detuned and resonant Kerr-cat qubits at large cat sizes \cite{qing2024benchmarking}, enabling their use as robust computational states in either regime. We propose SQUID-based circuits to implement Kerr-cat qubits in a superconducting circuit platform that achieves a Kerr-cat Hamiltonian with strong tunable two-photon driving and weak Kerr nonlinearity. We compare two designs: a double-SQUID loop STS and a single SQUID. As shown in subsequent sections, the STS design 1) can be operated at a high drive sensitivity point, enhancing two-photon driving strength under identical flux-pumping conditions, and 2) has an extra middle branch for the dilution of Kerr coefficient, making it superior for stabilizing Kerr-cat qubits.
\ \\

\subsection{Symmetrically Threaded SQUIDs (STS) Hamiltonian}
\label{sec:ham}
To realize the Hamiltonian in Eq.~(\ref{eq:Kerrcat}), we consider an alternative circuit for the Kerr-cat qubit based on the STS. The circuit design is shown in Fig.~\ref{fig:sketch_design}(a). For the sake of simplicity, we will consider only one STS circuit with single junctions and later discuss the implications of having multiple junctions in the middle branch and multiple STS in series. An external flux $\phi_{\rm e}~(\phi_{\rm e}^\prime)$, in units of flux quantum $h/2e$, is threaded through the SQUID loop formed by Josephson junctions ${\rm J}_{2}$ and ${\rm J}_{1}~({\rm J}_{3})$. A capacitor $C_{\rm T}$ is connected in parallel to the junction arrays. The Hamiltonian for the circuit in the lab frame can be written as \cite{misc_koch_ckt_quant}
\begin{align}
\label{eq:circ_ham1}
&\hat{H}_{\rm lab} = 4E_C \hat{n}^2 - E_{\rm J1}\cos(\hat{\varphi}+ \frac{2}{3}\phi_e + \frac{1}{3}\phi_e^\prime)\\
&-E_{\rm J2}\cos\left(\hat{\varphi}+ \frac{1}{3}\phi_\Delta\right)-E_{\rm J3}\cos(\hat{\varphi}- \frac{1}{3}\phi_e - \frac{2}{3}\phi_e^\prime),\nonumber
\end{align}
where we considered equal junction capacitances. $E_{{\rm J}i}$, $i=1,2,3$, gives the Josephson energy of the Josephson junction ${\rm J}_{i}$ and $E_{\rm C}$ is the effective capacitive energy. In deriving Eq.~(\ref{eq:circ_ham1}), we assumed geometric inductance to be much smaller than the Josephson inductance \cite{lu2023high}.

When the flux is asymmetrically threaded between the two loops, it has been shown that one can implement efficient 4-wave mixing processes in the ATS (Asymmetrically Threaded SQUID) \cite{lescanne2020exponential}. However, the ATS dissipative cat qubit differs from the STS Kerr-cat qubit both in the mode of stabilization of the cat qubit and operation \cite{ahmed}. Unlike the ATS, the STS circuit relies on its intrinsic
Kerr non-linearity to host the cat states. In the STS circuit design, since we are interested in doing efficient 3-wave mixing to generate the two-photon drive, we instead thread the flux symmetrically using $\phi_\Delta = \phi_e^\prime -\phi_e=0$. Eq.~(\ref{eq:circ_ham1}) reduces to
\begin{multline}
    \hat{H}_{\rm lab}=4E_{\rm C}\hat{n}^2-E_{\rm J2}\cos{\hat{\varphi}}-2E_{\rm J\Sigma}\cos\phi_e\cos{\hat{\varphi}}\\
    +2E_{\rm J\Delta}\sin\phi_e\sin{\hat{\varphi}},
    \label{eq:ham_unexp}
\end{multline}
where we defined $E_{\rm J\Sigma}=({E_{\rm J1}+E_{\rm J3}})/{2}$ and $E_{\rm J\Delta}=({E_{\rm J1}-E_{\rm J3}})/{2}$. For the sake of simplicity, we consider $E_{\rm J\Delta}=0$ in this section. Note that the effective capacitance (including the junction capacitance as well as the external capacitor $C_{\rm T}$) along with the junction ${\rm J}_2$ will act as a transmon, providing the necessary bound states and the Kerr nonlinearity for the oscillator. The lab Hamiltonian in Eq.~(\ref{eq:ham_unexp}) consists of a transmon type contribution (the first two terms) and a SQUID type contribution (the last two terms). The transmon part is obtained from the middle branch of the STS, whereas the SQUID contribution is obtained from the outer branches. From now on, we will call the middle branch with Josephson junction ${\rm J}_{2}$ as the ``transmon branch," whereas the other branches will be referred to as the ``SQUID branch." 

To generate a large two-photon drive, we symmetrically DC bias the SQUID around $-\pi/2$, then apply an AC modulation tone according to $\phi_e=\phi_e^\prime = -\pi/2 + \delta \phi \cos(\omega_d t)$. The DC bias point is chosen to create the highest first-order sensitivity on the modulation depth $\delta \phi$ while removing the parasitic even harmonics of the drive (see Appendix~\ref{app:Hamil}). The external flux in the two loops must be the same, which means we only modulate the common mode of those two flux loops, leaving the differential mode untouched. This requirement can be realized experimentally by designing the flux lines symmetrically with respect to the two SQUID loops or by carefully controlling the interference between the two flux lines. Expanding the Hamiltonian in Eq.~(\ref{eq:ham_unexp}) to fourth order in $\hat{\varphi} = \varphi_{\rm zps} (\hat{a}^\dagger + \hat{a})$ with $\varphi_{\rm zps} = \left(2E_{\rm C}/E_{\rm J}\right)^{1/4}$ being the zero point phase spread, we obtain
\begin{multline}
\hat{H}_{\rm lab} = \epsilon_c \hat{a}^\dagger \hat{a} - K \hat{a}^{\dagger^2}\hat{a}^2 -2E_{\rm J\Sigma} 
\sin{\left(\delta\phi \cos(\omega_d t)\right)}\\
\sum_{n=1,2} \frac{(-1)^n}{(2n)!}\left[\varphi_{\rm zps}(\hat{a}+\hat{a}^\dagger) \right]^{2n},
\label{eq:kerr_cat_ham_dr}
\end{multline}
where $\epsilon_c = \sqrt{8 E_{\rm C}E_{\rm J2}}$ and $K= E_{\rm C}/2$. In order to obtain the Kerr-cat Hamiltonian of Eq.~(\ref{eq:Kerrcat}), we look for the static effective representation of $\hat{H}_{\rm lab} $ under the condition $\hbar\omega_d\approx 2\epsilon_c$. 

Going to the rotating frame at frequency $\omega_{\rm d}/2$ following the transformation $\hat{a}\rightarrow \hat{a}e^{-i\omega_{\rm d} t/2}$ and using the generalized Schrieffer-Wolff transformation generated by $\hat{S}(t)$ (see Appendix~\ref{app:Hamil} for details), the effective Hamiltonian for the STS is given by ${\hat{\cal H}}_{\rm S} = e^{\hat{S}/i\hbar}\hat{H}_{\rm lab}(t) e^{-i\hat{S}/i\hbar} - i\hbar e^{\hat{S}/i\hbar}\partial_te^{\hat{S}/i\hbar}$. Up to ${\cal O}(\varphi_{\rm zps}^4)$, the static effective Hamiltonian is given by
\begin{equation}
\hat{{\cal H}}_{\rm S}= \Delta \hat{a}^\dagger \hat{a} + \epsilon_2 (\hat{a}^{\dagger^ 2}+\hat{a}^2) - K \hat{a}^{\dagger^ 2}\hat{a}^2+
\Lambda (\hat{a}^{\dagger} \hat{a}^3+\hat{a}^{\dagger^3}\hat{a}),
\label{eq:ham_main}
\end{equation}
where for STS with single junctions of energy $E_{\rm J1} = E_{\rm J2}=E_{\rm J3} = E_{\rm J}$ and up to first order in the modulation depth $\delta \phi$,
\begin{equation}
\epsilon_2 = \delta\phi/2\left(\sqrt{2E_{\rm C}E_{\rm J}}-E_{\rm C}\right) = \delta\phi/4 \left(\epsilon_{\rm c} - 2E_{\rm C}\right),
\end{equation}
$\Delta = \epsilon_{\rm c}-\hbar\omega_{\rm d}/2-2K-\delta\phi^2\epsilon_{\rm c}^2/8\hbar\omega_{\rm d}$, $K = E_{\rm C}/2$ and $\Lambda =-\delta\phi E_{\rm C}/3$. 
\begin{figure}[t]
\includegraphics[width=\columnwidth]{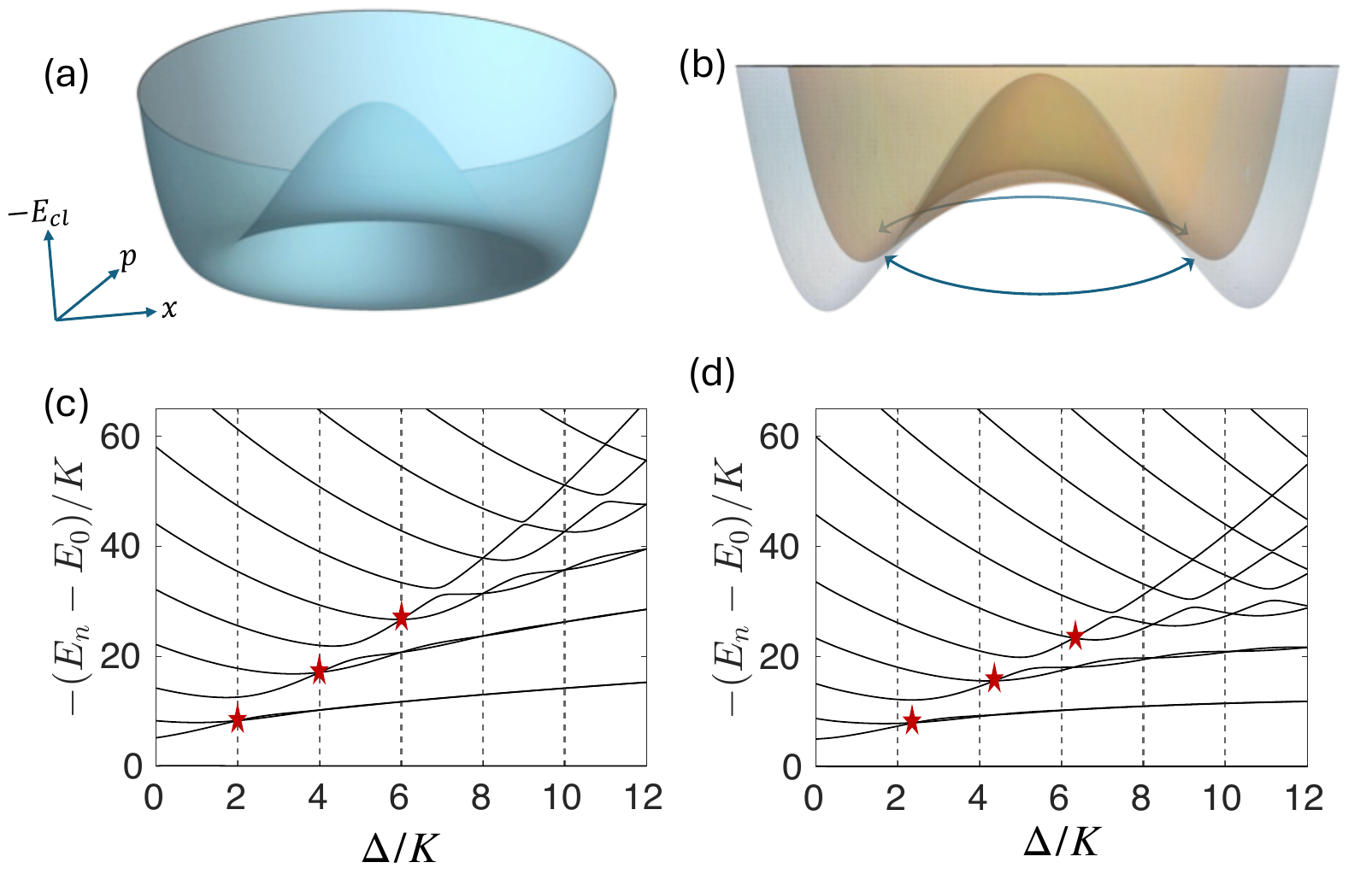}
 \caption{Classical phase space surface of the total effective energy of the Kerr-cat Hamiltonian in Eq.~(\ref{eq:ham_main}) for $\Delta/K = 4$, (a) $\epsilon_2/K=\Lambda/K=0$,  and 
(b) $\epsilon_2 /K= 0.3 $, $\Lambda = 0$ (yellow plot) and $\Lambda/K = 0.2 $ (grey plot). Energy spectrum of Eq.~(\ref{eq:ham_main}) as a function of $\Delta$ for $\epsilon_2/K=2$ and (c) $\Lambda/K = 0$ and (d) $\Lambda/K = 0.12$. Red stars denote the first degeneracies of the first three excited state pairs with opposite parity.} 
\label{fig:phase_deg}
\end{figure}
Note that the two-photon drive strength $\epsilon_2$ is linear in the modulation depth, the oscillator frequency, and the charging energy $E_{\rm C}$. The STS circuit achieves 2-photon driving at the second order of the zero-point phase spread ($\varphi_{\rm zps}$) compared to the SNAIL-based Kerr-cat qubit where one needs to go to the third order in non-linearity, resulting in a much larger two-photon drive strength with a similar pumping in the STS compared to SNAIL. Compared with the usual Kerr-cat Hamiltonian studied in Refs.~\cite{puri2017,grimm2020} and given in Eq.~(\ref{eq:Kerrcat}), the static effective Hamiltonian in Eq.~(\ref{eq:ham_main}) for STS has an extra term proportional to $\Lambda$. This extra term takes the qubit away from the Kerr-cat regime and is detrimental to the qubit's coherent lifetime. In Sec.~\ref{sec:lifetime_resonant}, we will investigate the decrease in lifetime due to this extra term and ways to mitigate its effects.

In Fig.~\ref{fig:phase_deg}(a), we show the total effective classical energy ($E_{\rm cl}$) of a non-linear oscillator ($\epsilon_2=0$) in the phase space. We observe a circularly symmetric energy well in the position $(x)$ and momentum $(p)$ space. The inter-well tunneling can only take place through the barrier in the middle. However, in the case of Kerr-cat qubit (see Fig.~\ref{fig:phase_deg}(b)), a classically forbidden region develops in between the double well. Further, two saddle points connect the two wells. The tunneling through these saddle points can destructively interfere, giving rise to robust qubits. In Ref.~\cite{jaya2023}, it was shown that destructive interference happens when the detuning is an even integer multiple of the Kerr coefficient, $\Delta/K = 2m$, where $m$ is a non-negative integer. 

In Fig.~\ref{fig:phase_deg}(c), we plot the energy spectra of the Kerr-cat qubit as a function of detuning for $\Lambda = 0$, which is equivalent to a SNAIL Kerr-cat Hamiltonian \cite{venkatraman2022}. For $\Delta/K = 2m$, we observe $m+1$ degeneracy points; note that the ground state energies of a Kerr-cat Hamiltonian, which are subtracted from the eigenenergies, are always degenerate. The main difference between our analysis based on STS and Ref.~\cite{puri2017} can be seen in Fig.~\ref{fig:phase_deg}(d) where we plot the energy spectra for $\Lambda/K= 0.12 $, recalling that, $\Lambda$ is absent in the SNAIL Kerr-cat Hamiltonian. Although the degeneracy points are independent of the values of $\epsilon_2$ in SNAIL Kerr-cats, we find that with STS they get shifted from $\Delta/K = 2m$ for finite values of $\Lambda$. Since $\Lambda$ and $\epsilon_2$ both depend on the modulation depth $\delta\phi$, the shift in degeneracy varies as a function of $\epsilon_2$. We can thus apply an external drive-dependent detuning to cancel the shift in degeneracy points.

In order to dilute the Kerr (necessary to mitigate multi-photon heating effects), we will consider multiple STS connected in series. Let us consider $M$ number of STS with a single junction in the SQUID branch. In that case, if $\tilde{N}$ number of junctions are used in series in the transmon branch of each STS (each with Josephson energy $E_{\rm J2}^\prime=N E_{\rm J2}$, where $N=M\tilde{N}$), the Kerr coefficient gets diluted to ${K}\rightarrow K/N^2$ (see Appendix~\ref{app:dilution} for details). Since the Kerr coefficient and the two-photon drive strength compete to stabilize the cat states, one can enhance the effect of pumping without increasing the modulation depth by just diluting the Kerr coefficient. This is desirable because a diluted Kerr coefficient leads to a stronger stabilization of the cat qubit for a similar drive strength, leading to faster qubit initialization and high-fidelity gates.

\subsection{Accuracy of the STS Static Effective Hamiltonian}
\label{sec:accuracy_STS}
\begin{figure}[hbt!]
\includegraphics[width=\columnwidth]{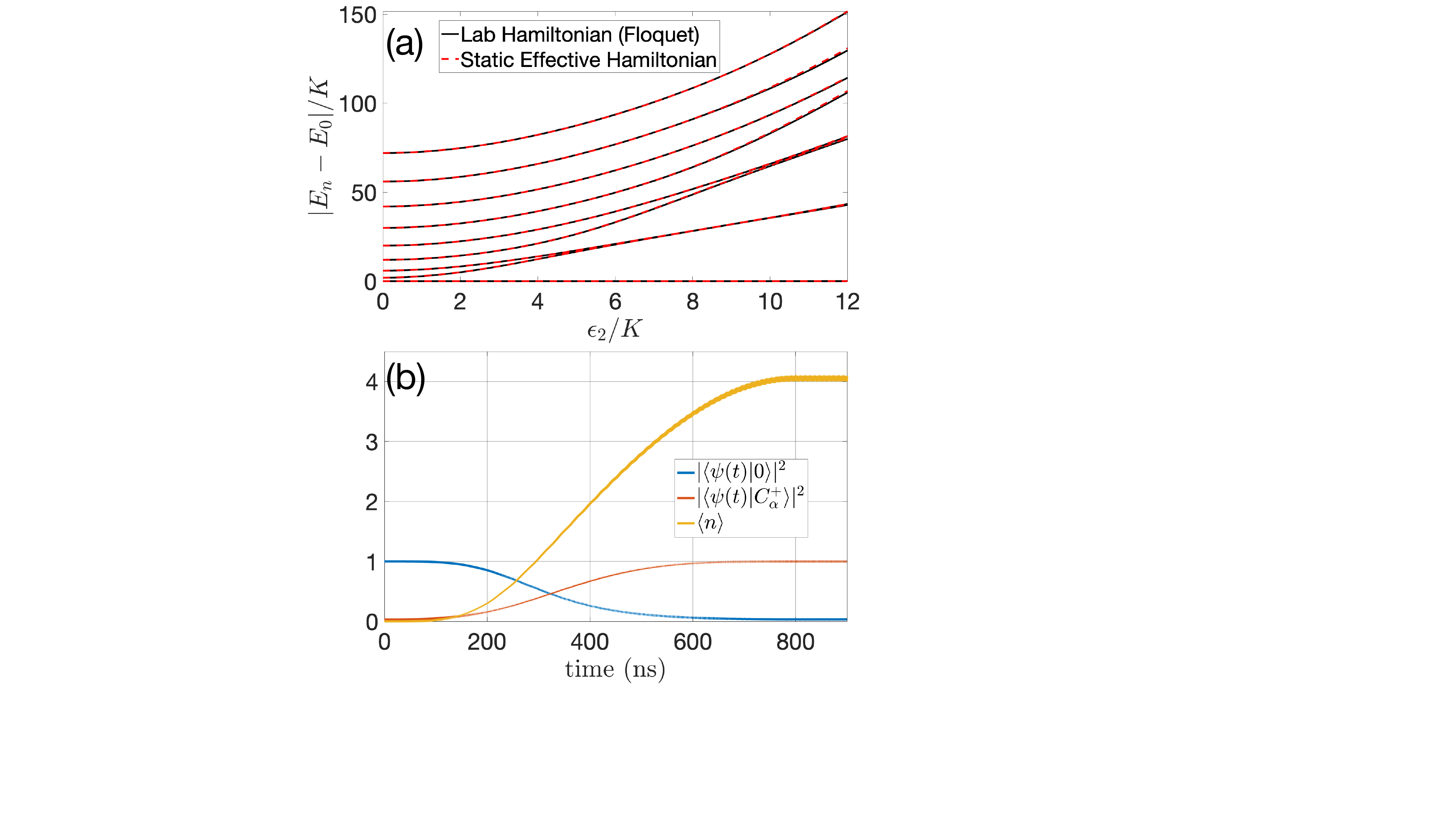}
 \caption{ (a) The Floquet quasi-energy level diagram of the STS lab Hamiltonian of Eq.~(\ref{eq:ham_unexp}) (solid black curves) compared to the eigenenergy level diagrams of the STS effective Hamiltonian of Eq.~(\ref{eq:ham_main}) (dashed red curves) plotted as a function of the two-photon drive strength for $K/h=7.81$ MHz. (b) The overlap between the eigenstates of the STS lab Hamiltonian and the cat states (eigenstates of the static effective Hamiltonian) plotted as a function of time when the two-photon drive is adiabatically switched on from $\epsilon_2/K = 0$ to $\epsilon_2/K = 4$ (solid red curve). The solid orange curve gives the photon population in the STS at each time $t$.}
\label{fig:accuracy}
\end{figure}
In this section, we investigate the accuracy of the static effective Hamiltonian of Eq.~(\ref{eq:ham_main}) by comparing its eigenenergies and eigenstates to that of the lab Hamiltonian of Eq.~(\ref{eq:kerr_cat_ham_dr}). Following Ref.~[\onlinecite{GarciaMata2024}], we use the Floquet formulation to derive the quasienergies of the lab Hamiltonian and compare them with the eigenenergies of the STS static effective Hamiltonian. In Fig.~\ref{fig:accuracy}(a), the solid black curves are the Floquet quasienergies ($|\epsilon_n-\epsilon_0|/K$, $n=1,2,3,..$) of the lab Hamiltonian whereas the dashed red curves are eigenenergies ($|E_n-E_0|/K$) of the STS static effective Hamiltonian for $K/h=7.81$ MHz. We find almost perfect agreement between the Floquet and Kerr-cat energy spectra. Therefore, the STS design behaves as an effective Kerr-cat qubit even for a high Kerr coefficient of $K/h=7.81 $ MHz.

In Fig.~\ref{fig:accuracy}(b), to study the overlap between the STS static effective Hamiltonian eigenstates and the eigenstates of the lab Hamiltonian, we adiabatically switch on the two-photon drive from $\epsilon_2/K=0$ to $\epsilon_2/K= 4$, and plot the overlap as a function of time, considering $K/h=7.81$ MHz. Starting from the ground state, we find that the overlap between the cat eigenstate $(\ket{C_\alpha^+})$ and the eigenstate of the lab Hamiltonian $(\ket{\psi})$ approaches 1 with increasing time (in ns). Further, the photon number (solid orange curve) asymptotically approaches $\approx \epsilon_2/K$, neglecting a small contribution proportional to $\Lambda$.

\subsection{Single SQUID Hamiltonian}
\label{sec:single_squid}
This section will study the SQUID-based Kerr-cat qubit and compare it against an STS Kerr-cat qubit. To realize a Kerr-cat qubit, we pump a capacitor shunted SQUID with an external flux, $\tilde{\phi}_{\rm e} = \pi/4 + \delta \phi \cos(\omega_{\rm d} t)$ (see Appendix~\ref{app:squid} for details). The pumping is done around $\pi/4$ since, unlike STS, a single SQUID behaves as a capacitor at $\pi/2$. Doing so, we obtain the following static effective Hamiltonian up to ${\cal O}(\varphi_{\rm zps}^4)$
\begin{multline}
\hat{{\cal H}}_{\rm SQ} = \tilde{\Delta}\hat{a}^\dagger \hat{a}+\tilde{\epsilon}_2(\hat{a}^{\dagger^2}+\hat{a}^2)-\tilde{K}\hat{a}^{\dagger^2}\hat{a}^2\\
+\tilde{\Lambda}\left(\hat{a}^{\dagger^3}\hat{a} + \hat{a}^{\dagger}\hat{a}^3\right) +\tilde{\Theta}\left(\hat{a}^{\dagger^4}+\hat{a}^4\right),
\label{eq:ham_squid}
\end{multline}
where for a single junction SQUID $\tilde{K}=E_{\rm C}/2$, $\tilde{\epsilon}_2 = -\frac{\delta\phi}{2\sqrt{2}}\left(\sqrt{2E_{\rm C}E_{\rm J}}-E_{\rm C}\right)+\delta\phi^3E_{\rm C}E_{\rm J}/4\omega_{\rm d}$ (for the details of the calculations and for the dependence of $\tilde{\Delta},\tilde{\Lambda}$ and $\tilde{\Theta}$ on the circuit parameters, see Appendix~\ref{app:squid}).
\begin{center}
\begin{table}
\renewcommand{\arraystretch}{1.9}
\begin{tabular}{|c|c|c|c| } 
\hline
\hline
\multirow{2}{6em} & SQUID& SNAIL\cite{frattini2022} & STS \\
\hline 
\hline
Drive & Flux & Charge & Flux \\ 
\hline
RWA & ${\cal O}(\varphi_{\rm zps}^0)$ & ${\cal O}(\varphi_{\rm zps}^0)$ & ${\cal O}(\varphi_{\rm zps}^0)$ \\ 
\hline
\multirow{2}{5.2em}{Two~photon dissipation}  & ${\cal O}(\varphi_{\rm zps}^3)$  & ${\cal O}(\varphi_{\rm zps}^1)$  & ${\cal O}(\varphi_{\rm zps}^3)$  \\ 
&$\propto E_{\rm J\Delta}$  & ~$\propto g_3$ & $\propto E_{\rm J\Delta}$ \\ 
\hline
{$\epsilon_2$}  & $-\frac{\delta\phi}{2\sqrt{2}}E_{\rm J\Sigma}\varphi_{\rm zps}^2$ & $g_3 \frac{4\Omega_{\rm d}}{3\omega_{\rm d}}$ & $\frac{\delta\phi}{2}E_{\rm J\Sigma}\varphi_{\rm zps}^2$\\ 
\hline
$K$ & $E_{\rm C}/2$ & $-\frac{3g_4}{2}+\frac{10g_3^2}{3\omega_{\rm d}}$ & $E_{\rm C}/2$\\ 
\hline
 {$T_\alpha $} & 188 $\mu$s & 2.58 $\mu$s & 1.22 ms\\ 
\hline
\end{tabular}
\caption{Comparison of three different circuit designs for Kerr-cat qubit. $\Omega_{\rm d}$ is the periodic drive strength and $g_m$ is the $m$-th order nonlinearity coefficient in the SNAIL Hamiltonian \cite{frattini2022}. We compare only the leading order contributions. The lifetime $T_\alpha$ was calculated for $\epsilon_2/K=8$ (for STS and SNAIL) and $K/h=14.4{\rm MHz}$ (see Fig.~\ref{fig:snailvssquid}). The two-photon drive strength for the single SQUID, $\tilde{\epsilon}=-\epsilon_2/\sqrt{2}$.}
\label{tab:diff_design}
\end{table}
\end{center}
The two notable differences between the single SQUID Kerr-cat Hamiltonian and the STS Hamiltonian in Eq.~(\ref{eq:ham_main}) is the negative sign in front of the two-photon drive strength $\tilde{\epsilon}_2$ and the final quartic term in $\hat{a}$ and $\hat{a}^\dagger$ which is absent in the STS case. However, $\tilde{\Theta}$ is proportional to $\sim \varphi_{\rm zps}^4/100$, giving a negligible contribution for a sufficiently diluted Kerr coefficient. Comparing the Kerr and two-photon driving strength with the STS case, we find $\tilde{K} = K$ and $\tilde{\epsilon}_2 = -\epsilon_2/\sqrt{2}$ up to the leading order in zero point phase spread. Note that when all the parameters are considered the same, the Kerr coefficient for the SQUID is the same as the Kerr coefficient for the STS, whereas the two-photon drive for the same value of modulation depth is largely reduced compared to the STS. This results in a reduced $T_\alpha$ for SQUID in comparison to the STS (see Fig.~\ref{fig:single_vs_double} in Appendix~\ref{app:squid}). 

In Table~\ref{tab:diff_design}, we compare different properties of the three different proposed designs of the Kerr-cat qubit, namely the SQUID, SNAIL, and STS designs. Compared to single-SQUID designs, the STS exhibits two key advantages: first, its dedicated middle branch enables independent control of Kerr dilution, which single SQUID lacks; second, the STS operates effectively at the sensitive flux bias point of $\phi_{\rm e} = \pi/2$, whereas single SQUIDs cannot function optimally near $\phi_{\rm e} = \pi/2$ and must instead operate at alternative flux points. This constraint in single SQUID reduces its two-photon driving efficiency under comparable flux modulation, limiting its performance. Further, note that the two-photon dissipation, which enters at ${\cal O}(\varphi_{\rm zps}^1)$ for the SNAIL design, only gets introduced at ${\cal O}(\varphi_{\rm zps}^3)$ for SQUID-based designs. This leads to robust $T_\alpha$ for the STS design even in the high Kerr limit compared to the SNAIL (see Appendix~\ref{app:snail_vs_squid} for details). Together, these features position the STS as a promising platform for high-fidelity Kerr-cat qubit operations.

The STS circuit architecture offers a simplified flux-pumped design that addresses critical challenges in Kerr-cat qubit operation. By engineering the flux operator of the drive Hamiltonian to generate only even-order harmonics, the STS suppresses two-photon dissipation (see Sec.~\ref{sec:lifetime_resonant}), a dominant loss mechanism in high-Kerr regimes, while enabling access to the strongly non-linear regime for faster gate operations. Furthermore, the STS’s middle branch facilitates Kerr coefficient dilution through multiple junctions, achieving a weak non-linear oscillator as required for qubit operation, without diminishing the two-photon drive strength. This satisfies the criteria outlined in Ref.~\cite{hua2024engineering} for practical applications: (1) a static Hamiltonian with diluted Kerr nonlinearity and (2) a drive Hamiltonian restricted to even harmonics, ensuring robust two-photon driving without dissipation.

\section{Kerr-cat Noise Model}
\label{sec:master_equation}
In this section, we consider an open quantum system to study the lifetime of the coherent states of the STS Kerr-cat qubit in the presence of an external environment. The environment is considered to be a macroscopic system at thermal equilibrium with temperature $T$ composed of a bath of linear oscillators with continuous modes. The Hamiltonian for the bath is given by
\begin{equation}
\hat{H}_{\rm B} = \sum_j \hbar \omega_{j}\hat{b}_j^\dagger \hat{b}_j,
\end{equation}
where $\hat{b}_j$ and $\hat{b}_j^\dagger$ are, respectively, the annihilation and creation operators of an excitation of energy $\hbar \omega_j$ in the environment. The system environment coupling Hamiltonian in the rotating frame takes the following form
\begin{equation}
\hat{H}_{\rm SB}(t)=i\left(\hat{a}e^{-i\omega_dt/2}-\hat{a}^\dagger e^{i\omega_dt/2}\right)\hat{B}(t),
\end{equation}
where $\hat{ B}(t)=\sum_j ih_j\left(\hat{b}_je^{-i\omega_jt}-\hat{b}_j^\dagger e^{i\omega_jt}\right)$ and $h_j$ gives the coupling strength.

For a system that is weakly coupled to a thermal environment with fast dynamics such that any excitation in the environment induced by the system is quickly carried away, the dynamics of the system are given by the Gorini-Kossakowski-Sudarshan-Lindblad (GKLS) master equation,
\begin{equation}
\frac{d\hat{\rho}_{\rm S}}{dt}=-\frac{i}{\hbar}\left[\hat{\cal H}_{\rm S},\hat{\rho}_{\rm S}\right]+\sum_l \gamma_l{\cal D}[\hat{ O}_l]\hat{\rho}_{\rm S},
\label{eq:lin_general}
\end{equation}
where ${\cal D}[\hat{ O}_l]\hat{\rho}_{\rm S} = \hat{O}_l\hat{\rho}_{\rm S} \hat{O}_l^\dagger -\left(\hat{ O}_l^\dagger \hat{ O}_l\hat{\rho}_{\rm S} + \hat{\rho}_{\rm S} \hat{O}_l^\dagger\hat{ O}_l\right)/2$ for the jump operator $\hat{ O}_l$ and $\gamma_l$ is the corresponding transition rate. The first term in the right-hand side of Eq.~(\ref{eq:lin_general}) gives the unitary dynamics due to the system Hamiltonian, and the second term gives the dissipation introduced by the environment. 

In the frame introduced by the Schrieffer-Wolff transformation generated by $\hat{S}(t)$, although the system Hamiltonian is static, the transformed system environment Hamiltonian would not necessarily be static. The transformed system environment coupling up to ${\cal O}(\varphi_{\rm zps}^2)$ is given by (see Appendix~\ref{app:master} for details)
\begin{multline}
    \hat{\mathcal{H}}_{\rm SB}^{(2)}(t)= \hat{H}_{\rm SB}(t) + \frac{iG_{\rm 2,S}}{\hbar\omega_{\rm d}}\bigg\{3(-\hat{a}^\dagger e^{i3\omega_{\rm d}t/2}+\hat{a} e^{-i3\omega_{\rm d}t/2})\\
    +2(-\hat{a} e^{i\omega_{\rm d}t/2}+\hat{a}^\dagger e^{-i\omega_{\rm d}t/2})\bigg\}\hat{B}(t),
\end{multline}
where $G_{\rm 2,S}=\delta\varphi E_{\rm J\Sigma}\varphi_{\rm zps}^2/2$. The corresponding GKLS master equation is given by
\begin{multline}
    \frac{\partial\hat{\tilde{\rho}}_{\rm S}(t)}{\partial t}= \bigg\{\gamma(\omega_{\rm d}/2)\mathcal{D}[\hat{\bar{a}}^\dagger]+\Upsilon(\omega_{\rm d}/2) \mathcal{D}[\hat{\bar{a}}]\bigg\}\hat{\rho}_{\rm S}(t)\\
    +\bigg(\frac{3G_{\rm 2,S}}{\hbar\omega_{\rm d}}\bigg)^2\bigg\{ \gamma(3\omega_{\rm d}/2)\mathcal{D}[\hat{a}^\dagger]+\Upsilon(3\omega_{\rm d}/2)\mathcal{D}[\hat{a}]\bigg\}\hat{\rho}_{\rm S}(t),
    \label{eq:master2}
\end{multline}
where $\frac{\partial\hat{\tilde{\rho}}_{\rm S}(t)}{\partial t} = \frac{\partial\hat{\rho}_{\rm S}(t)}{\partial t} - \frac{1}{i\hbar}[\hat{\mathcal{H}}_{\rm S},\hat{\rho}_{\rm S}(t)]$ and $\hat{\bar{a}} =\hat{a}+\frac{2G_{\rm 2,S}}{\hbar\omega_d}\hat{a}^\dagger$. $\gamma(\omega) = \hbar^{-1}\kappa(\omega) n(\omega)$ and $\Upsilon(\omega) = \hbar^{-1}\kappa(\omega) (1+n(\omega))$ are the incoming (from the environment to the qubit) and the outgoing (from the qubit to the environment) transition rates, respectively. $\kappa(\omega) = 2\pi \hbar^{-1}\sum_j |h_j|^2\delta(\omega-\omega_j)$ is the spectral density and $n(\omega) = [e^{\hbar\omega/k_{\rm B}T}-1]^{-1}$ is the Bose-Einstein distribution function of the environment. The system-environment coupling and the master equation up to ${\cal O}(\varphi_{\rm zps}^4)$ are calculated in Appendix~\ref{app:master}.

The first line on the right-hand side of Eq.~(\ref{eq:master2}) gives the master equation under the rotating wave approximation (RWA) for $\hat{\bar{a}}\rightarrow \hat{a}$. Unlike the SNAIL case \cite{venkatraman2022}, we observe that up to the leading order beyond RWA the master equation contains only single-photon effects. For symmetric SQUID junctions, we do not observe two-photon heating and cooling effects, even at higher orders of $\varphi_{\rm zps}$. The next order terms include three-photon effects at ${\cal O}(\varphi_{\rm zps}^4)$ (see Appendix~\ref{app:master}). However, the asymmetry in SQUID junctions can result in two-photon heating and cooling effects, which we discuss in more detail in Appendix \ref{app:master}. Note that we undergo several approximations to derive the GKLS master equation: 1) Born approximation which demands weak system-environment coupling, i.e. $V_k \ll \epsilon_c\varphi_{\rm zps}^2$, 2) Fast bath dynamics compared to the system relaxation time (Markov approximation), and 3) weak zero point phase spread $(\varphi_{\rm zps})$ and modulation depth $(\delta\phi)$. To undergo perturbation around the RWA, all the parameters in the lab Hamiltonian (expressed in the rotating frame) should be smaller than the drive frequency. Since the drive frequency $\omega_{\rm d}\approx 2\epsilon_{\rm c}/\hbar$ is fixed, this will imply that the RWA, which is ${\cal O}(\varphi_{\rm zps}^{0})$, will be valid for $\varphi_{\rm zps}\ll 1$. More details regarding the limits set by $\varphi_{\rm zps}$ and $\delta\phi$ are given in Appendix \ref{app:bounds}. The derived master equation in this section only includes the linear tunneling Hamiltonian and does not account for dephasing. We can add a term proportional to the number operator in the Lindbladian to study the effect of dephasing (see Appendix.~\ref{app:dephasing}).

With symmetric SQUIDs, we only get terms proportional to even powers of the $\varphi_{\rm zps}$ in both the static effective Hamiltonian and the master equation. However, when the asymmetry in the SQUID junctions is taken into account, all orders in $\varphi_{\rm zps}$ contribute. This introduces extra terms in the static effective Hamiltonian and two-photon heating and cooling effects in the master equation, which are absent in the symmetric case (see the ${\cal O}(\varphi_{\rm zps}^3)$ master equation in Appendix~\ref{app:master}). We find that the multi-photon dissipative effects due to $5\%$ asymmetry in the SQUID junction result in a negligible reduction of $T_\alpha$ for $K/h< 10 {\rm MHz}$. Further, two-photon drive strength ($\epsilon_2/K$) depends on both the Kerr coefficient and the modulation depth ($\delta\phi$). In the high Kerr coefficient limit, stronger modulation is required to obtain large enough two-photon drive strength; the first order approximation, $\sin\left(\delta\phi \cos(\omega_{\rm d}t)\right)\approx \delta\phi \cos(\omega_{\rm d}t)$, breaks down and one has to consider higher order terms in the modulation depth, $\delta\phi$. We do the calculation for stronger modulation depth in Appendix~\ref{app:strong}. The result is a longer first plateau in $T_\alpha$ plots, which we show in more detail in Appendix \ref{app:strong}.
\ \\
\subsection{Lifetime of Resonant Kerr-cat Qubit}
\label{sec:lifetime_resonant}

\begin{figure}[t]
\includegraphics[width=\columnwidth]{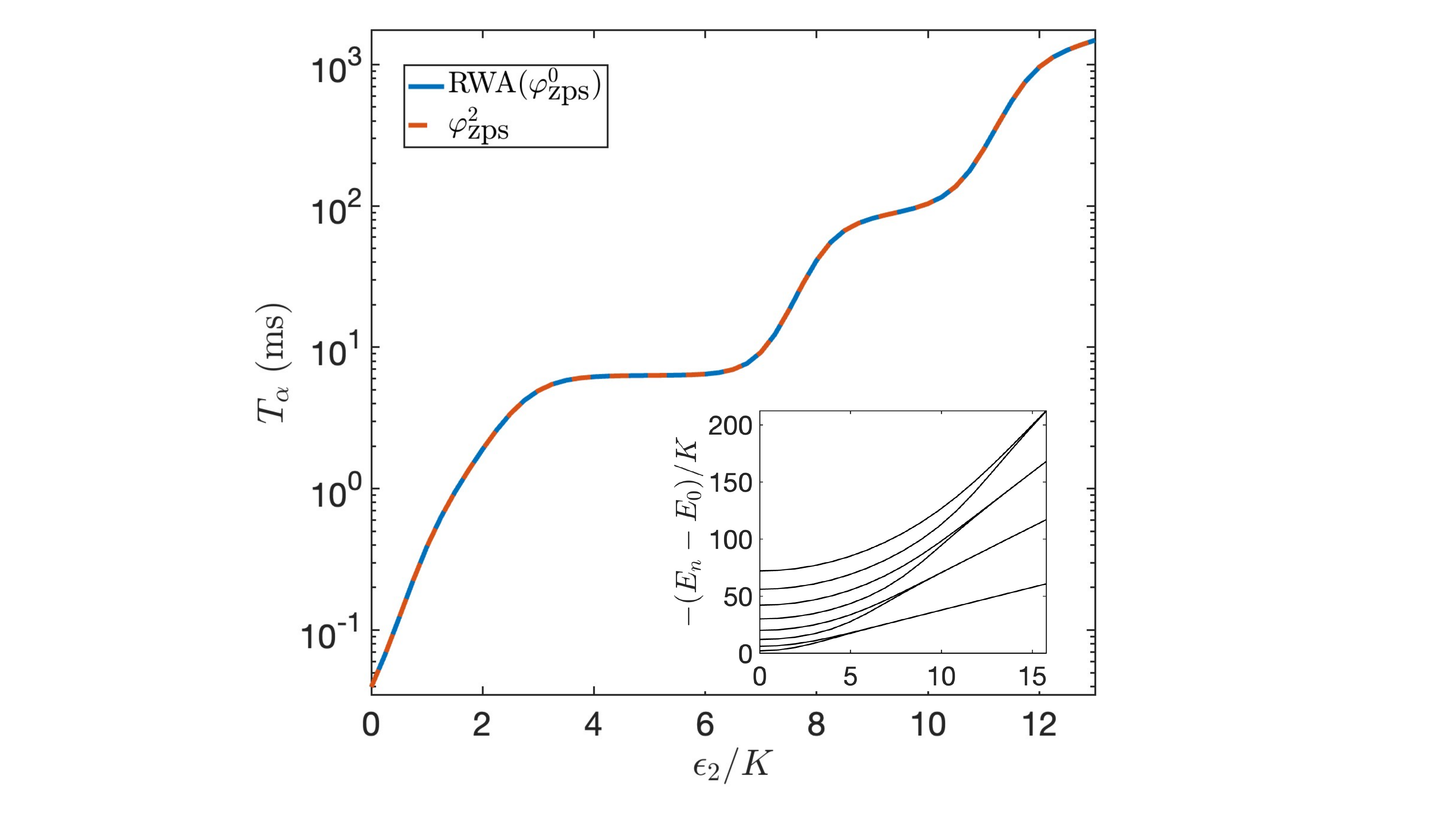}
 \caption{$T_\alpha$ of the Kerr-cat qubit as a function of the two-photon drive strength $\epsilon_2/K$ for 10 STS connected in series with a single junction in each branch setting the Kerr nonlinearity to $K/h=1.25\textrm{~MHz}$. The RWA calculation, which takes the system environment coupling up to order $\varphi_{\rm zps}^{0}$ and keeps only single-photon processes in the master equation, is given by the solid blue line. The dashed orange curve gives the leading order beyond RWA ${\cal O}(\varphi_{\rm zps}^{2})$ calculation and keeps multiphoton processes in the master equation. In the inset, we plot the energy spectra of the Kerr-cat qubit as a function of the two-photon drive strength $\epsilon_2/K$.} 
\label{fig:lifetime_phizps}
\end{figure}

 The staircase type pattern as a function of the two-photon drive strength is the most remarkable feature of the $T_\alpha$ plot in Fig.~(\ref{fig:lifetime_phizps}). This pattern has been analyzed and explained in detail in Refs.~\cite{ venkatraman2022,frattini2022}. For $\epsilon_2/K\rightarrow 0$, $T_\alpha$ of the Kerr-cat qubit is effectively given by the lifetime of the coherent superposition, $(\left(\ket{0}\pm \ket{1}\right)/\sqrt{2})$, where $\ket{0}(\ket{1})$  are the zeroth Fock (first Fock) state of the Kerr nonlinear oscillator. The Kerr nonlinearity, along with the two-photon drive, creates a double-well potential in the phase space (see Fig.~\ref{fig:phase_deg} (a) and (b)). The exponential growth of $T_\alpha$ with increasing two-photon drive strength $(\epsilon_2/K)$ can be attributed to the inclusion of energy levels inside the double-well meta-potential. The number of bound states inside the double-well potential is approximately given by ${\cal N}=\epsilon_2/\pi K$ \cite{frattini2022} with a new pair of excited states entering the potential well every time ${\cal N}$ takes an integer value. As shown in the inset of Fig.~(\ref{fig:lifetime_phizps}), the splitting between the energy levels entering the potential well becomes smaller with increasing $\epsilon_2$. Consequently, the quantum tunneling between the two levels inside the double-well gets suppressed with decreasing splitting and becomes maximally protected from tunneling for degenerate states. This leads to exponential growth in lifetime as a function of the two-photon drive strength. However, in order to explain the plateau in Fig.~(\ref{fig:lifetime_phizps}), we would have to consider the effect of the environment. $T_\alpha$ increases as a function of decreasing splitting until the rate of dissipation ($\gamma$) to the environment overcomes it. The tunneling between the two nearly degenerate states and the dissipation to the environment compete with each other, resulting in lifetime saturation at $T_\alpha = (\gamma n_{\rm th})^{-1}$ when only single photon heating and cooling effects are considered. As long as one takes the RWA, the plateau occurs at the same lifetime ($T_\alpha$). However, non-RWA terms can result in the lowering of the lifetime plateau. The second exponential growth starts when the next set of excited states fall into the meta-potential, and the splitting between them decreases as a function of the two-photon drive strength. Unlike the SNAIL qubit, we observe that the STS qubit is well approximated in the RWA regime \cite{venkatraman2022}. This is because the STS master equation does not contain terms on the order of $\varphi_{\rm zps}^1$ for symmetric junctions. Therefore, the next largest contribution is on the order of $\varphi_{\rm zps}^2$, which is much smaller than the terms that survive RWA. Hence, the multi-photon heating and cooling effect is largely suppressed in the STS Kerr-cat qubits even when asymmetry between the Josephson energies is properly accounted for. For the simulation parameters, we take the coupling to the external environment, $\gamma(\omega_{\rm d}/2)/h=\gamma(3\omega_{\rm d}/2)/h=8~{\rm kHz}$. For the temperature of the environment, we choose $T_{\omega_{\rm d}/2}=T_{3\omega_{\rm d}/2}=50\textrm { mK}$. Moreover, the Josephson energy for the transmon and SQUID junctions are considered to be of similar magnitude. The Josephson and capacitive energy are given by $E_{\rm J}/h=80 \textrm{ GHz}$ and $E_{\rm C}/h=250\textrm{ MHz}$ respectively, and consequently the Kerr coefficient for a single junction is $K/h=125~{\rm MHz}$. Further, the drive frequency is given by $\omega_{\rm d}=24\pi\textrm{ GHz}$. Unless mentioned otherwise, we consider the detuning, $\Delta/K=0$. These parameters were chosen to minimize decoherence in the static circuit~\cite{wang2022towards,ahmed,frattini2022}.
 
\begin{figure}[t]
\includegraphics[width=\columnwidth]{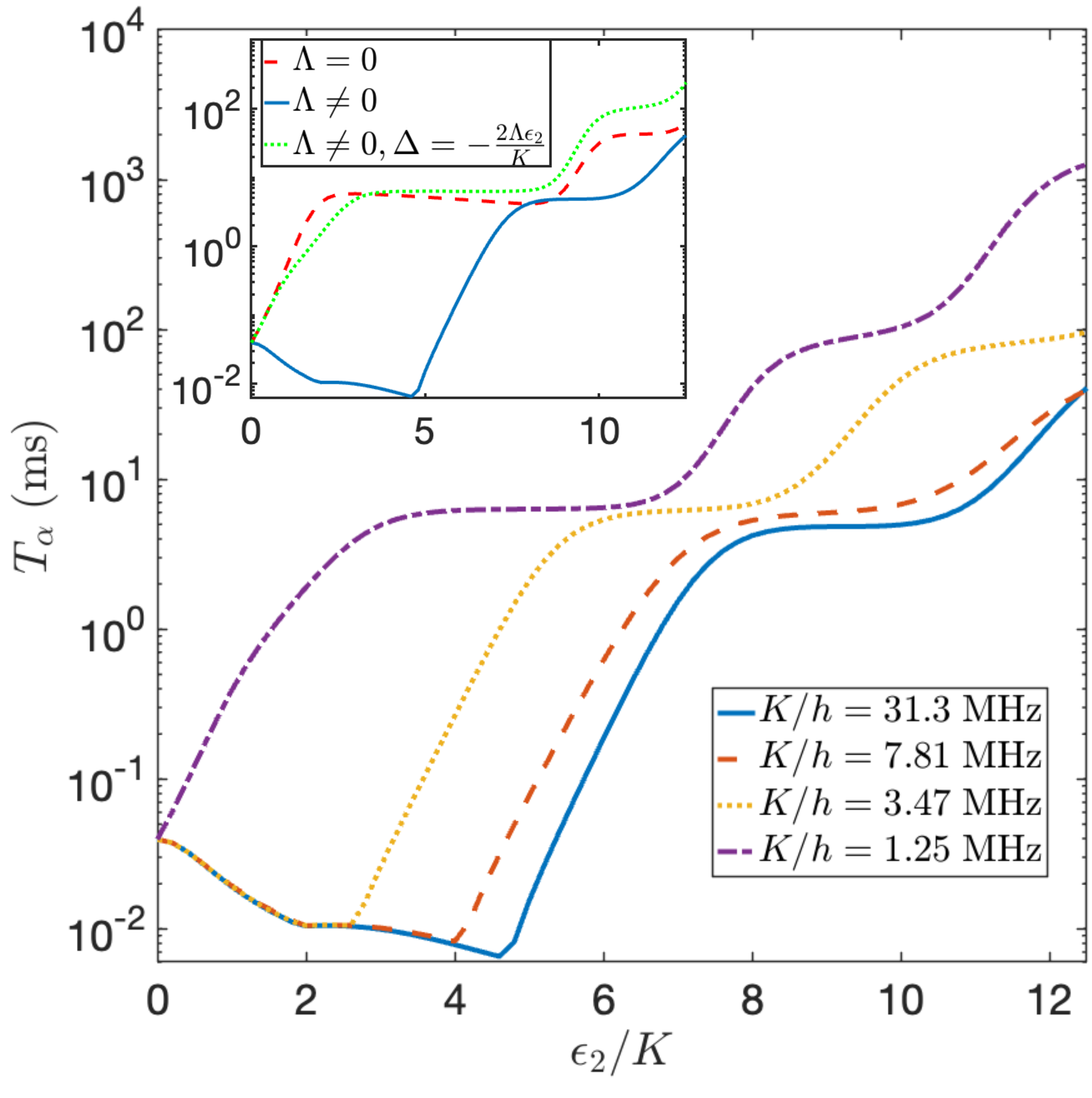}
 \caption{$T_\alpha$ of the Kerr-cat qubit as a function of the two-photon drive strength $\epsilon_2/K$ for different values of Kerr nonlinearity. In order to obtain the mentioned Kerr coefficient, we took $M=2,N=2$ (solid blue curve), $M=2,N=4$ (dashed orange curve) and $M=3,N=6$ (dotted yellow curve) and $M=10,N=10$ (dot-dashed purple curve). In the inset, we show how the initial dip in the lifetime observed for $K/h = 31.3 {\rm MHz}$ can be actively canceled by introducing a drive-dependent detuning.} 
\label{fig:lifetime_kerr2}
\end{figure}
 
The main difference between the usual Kerr-cat Hamiltonian \cite{puri2017} and the static effective Hamiltonian in Eq.~(\ref{eq:ham_main}) is the presence of the last term, $
\Lambda (\hat{a}^{\dagger} \hat{a}^3+\hat{a}^{\dagger^3}\hat{a})$. The most dominant contribution to $\Lambda$ is ${\cal O}(\varphi_{\rm zps}^4)$, hence it is significantly smaller for $E_{\rm C}\ll E_{\rm J2}$. However, for larger values of Kerr coefficient, which is also ${\cal O}(\varphi_{\rm zps}^4)$, $\Lambda$ becomes significant.  The extra term proportional to $\Lambda$ is responsible for the observed dip in $T_\alpha$ for $K/h=31.3 {~\rm MHz}$ (solid blue curve in Fig.~\ref{fig:lifetime_kerr2} where we consider only two junctions in the transmon branch). The average effect of this term is to introduce an additional detuning given by $\Lambda \left(\hat{a}^{\dagger} \hat{a}^3+\hat{a}^{\dagger^3}\hat{a}\right)\approx \left(2\Lambda \epsilon_2/K\right) \hat{a}^\dagger \hat{a}$. This additional detuning takes the Kerr-cat qubit away from the sweet spot where the detuning is an even multiple of the Kerr coefficient. Apart from the dip in $T_\alpha$, we also observe a decrease in the plateau height for $K/h = 31.3 {~\rm MHz}$, which results from the non-RWA terms in the master equation. Increasing the Kerr coefficient eventually increases the effect of higher order in $\varphi_{\rm zps}$ terms. The above behavior motivates us to consider the case where the Kerr coefficient can be significantly reduced, hence mitigating the unwanted contribution from the extra term in the STS Hamiltonian and non-RWA multi-photon effects. One possible way, as suggested in the Sec.~\ref{sec:ham}, is to consider the case with multiple junctions instead of a single junction in the transmon branch. This leads to the dilution of Kerr coefficient by $K\rightarrow K/(M\tilde{N})^2$, where $\tilde{N}$ is defined through $N=M\tilde{N}$. $M$ is the number of STS in series, and $\tilde{N}$ is the number of junctions in the transmon branch of each STS (see Appendix~\ref{app:dilution} for details). In Fig.~\ref{fig:lifetime_kerr2}, the smaller Kerr coefficients are obtained by considering different numbers of STS in series, namely $M=2$ (dashed orange curve), $M=3$ (dotted yellow curve), and $M=10$ (dot-dashed purple curve). We find that increasing the number of STS in series decreases the value of $\epsilon_2/K$ required for initiating the exponential growth of lifetime ($\epsilon_2/K\approx 4.5$ for $M=2$, whereas $\epsilon_2/K\approx 2.5$ for $M=3$). Further, with sufficient number of STS which ensures sufficient reduction in the Kerr coefficient (from $K/h = 31.3 {\rm ~MHz}$ for $M=N=2$ to $K/h = 1.25 {\rm ~MHz}$ for $M=N=10$), one can get rid of the dip in $T_\alpha$ as well as the lowering of the plateau altogether. We further observe that although the $T_\alpha$ curve still has a dip for $K/h = 3.47 {\rm ~MHz}$ (dotted yellow curve), the plateau height has been restored to the level of the purple curve. This implies that the multi-photon dissipative effects decay faster as a function of decreasing Kerr coefficient compared to the additional detuning introduced by $\Lambda$.

We find that one has to increase the number of STS to almost 10 to eliminate the dip in the $T_\alpha$ due to the extra term in the STS Kerr-cat Hamiltonian. Experimentally, having so many STS in a series will be challenging. Instead of adding STS, one can think of actively canceling the effect of the extra term. In the inset of Fig.~\ref{fig:lifetime_kerr2}, we propose a strategy to cancel the dip in the $T_\alpha$ for $M=2,N=2$ case (solid blue curve in the main plot). The idea is to add an extra drive-dependent detuning in the circuit which is opposite in sign compared to the extra term proportional to $\Lambda$ which results in a STS Kerr-cat Hamiltonian of Eq.~(\ref{eq:ham_main}) with $\Delta = -2\Lambda\epsilon_2/K$ since $\Lambda \left(\hat{a}^{\dagger} \hat{a}^3+\hat{a}^{\dagger^3}\hat{a}\right)\approx \left(2\Lambda \epsilon_2/K\right) \hat{a}^\dagger \hat{a}$. We find that with this additional drive-dependent detuning, the $T_\alpha$ plot for $M=2,N=2$ (green dotted curve) has no initial dip and almost overlaps the plot done for $\Lambda=0$ (dashed red curve), recovering the staircase-type pattern observed for low Kerr coefficient. An alternative approach to reducing the Kerr nonlinearity could involve lowering $E_{\rm 
C}$ rather than employing multiple STS circuits connected in series. However, reducing $E_{\rm C}$ also reduces the oscillator gap ($\epsilon_c = \sqrt{8E_{\rm C}E_{\rm J}}$) and the two-photon drive strength $\epsilon_2 = \delta\phi/2 (\sqrt{2E_{\rm C}E_{\rm J}}-E_{\rm C})$, necessitating stronger modulation depth to achieve comparable two-photon drive strength. There are three key issues associated with stronger modulation depth: 1) the weak modulation depth approximation used to derive the static effective Hamiltonian and the master equation breaks down, 2) including higher order in modulation depth terms leads to the suppression of the lifetime $T_\alpha$ (see Appendix~\ref{app:strong}), and 3) the dynamical dephasing rate is directly proportional to the modulation depth leading to the further suppression of the lifetime (see Eq.~(\ref{eq:mas_eq_deph}) and Appendix~(\ref{app:dephasing})). By contrast, employing STS with multiple junctions in the middle branch or multiple STS in series reduces the Kerr coefficient without affecting the two-photon drive strength.

Using the master equation framework, robust agreement with experimental data for $T_\alpha$ as a function of $\epsilon_2$ was shown in Ref.~\cite{ahmed,qing2024benchmarking} for SNAIL-based design by systematically accounting for distinct decoherence channels. While standard approximations, such as static thermal distributions and weak system-environment coupling with energy-independent tunneling rates, hold for weak drives, deviations emerge at stronger two-photon drive strengths or detuning. This necessitates dynamic adjustments to the thermal population and energy-dependent couplings. Notably, it was observed that single-photon dissipation alone fails to capture key experimental features: inclined plateaus in $T_\alpha$, overshooting lifetimes, and delayed rise in $T_\alpha$. Although including two-photon heating and dephasing effects led to a better agreement with the lifetime, it also led to a substantial decrease in the lifetime, and we predict a further decrease for higher Kerr coefficient (See Figure~\ref{fig:snailvssquid} and Appendix~\ref{app:snail_vs_squid}). In contrast, the STS architecture suppresses these losses through its symmetric circuit design: two-photon dissipation enters at ${\cal O}(\varphi_{\rm zps}^3)$ (vs ${\cal O}(\varphi_{\rm zps}^1)$ in SNAILs) and is proportional to junction asymmetry, and dynamical dephasing enters at ${\cal O}(\varphi_{\rm zps}^4)$ (vs ${\cal O}(\varphi_{\rm zps}^2)$ for SNAILs), enabling robust operation in high-Kerr regimes. This hierarchy explains why STS-based designs are projected to outperform SNAILs for $K/h\sim 10 $ MHz, achieving $T_\alpha \sim 10 $ ms.

\begin{figure}[t]
\includegraphics[width=\columnwidth]{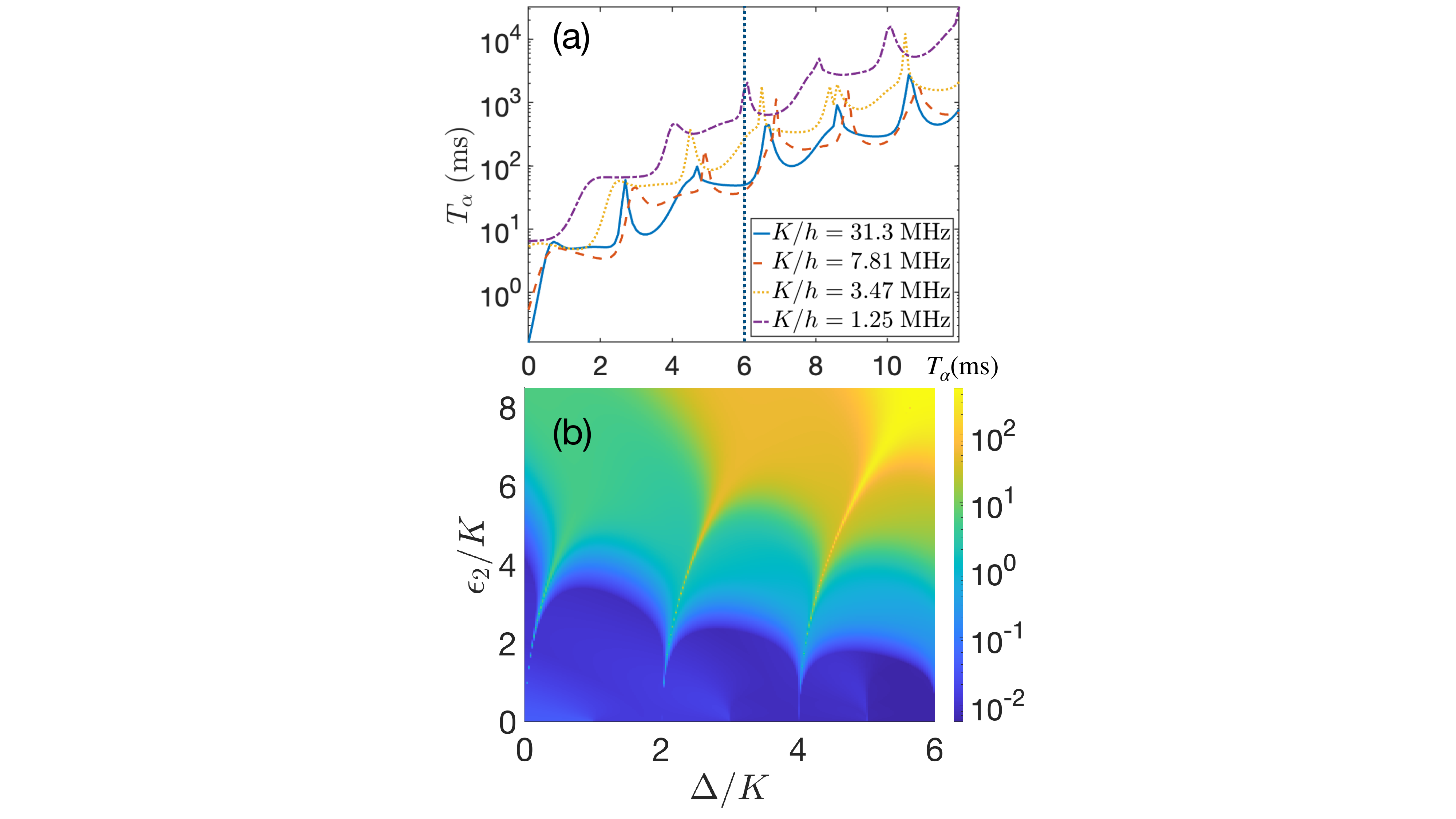}
 \caption{$T_\alpha$ of the Kerr-cat qubit (a) as a function of $\Delta/K$ for $\epsilon_2/K=6$ and for different values of Kerr coefficient (same as Fig.~\ref{fig:lifetime_kerr2}) and (b) as a function of both $\epsilon_2/K$ and $\Delta/K$ for $M=2,N=4\,(K/h = 7.81 {\rm MHz})$.} 
\label{fig:lifetime_3D}
\end{figure}
\ \\
\subsection{Lifetime of Detuned Kerr-cat Qubit}
\label{sec:lifetime_detuned}
For a small Kerr coefficient, we observed that $T_\alpha$ of the STS Kerr-cat qubit depends only on the two-photon drive strength, i.e., the ratio $\epsilon_2/K$. We further observed that as the two-photon drive strength increases, the number of bound states inside the double-well and the number of degenerate states increase, leading to enhancement of $T_\alpha$. However, so far, we have not looked into the effect of detuning ($\Delta$) on $T_\alpha$ of the Kerr-cat qubit. In Fig.~\ref{fig:lifetime_3D}(a), we study the lifetime of a detuned Kerr-cat qubit given by the Hamiltonian in Eq.~(\ref{eq:Kerrcat}). As mentioned in Sec.~\ref{sec:ham}, the inter-well tunneling through the saddle points destructively interferes, giving $m+1$ degeneracies for $\Delta /K= 2m$. These degeneracies lead to a sharp spike in the $T_\alpha$ plots for even integer values of $\Delta/K$ as shown in Fig.~\ref{fig:lifetime_3D}(a) \cite{jaya2023}. Note that the spikes are shifted a bit from the even integer values of $\Delta/K$ for larger values of the Kerr coefficient (follow the black dashed line), where the extra detuning introduced by $\Lambda$ is stronger. This shift results from the shift in degeneracy points in the energy spectra for finite values of $\Lambda$ (see Fig.~\ref{fig:phase_deg}(d)). In Fig.~\ref{fig:lifetime_3D}(b), we plot $T_\alpha$ as a function of both the detuning $\Delta/K$ and the two-photon drive strength $\epsilon_2$. In Ref.~\cite{jaya2023}, it was shown that the number of excited states entering the double-well increases faster when both $\epsilon_2/K$ and $\Delta/K$ are increased instead of just the two-photon drive strength. We observe similar features in Fig.~\ref{fig:lifetime_3D}(b). One can analyze the lifetime plot fixing $\Delta/K$ to 1 and 6 and observe that lifetime increases significantly as a function of $\epsilon_2/K$ for $\Delta/K=6$. Also note the sharp spikes in the lifetime around even integer values of $\Delta/K$ as observed in Fig.~\ref{fig:lifetime_3D}(a).  

The lifetime peak as a function of detuning was experimentally observed and theoretically modeled for SNAIL Kerr-cat qubits using master equations in Ref. \cite{qing2024benchmarking}. While we assumed energy-independent spectral densities (fixed photon loss rates) for simplicity, this framework could only reproduce one peak in the experimental $T_\alpha(\Delta)$ data. This limitation stems from the spectral density’s dependence on the drive frequency $\omega_{\rm d}$, which is related to detuning via $\Delta \approx \epsilon_{\rm c}-\hbar \omega_{\rm d}/2 $. This suggests energy-dependent spectral densities would be required to properly address the lifetime as a function of detuning.

\section{Qubit Initialization}
\label{sec:qbt_ini}
Our observations reveal that detuned Kerr-cat qubits exhibit significantly enhanced 
$T_\alpha$ compared to their resonant counterparts, making them ideal candidates for quantum operations. However, a critical challenge arises in initializing detuned Kerr-cat qubits. For resonant Kerr-cat qubits, the cat states emerge as eigenstates under sufficiently strong two-photon driving ($\epsilon_2$), and initialization can be achieved by adiabatically ramping $\epsilon_2$ from zero to a finite value. The adiabatic ramping maps the Fock states $\ket{0}$ and $\ket{1}$ to the eigenstates, i.e., the cat states, of the resonant Kerr-cat Hamiltonian.  This method fails for detuned Kerr-cat qubits because the computational basis states ($\ket{0}$ and $\ket{1}$) are no longer the ground states of the Fock Hamiltonian, despite the cat states remaining the ground states of the detuned Kerr-cat Hamiltonian. This prevents mapping the $\ket{0}$ and $\ket{1}$ states to the cat states through adiabatic ramping of $\epsilon_2$ in detuned Kerr-cat qubits. 
In this section, we will discuss how to initialize the detuned Kerr-cat qubits using single photon dissipation and readout.

\begin{figure}[t]
\includegraphics[width=\columnwidth]{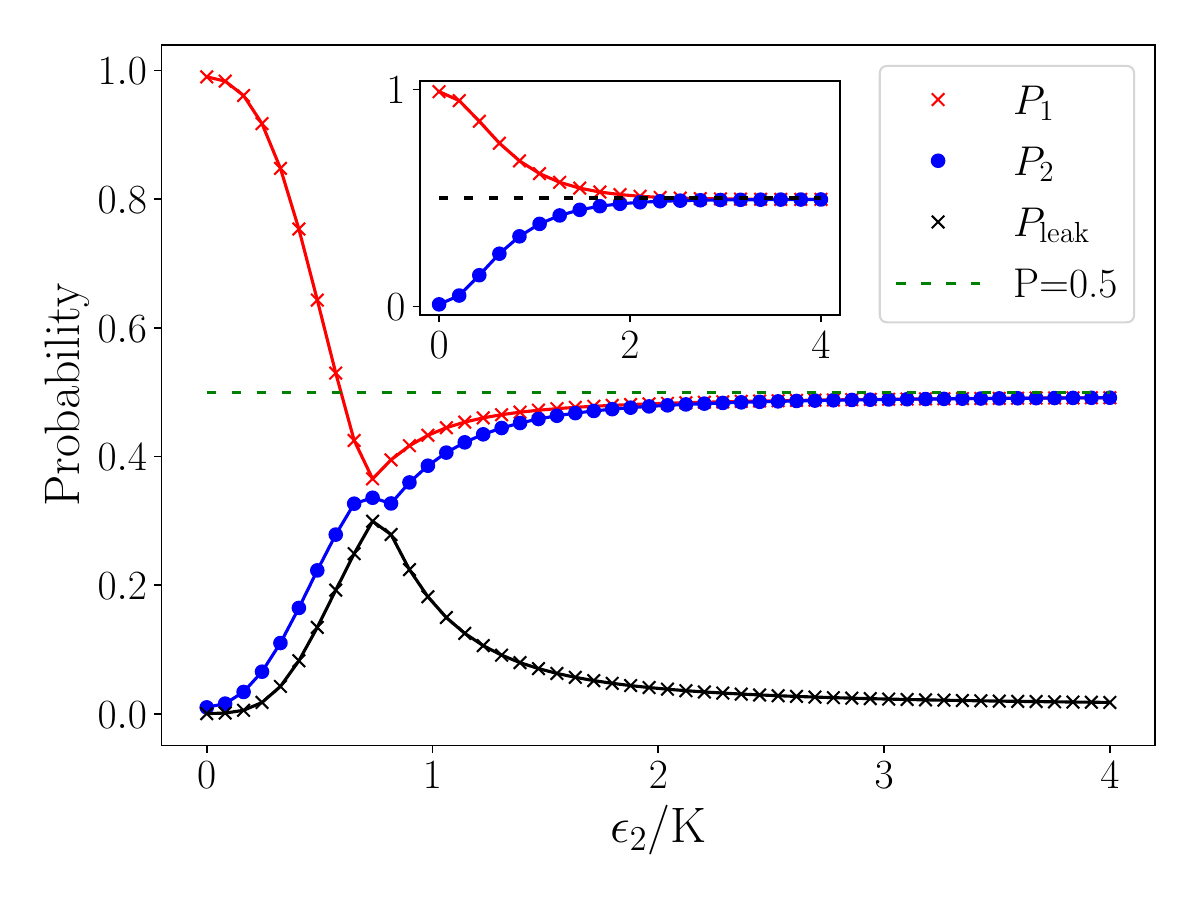}
 \caption{The spectrum of the steady state density matrix (red curve with cross and blue curve with dots for the first two eigenstates) as a function of the two-photon drive strength $\epsilon_2$ for detuning $\Delta/K = 2.1$. The black curve with crosses gives the leakage probability out of the first two eigenstates, $P_{\rm leak} =  1- P_{1}-P_{2}$. In the inset, we plot the $\Delta/K = 0$ case. We considered the single photon loss rate, $\gamma/K = 0.05 $, and the Bose-Einstein distribution, $n_{\rm th} = 0.01$.} 
\label{fig:qbt_ini_prob}
\end{figure}

\begin{figure}[hbt!]
\includegraphics[width=\columnwidth]{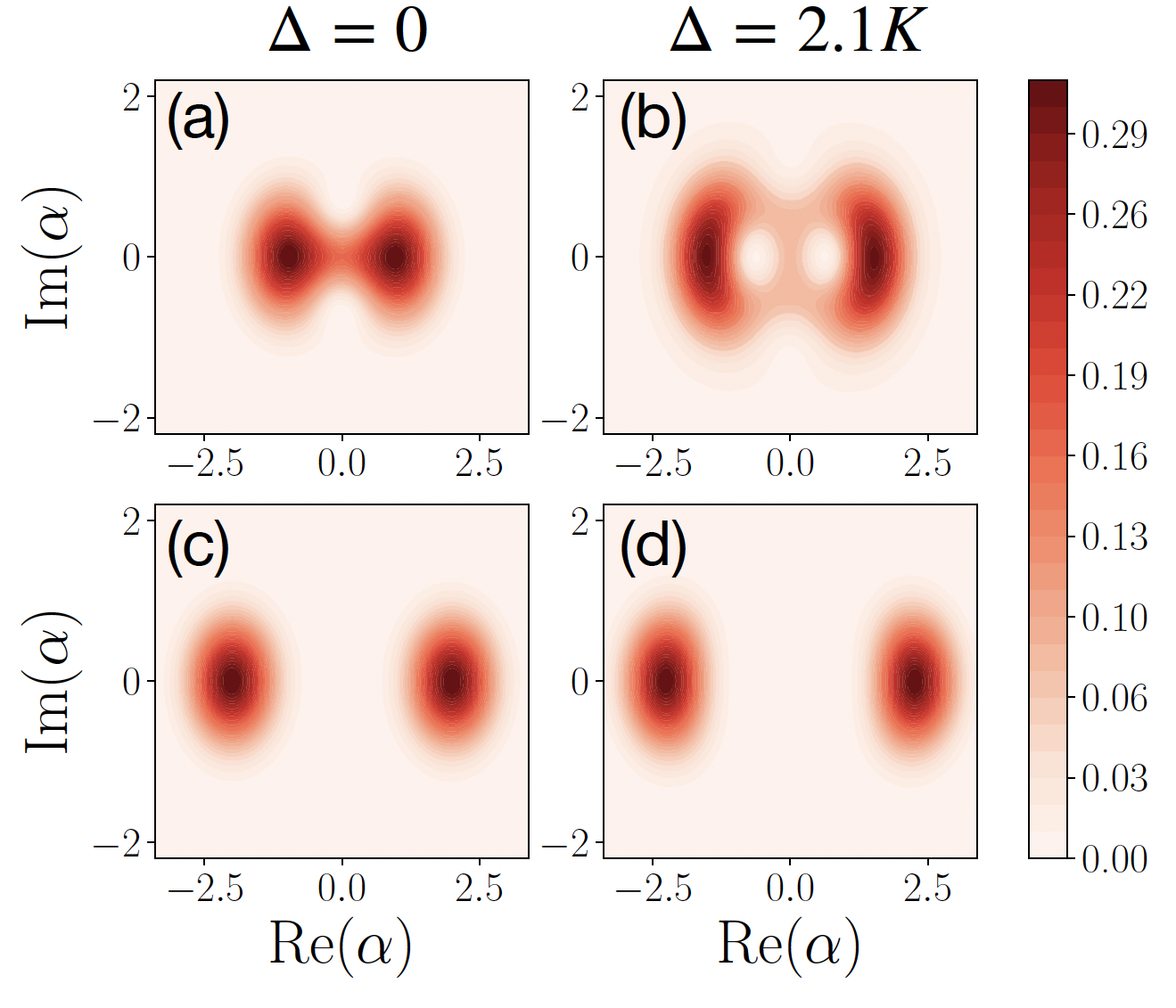}
 \caption{Wigner function plot for $\epsilon_2/K =1$ ((a) and (b)) and for $\epsilon_2/K = 4$ ((c) and (d)) for two different values of detuning. We considered the single photon loss rate, $\gamma /K= 0.05 $, and the Bose-Einstein distribution, $n_{\rm th} = 0.01$. } 
\label{fig:wigner}
\end{figure}

In Fig.~\ref{fig:qbt_ini_prob}, we plot the spectrum of the steady state density matrix of our system, $\hat{\rho}_{\rm st}=\sum_k P_k \ket{k}\bra{k}$, without detuning (the inset) and with detuning ($\Delta/K=2.1$) over a range of two-photon drive strength. In the case of no detuning, the coherent states $(|\pm\alpha\rangle)$ are the first two degenerate eigenstates of the Hamiltonian at any drive strength. However, coherent states are not orthogonal at weak drive (small cat size) limit. To stabilize the cat states and, hence, to initialize a noise-biased Kerr-cat qubit, we require a stronger two-photon drive giving orthogonal coherent states $(|\pm\alpha\rangle)$. For stronger drives, the coherent states become orthogonal as the red and blue curves converge at $P=0.5$, and the density matrix is a classical mixture of the coherent states with equal probability. Since the steady state can be expressed as a superposition of the coherent and cat states simultaneously, it is bimodal \cite{Bartolo_2017}. However, for $\Delta /K=2.1 $, the steady state density matrix has contributions beyond the first two eigenstates, causing leakage. Therefore, when we look at the weak two-photon drive regime for $\epsilon_2/K$ between 0 and $\Delta/K$, we observe a finite leakage probability $P_{\rm leak}=1-P_{1}-P_{2}$ (black curve), where $P_{1}$ (red curve) and $P_{2}$ (blue curve) are the first two eigenvalues of the steady state density matrix. These eigenstates correspond to the coherent states in the absence of detuning (the inset of Fig.~\ref{fig:qbt_ini_prob} where the leakage probability is zero even for weak two-photon drives). For $\Delta/K = 0$, all the eigenenergies increase monotonously with increasing $\epsilon_2/K$ whereas, for $\Delta /K\approx 2m$, all degenerate energy levels, other than the first two eigenenergies, show non-monotonous behavior as a function of the two-photon drive strength (see Appendix~\ref{app:detuned_leak_energy}). Further, for $\Delta/K\approx2$, the fourth level with energy $E_3$ moves much closer to $E_0$, compared to $\Delta/K=0$. The non-monotonous behavior of the degenerate energy levels, as well as the closing of the energy gap, leads to a substantial leakage out of the first two density matrix eigenstates as shown in Fig.~\ref{fig:qbt_ini_prob}.

In Fig.~\ref{fig:wigner}, we show the Wigner plots for the detuned and non-detuned qubits at different drive strengths. We can see that at weak drive, $\epsilon_2/K=1$, coherent states are not stabilized enough to be fully separated in both Figs.~\ref{fig:wigner} (a) and (b). However, in the detuned case in Fig.~\ref{fig:wigner} (b), the coherent states seem to be more distorted than the non-detuned case in Fig.~\ref{fig:wigner} (a). On the other hand, for a strong drive, $\epsilon_2/K=4$, we can see in both Figs.~\ref{fig:wigner} (c) and (d) that the coherent states are fully separated and have equal probability. This supports our previous argument in Fig.~\ref{fig:qbt_ini_prob}: for a strong enough two-photon drive and long enough relaxation time, regardless of the detuning, we get a classical mixture of coherent states $|\pm\alpha\rangle$ at equal steady-state probability.

Consequently, we can initialize the detuned Kerr-cat qubit (that satisfies the drive and relaxation criteria mentioned previously) by performing a readout along the qubit's z-axis and preparing it in the coherent state $|\pm\alpha\rangle$\cite{frattini2022}, based on the outcome of the readout. Finally, it is important to distinguish that the initialization time of a detuned Kerr-cat qubit depends strongly on the relaxation time rather than the ramp time of the two-photon drive, unlike the non-detuned case.

\section{Conclusion}
\label{sec:conc}
We presented the Symmetrically Threaded SQUID (STS) architecture as an alternate approach to realizing high-performance Kerr-cat qubits, based on a flux-pumped superconducting circuit design, compared to the traditionally considered charge-pumped SNAIL design. By engineering the flux drive to generate only even-order harmonics, we observed that the STS has the added benefit of suppressing two-photon dissipation, a dominant loss mechanism in high Kerr coefficient regimes while maintaining robust two-photon driving. Its design and symmetric flux-driving elevate multi-photon dissipation and dephasing to higher orders in zero-point phase fluctuations, ensuring resilience against decoherence even in Kerr-cat qubits with stronger Kerr nonlinearities ($\sim 10$ MHz).

While single SQUID offers a simpler flux-pumped design, we observed that it suffers from two critical limitations: (1) inability to dilute Kerr nonlinearity without compromising two-photon drive strength, and (2) incompatibility with operation at the high-sensitivity flux bias point $\phi_{\rm e} = \pi/2$. The STS overcomes these challenges via its dedicated middle branch, which enables independent Kerr dilution and supports operation at $\phi_{\rm e} = \pi/2$ through symmetric driving of the two flux loops. Thus, the STS design serves as a flux-pumped counterpart to the charge-driven SNAIL architecture, with the added advantage of enhanced resistance to two-photon heating and dynamical dephasing. These properties position the STS as an ideal platform for operating Kerr-cat qubits in high Kerr coefficient regimes. We asserted this by studying the coherent state lifetime of both resonant and detuned STS Kerr-cat qubits. We further demonstrated that detuned STS Kerr-cat qubits (with $\Delta/K \approx 2m $, $m$ being an integer) achieve significantly enhanced $T_\alpha$ compared to resonant configurations. However, detuned qubits can not be adiabatically initialized, as Fock states do not map directly to cat states via drive ramping. We addressed this by leveraging single-photon dissipation and readout along the z-axis to prepare the qubit in the coherent state. Future work will focus on experimentally validating STS-based Kerr-cat qubits, particularly their robustness against multi-photon dissipation, and exploring broader applications of the STS circuit design, including parametric amplifiers.

\section{Acknowledgement}
We thank Jayameenakshi Venkatraman, Michel Devoret, and Tathagata Karmakar for valuable discussions. This work was supported by the U. S. Army Research Office under grant W911NF-22-1-0258.

\appendix

\begin{widetext}
\section{STS - Hamiltonian}
\label{app:Hamil}
We consider the STS circuit shown in Fig.~\ref{fig:sketch_design}, where three Josephson junctions ${\rm J}_i$, $i=1,2,3$ are placed in parallel to each other. The STS is shunted by a capacitor of capacitance $C_{\rm T}$. The Josephson energy associated with the junction ${\rm J}_i$ is given by $E_{{\rm J}i}$. Moreover, $C_i$ gives the self-capacitance of the junction ${\rm J}_i$. Following Ref.~\cite{You2019}, the Hamiltonian for the circuit can be written as
\begin{multline}
    \hat{H}_{\rm lab}=4E_{\rm C}\hat{n}^2-E_{\rm J1}\cos\bigg(\hat{\varphi}+\frac{C_2+C_3}{C_\Sigma}\phi_{\rm e}+\frac{C_3}{C_\Sigma}\phi_{\rm e}^{\prime}\bigg)-\\E_{\rm J2}\cos\bigg(\hat{\varphi}-\frac{C_1}{C_\Sigma}\phi_{\rm e}+\frac{C_3}{C_\Sigma}\phi_{\rm e}^\prime\bigg)
-E_{\rm J3}\cos\bigg(\hat{\varphi}-\frac{C_1}{C_\Sigma}\phi_{\rm e}-\frac{C_1+C_2}{C_\Sigma}\phi_{\rm e}^\prime\bigg),
\end{multline}
where $C_\Sigma = C_1+C_2+C_3$ and the charging energy, $E_{\rm C} = e^2/{2C},~C = C_{\Sigma} + C_{\rm T}$. $\phi_{\rm e},~\phi_{\rm e}^\prime$, $i=1,2$ gives the external flux threading the two SQUID loops in units of the flux quantum, $\phi_0=h/2e$. The canonical variables $\hat{\varphi}$ and $\hat{n}$ are the phase and charge operators that satisfy the commutation relation, $[\hat{\varphi},\hat{n}]=i$, where $i$ is the imaginary unit. In our analysis, we will consider equal self-capacitances for each Josephson junction, i.e. $C_{1} = C_{2} = C_{3}$. The Hamiltonian reduces to
\begin{equation}
    \hat{H}_{\rm lab}=4E_{\rm C}\hat{n}^2-E_{\rm J1}\cos\bigg(\hat{\varphi}+\frac{2}{3}\phi_{\rm e}+\frac{1}{3}\phi_{\rm e}^\prime \bigg)-E_{\rm J2}\cos\bigg(\hat{\varphi}-\frac{1}{3}\phi_{\rm e}+\frac{1}{3}\phi_{\rm e}^\prime\bigg)
-E_{\rm J3}\cos\bigg(\hat{\varphi}-\frac{1}{3}\phi_{\rm e}-\frac{2}{3}\phi_{\rm e}^\prime\bigg).
\end{equation}
Taking $\phi_{\rm e} = \phi_{\rm e}^\prime = \phi_{\rm e}$ along with $E_{\rm J\Sigma}=({E_{\rm J1}+E_{\rm J3}})/{2}$ and $E_{\rm J\Delta}=({E_{\rm J1}-E_{\rm J3}})/{2}$, the Hamiltonian becomes
\begin{equation}
    \hat{H}_{\rm lab}=4E_{\rm C}\hat{n}^2-E_{\rm J2}\cos{\hat{\varphi}}-2E_{\rm J\Sigma}\cos\phi_e\cos{\hat{\varphi}}+2E_{\rm J\Delta}\sin\phi_e\sin{\hat{\varphi}},
    \label{eq:ham_double_squid}
\end{equation}
where the first two terms on the right-hand side represent a transmon and the last two terms are driven by SQUID. Hence, under the approximations considered, an STS circuit is equivalent to having a SQUID and a transmon sharing the same node. For symmetric junctions, one gets contributions that are even in the order of the phase operator contributions, whereas the junction asymmetry leads to odd orders. We take the external flux drive of the form, $\phi_e=-\frac{\pi}{2}+\delta\varphi\cos{(\omega_d t)}$, where the DC bias $-\pi/2$ is chosen so as to attain highest first-order sensitivity to the modulation depth while removing the parasitic even harmonics of the drive. The even harmonics of the drive do enter through the asymmetric SQUID term (last term in Eq.~(\ref{eq:ham_double_squid})); however, its effect can be mitigated by making the SQUID junctions as symmetric as possible. The Hamiltonian in Eq.~(\ref{eq:ham_double_squid}) reduces
\begin{equation}
\hat{H}_{\rm lab}=4E_{\rm C}\hat{n}^2-E_{\rm J2}\cos{\hat{\varphi}}-2E_{\rm J\Sigma}\sin(\delta\phi\cos{(\omega_d t)})\cos{\hat{\varphi}}
-2E_{\rm J\Delta}\cos(\delta\phi\cos{(\omega_d t)})\sin{\hat{\varphi}}
\label{eq:main_H_non_rot}
\end{equation}
We can represent the first two terms on the right-hand side of the above Hamiltonian as the transmon Hamiltonian $(\hat{H}_{\rm lab,T})$, the next term as symmetric SQUID Hamiltonian $(\hat{H}_{\rm lab,SQ}^{\rm sym})$ and final term as the asymmetric SQUID contribution $(\hat{H}_{\rm SQ}^{\rm lab,asym})$. Hence,
\begin{equation}
\hat{H}_{\rm lab} = \hat{H}_{\rm lab,T} + \hat{H}_{\rm lab,SQ}^{\rm sym} + \hat{H}_{\rm lab,SQ}^{\rm asym}
\label{eq:Tsymasym}
\end{equation}
Using $\hat{\varphi} = \varphi_{\rm zps}(\hat{a}^\dagger + \hat{a})$ and $\hat{n}=-i(\hat{a}-\hat{a}^\dagger)/(2\varphi_{\rm zps})$, where $\varphi_{\rm zps}=\left(2E_{\rm C}/E_{\rm J2}\right)^{1/4}$ is the zero point spread of the phase operator around the Josephson junction and $\hat{a}(\hat{a}^\dagger)$ is the bosonic annihilation(creation) operator, we obtain
\begin{equation}
\hat{H}_{\rm lab,T} = \epsilon_{\rm c}\hat{a}^\dagger \hat{a} - K \hat{a}^{\dagger^2} \hat{a}^2,
\end{equation}
where $\epsilon_{\rm c}=\sqrt{8 E_{\rm C} E_{\rm 2J}}$ and $K=E_{\rm C}/2$. Similarly, the symmetric SQUID Hamiltonian can be written as
\begin{equation}
\hat{H}_{\rm lab,SQ}^{\rm sym}=-2E_{\rm J\Sigma}
\sin{\left(\delta\phi \cos(\omega_d t)\right)}\sum_{n=1,2} \frac{(-1)^n}{(2n)!}\left[\varphi_{\rm zps}(\hat{a}+\hat{a}^\dagger)\right]^{2n}.
\end{equation}
We will seek a static effective Hamiltonian under the condition $\hbar\omega_{\rm d} \approx 2\epsilon_{\rm c}$. Going to rotating frame of the qubit defined through $a\rightarrow a e^{-i\epsilon_{\rm c}t/\hbar}$ and  $\hat{H}_{\rm lab}\rightarrow \hat{H}_{\rm T}+\hat{H}_{\rm SQ}^{\rm sym}+\hat{H}_{\rm SQ}^{\rm asym}$, we obtain
\begin{equation}
\hat{H}_{\rm T} =\delta \hat{a}^\dagger \hat{a} - K\hat{a}^{\dagger^2}\hat{a}^2,
\label{eq:tran_rot}
\end{equation}
and
\begin{equation}
\hat{H}_{\rm SQ}^{\rm sym}=-2E_{\rm J\Sigma}
\sin{\left(\delta\phi \cos(\omega_d t)\right)}\sum_{n=1,2} \frac{(-1)^n}{(2n)!}\left[\varphi_{\rm zps}(\hat{a}e^{-i\omega_d t/2}+\hat{a}^\dagger e^{i\omega_d t/2})\right]^{2n},
\label{eq:sym_rot}
\end{equation}
where $\delta = \epsilon_c - \hbar\omega_d/2$. For weak modulation depth such that $\delta \phi \approx \varphi_{\rm zps}$, we can make the following approximation, $\sin(\delta\varphi \cos(\omega_d t))\approx \delta\varphi \cos(\omega_d t)$. Doing so, we will keep only up to ${\cal O}(\varphi_{\rm zps}^4)$ contributions in our Hamiltonian and master equation calculations. The symmetric SQUID Hamiltonian reduces to
\begin{equation}
    \hat{H}_{\rm SQ}^{\rm sym}=\sum_{n=2,4}G_{n,{\rm S}}(e^{i\omega_d t}+e^{-i\omega_d t})\left(\hat{a}e^{-i\omega_dt/2}+\hat{a}^\dagger e^{i\omega_dt/2}\right)^n,
\end{equation}
where $G_{\rm 2,S}={\delta\varphi}E_{\rm J\Sigma} \varphi_{\rm zps}^2/2$ and $ G_{\rm 4,S}=- {{\delta\varphi}E_{\rm J\Sigma}\varphi_{\rm zps}^4}/{24}$.
Now, the asymmetric SQUID contribution in Eq.~(\ref{eq:Tsymasym}) can be written as
\begin{equation}
    \hat{H}_{\textrm{SQ}}^{\rm asym}=-2E_{\rm J\Delta}\bigg[1-{\delta\phi^2}\cos^2(\omega_dt)/2\bigg]\sin{\hat{\varphi}},
\end{equation}
where we made the following approximation, $\cos(\delta\phi \cos(\omega_d t))\approx 1-\delta\phi^2 \cos^2(\omega_d t)/2$. Similar to the symmetric case, we will keep only up to ${\cal O}({\rm \varphi_{zps}^4})$ terms in our calculations. Taylor expanding $\sin{\hat{\varphi}}$, we obtain
\begin{equation}
    \hat{H}_{\textrm{SQ}}^{\rm asym}=
    -2E_{\rm J\Delta}\bigg[\bigg(1-\frac{\delta\phi^2}{4}\bigg)-\frac{\delta\phi^2}{8}(e^{2i\omega_dt}+e^{-2i\omega_dt})\bigg]\bigg(\hat{\varphi}-\frac{\hat{\varphi}^3}{3!}\bigg)
\end{equation}
Using $\hat{\varphi} = \varphi_{\rm zps}(\hat{a}^\dagger + \hat{a})$ and going to the rotating frame (see Eq.~(\ref{eq:sym_rot}) for reference), we obtain
\begin{multline}
    \hat{H}_{\textrm{SQ}}^{\rm asym}=
    -2E_{\rm J\Delta}\bigg[\bigg(1-\frac{\delta\phi^2}{4}\bigg)-\frac{\delta\phi^2}{8}(e^{2i\omega_dt}+e^{-2i\omega_dt})\bigg]\\
    \bigg[\varphi_{\rm zps}(\hat{a}^\dagger e^{i\omega_d t/2}+\hat{a} e^{-i\omega_d t/2})-\frac{\varphi_{\rm zps}^3}{3!}(\hat{a}^\dagger e^{i\omega_d t/2}+\hat{a} e^{-i\omega_d t/2})^3\bigg].
\end{multline}
Since we consider $\delta\phi\sim \varphi_{\rm zps}$,  terms containing $\varphi_{\rm zps}^3\delta\phi^2$ would have an order comparable to $\varphi_{\rm zps}^5$, which can then be dropped. If we drop all of these terms, we get
\begin{multline}
    \hat{H}_{\textrm{SQ}}^{\rm asym}=
    -2E_{\rm J\Delta}\bigg[\bigg(1-\frac{\delta\phi^2}{4}\bigg)-\frac{\delta\phi^2}{8}(e^{2i\omega_dt}+e^{-2i\omega_dt})\bigg]\\
    \bigg[\varphi_{\rm zps}(\hat{a}^\dagger e^{i\omega_d t/2}+\hat{a} e^{-i\omega_d t/2})\bigg]+2E_{\rm J\Delta}\bigg[\frac{\varphi_{\rm zps}^3}{3!}(\hat{a}^\dagger e^{i\omega_d t/2}+\hat{a} e^{-i\omega_d t/2})^3\bigg].
\end{multline}
We can rewrite the above Hamiltonian as
\begin{equation}
   \hat{H}_{\textrm{SQ}}^{\rm asym}=
    \left(G_{\rm 1,S} + \tilde{G}_{\rm 1,S}(e^{2i\omega_d t}+e^{-2i\omega_d t})\right)(\hat{a}^\dagger e^{i\omega_d t/2}+\hat{a} e^{-i\omega_d t/2})
    + G_{\rm 3,S}(\hat{a}^\dagger e^{i\omega_d t/2}+\hat{a} e^{-i\omega_d t/2})^3,
\end{equation}
where $G_{\rm 1,S}=-2E_{\rm J\Delta}\varphi_{\rm zps}\left(1-\delta\phi^2/4\right)$, $\tilde{G}_{\rm 1,S} = E_{\rm J\Delta} \varphi_{\rm zps}\delta\phi^2/4$ and $G_{\rm 3,S} = E_{J\Delta}\varphi_{\rm zps}^3/3$.

\subsection{Dilution of Kerr coefficient}
\label{app:dilution}
\begin{figure}[hbt!]
\includegraphics[width=0.7\columnwidth]{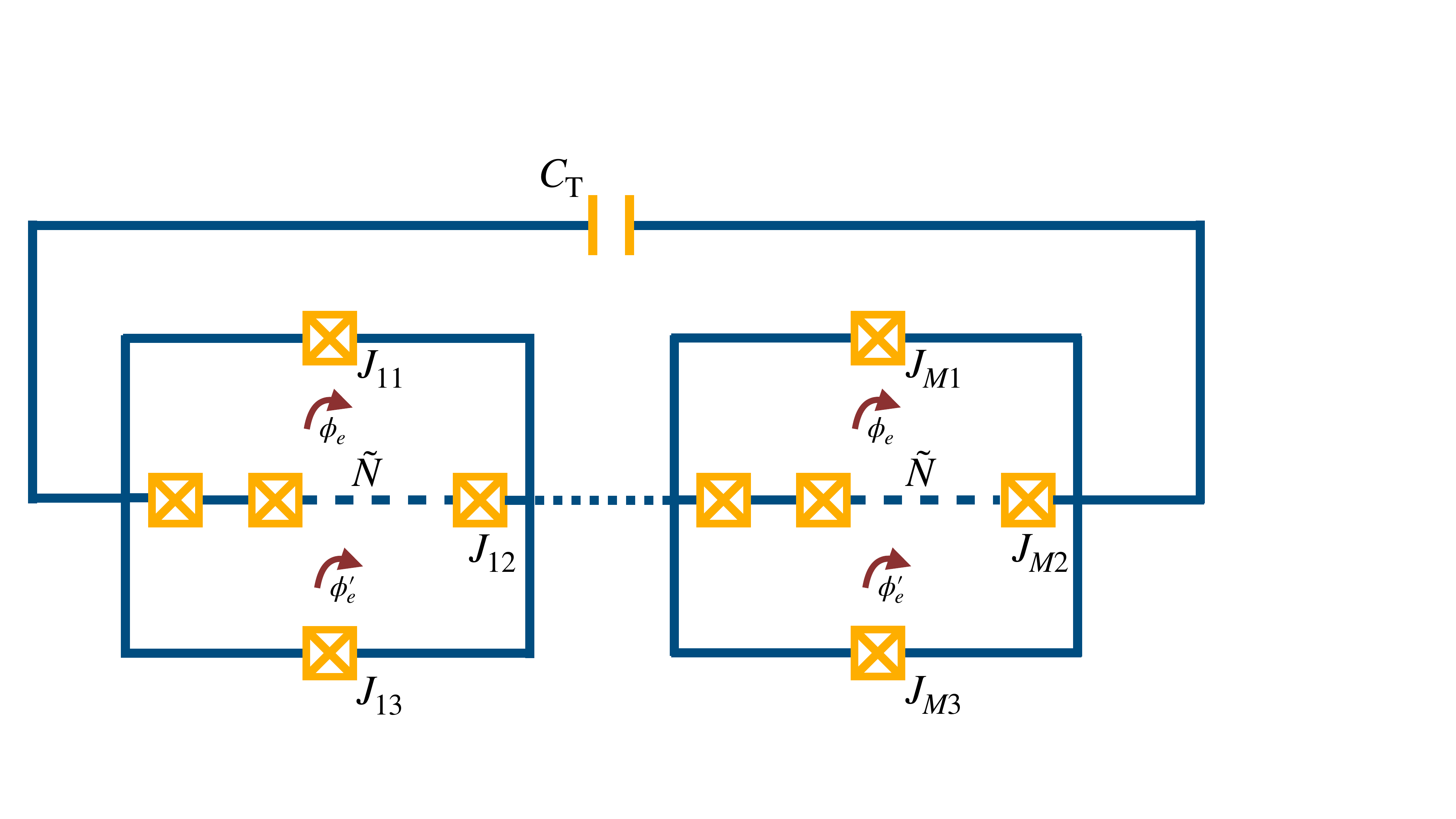}
 \caption{Multiple STS circuits in series.}
\label{fig:sketch_multiple}
\end{figure}
As shown in Fig.~\ref{fig:sketch_multiple}, we consider $M$ number of identical STS coupled in series. Each STS has $\tilde{N}$ number of identical junctions in the trasmon branch. We argue that the circuit is only a function of a single degree of freedom $\hat{\varphi}$ although it consists of $M(\tilde{N}+2)$ Josephson junctions. This can be done since the plasma frequency $\Omega_{\rm p}=\sqrt{8E_{\rm J}E_{\rm C_J}}/h$, where $E_{\rm 
 C_J}$ is the charging energy associated with the junction capacitance $C_{\rm J}$, is sufficiently higher than the frequency at which the dynamics we are interested in occurs. In this case, the dynamics coming from the extra degrees of freedom pertaining to each Josephson junctions will oscillate fast compared to that of the mode $\hat{\varphi}$ and can be integrated out \cite{thesis_frattini}. The Hamiltonian is given by
\begin{equation}
    \hat{H}_{\rm lab}=\hat{H}_{\rm lab}^{\rm sym}=4E_{\rm C}'\hat{n}^2-NE_{\rm J 2}^\prime\cos{\frac{\hat{\varphi}}{N}}-2ME_{\rm J}^\prime\cos\phi_e\cos{\frac{\hat{\varphi}}{M}},
\end{equation}
where $E'_{\rm C}\approx E_{\rm C}$ due to the large transmon capacitance and $N = M\tilde{N}$. Taylor expanding the sine and cosine up to fourth order in $\varphi_{\rm zps}$ for the phase operators and dropping scalars gives
\begin{equation}
    \hat{H}_{\rm lab}^{\rm sym}=4E_{\rm C}\hat{n}^2+NE_{\rm J2}^\prime\bigg[\frac{1}{2!}\bigg(\frac{\hat{\varphi}}{N}\bigg)^2-\frac{1}{4!}\bigg(\frac{\hat{\varphi}}{N}\bigg)^4\bigg]+2ME_{\rm J}^\prime\cos\phi_e\bigg[\frac{1}{2!}\bigg(\frac{\hat{\varphi}}{M}\bigg)^2-\frac{1}{4!}\bigg(\frac{\hat{\varphi}}{M}\bigg)^4\bigg].
\end{equation}
 If we set $E_{\textrm{J2}}=\frac{E_{\textrm{J2}}^\prime}{N}$ and $E_{\textrm{J}}=\frac{E_{\textrm{J}}^\prime}{M}$, we can rewrite the Hamiltonian in terms of $E_{J2}$ and $E_{\textrm{J}}$ as
 \begin{equation}
     \hat{H}_{\rm lab}^{\rm sym}=4E_{\rm C}\hat{n}^2+E_{\rm J2}\bigg[\frac{\hat{\varphi}^2}{2!}-\frac{1}{4!}\bigg(\frac{\hat{\varphi}^2}{N}\bigg)^2\bigg]+2E_{\rm J}\cos\phi_e\bigg[\frac{\hat{\varphi}^2}{2!}-\frac{1}{4!}\bigg(\frac{\hat{\varphi}^2}{M}\bigg)^2\bigg].
 \end{equation}
 We note that this is the same form as with the single junction case, except that the fourth order $\varphi_{\textrm{zps}}$ terms now have an extra $\frac{1}{M^2}$ or $\frac{1}{N^2}$ factor. The new expressions for the zero point phase spread, Kerr coefficient and two-photon drive strength are
 \begin{equation}
     \varphi_{\textrm{zps}}=\bigg(\frac{2E_\textrm{C}}{E_{\textrm{J2}}}\bigg)^{\frac{1}{4}}=N^{\frac{1}{4}}\varphi_{\textrm{zps}}^\prime,
 \end{equation}
 \begin{equation}
     K=-\frac{{E_{\textrm{J2}}}\varphi_{\textrm{zps}}^{4}}{4N^2}=\frac{K^\prime}{N^2},
 \end{equation}
 \begin{equation}
G_{\rm 2,S}=\frac{\delta\varphi}{2}E_\textrm{J} \varphi_{\textrm{zps}}^2=\frac{\sqrt{N}}{M}G_{2,S}^\prime,
\end{equation}
 and
 \begin{equation}
    G_{\rm 4,S}=- \frac{{\delta\varphi}E'_{\textrm{J}}\varphi_{\textrm{zps}}^4}{24M^2}= \frac{N}{M^3}G_{\rm 4,S}^\prime.
 \end{equation}
 
\subsection{Zero point phase spread and modulation depth}
\label{app:bounds}
In this appendix, we will discuss the limit set by weak zero point phase spread and modulation depth in the static effective Hamiltonian of the STS design. First, we start with the zero point phase spread. So far we have only considered each harmonic to the leading order in $\hat{\varphi}$ satisfying the RWA but there are  higher order terms suppressed by the zero point fluctuation, $(2E_{\rm C}/E_{\rm J2})^{1/4}$. For example, the term $\hbar \omega_{\rm d}\approx 2\epsilon_{\rm c}$ can stabilize the squeeze drive as well as terms coming from $\hat{\varphi}^4$ and  $\hat{\varphi}^6$. Those corrections, respectively, are given by
\begin{equation}
-\frac{\epsilon_{2}}{12}\left(\frac{2 E_{\rm C}}{E_{\rm J2}}\right)^{\frac{1}{2}}\left\{\hat{a}^{\dagger 2}+\hat{a}^{2}, 1+2 \hat{a}^{\dagger} \hat{a}\right\},
\end{equation}
and
\begin{equation}
\frac{\epsilon_{2}}{360}\left(\frac{2 E_{\rm C}}{E_{\rm J2}}\right)\left(\{\hat{a}^{\dagger 2}+\hat{a}^{2}, 1+2 \hat{a}^{\dagger} \hat{a}\right\}\left(1+2 \hat{a}^{\dagger} \hat{a}\right)+\left(\hat{a}^{\dagger 2} \hat{a}^{2}+\hat{a}^{2} \hat{a}^{\dagger 2}+4 \hat{a}^{\dagger} \hat{a} \hat{a}^{\dagger} \hat{a}+4 \hat{a}^{\dagger} \hat{a}+1\right)\left(\hat{a}^{\dagger 2}+\hat{a}^{2}\right)).
\end{equation}

Therefore, it is desired to increase the ratio $E_{\rm J2}/E_{\rm C}$ not only to increase the number of bound states but also to suppress those unwanted terms. The first term tends to compete with the squeeze drive (due to the alternating sign in the expansion of $\cos\hat{\varphi}$) and leads to a smaller cat. The second term tends to assist the squeeze drive. We can estimate their contribution by projecting them to the cat space and using $\hat{a}^{\dagger }\hat{a} \sim\alpha^2 $. However, their contribution is very small for $\varphi_{\rm zps}<<1$. The other correction we must consider is the sixth order non-linearity competing with the Kerr coming from the expansion of the transmon Hamiltonian. Keeping the sixth order term, Eq.~(\ref{eq:tran_rot}) becomes
\begin{equation}
\hat{H}_{T}\approx-K \hat{a}^{\dagger^2}\hat{a}^2+\frac{K}{9}\left(\frac{2E_{\rm C}}{E_{\rm J2}}\right)^{1/2}\hat{a}^{\dagger^3}\hat{a}^3,
\end{equation}
where we considered, $\delta = 0$. This term like all the higher order terms is suppressed by the the zero point fluctuation squared (i.e. $\sqrt{2E_{\rm C}/E_{\rm J}}$). However, this sets a limit on how big the cat size can be before this correction can be neglected. If $\frac{\alpha^2}{9}\left(\frac{2E_{\rm C}}{E_{\rm J}}\right)^{1/2} \sim 1$, then the anharmonicity terms are almost on equal footing, which sets a very large bar on the cat size.

In the case of modulation depth, the approximation $\sin(\delta\phi \cos(\omega_{\rm d} t))\approx \delta\phi \cos(\omega_{\rm d} t)$ is  valid for two reasons: 1) The modulation depth $\delta\phi$ ($\lesssim $0.2) is very small. 2) Most of the higher order terms do not satisfy the RWA. In general, we can use the Jacobi-Anger formula to examine the harmonics: 
\begin{equation}
\sin(\delta\phi \cos(\omega_{\rm d} t))=\sum_{n=0} \delta\phi^{(n)} \cos((2n+1)\omega_{\rm d} t),
\end{equation}
such that
\begin{equation}
 \delta\phi^{(n)}=2 (-1)^{n} J_{2n+1}(\delta\phi)=2 (-1)^{n} \sum_{k=0} \frac{(-1)^k}{k! \Gamma(2n+k+2)}(\delta\phi/2)^{2n+1+2k},
\end{equation}
where  $J_{m}$ is the m-th Bessel function of the first kind. Since only odd harmonics contribute to the expansion of $\delta\phi^{(n)}$ and $\cos(\hat{\varphi})$ is an even function, the leading order static terms for each harmonics are $\hat{a}^2,\hat{a}^6,...$. Focusing on the squeeze drive, we can calculate the amplitude to any desired order of the modulation depth as
\begin{equation}
\epsilon_2\rightarrow \epsilon_2 \left(1-\frac{(\delta\phi)^2}{8}+\frac{(\delta\phi)^4}{192}+\cdot \cdot \cdot \right).
\end{equation}
The second non-zero harmonic is $3\hbar\omega_{\rm d}=2\epsilon_{\rm c}$ which stabilizes $\hat{a}^6(\hat{a}^{\dagger ^6})$, giving a correction to the SQUID Hamiltonian
\begin{equation}
\hat{H}^{\rm corr}_{\rm SQ}\approx-\frac{4E_{\rm C}}{3E_{\rm J2}} \epsilon_2\delta\phi^2 (\hat{a}^6+\hat{a}^{\dagger 6}).
\end{equation}
For $\frac{4E_{\rm C}}{3E_{\rm J2}} \approx 10^{-3} \:\text{and} \:\delta\phi\approx (0.2) $, the correction from this term is around $10^{-5}$ times smaller than the leading squeeze drive strength. 

 \subsection{Static effective Hamiltonian}
 \label{app:static_eff}
The STS Hamiltonian (which includes contribution from transmon, the symmetric and asymmetric part of SQUID) takes the following form 
\begin{multline}
    \hat{H}=\delta \hat{a}^\dagger \hat{a}-K\hat{a}^{\dagger^2}\hat{a}^2+\sum_{n=2,4}G_{n,{\rm S}}(e^{i\omega_d t}+e^{-i\omega_d t})\left(\hat{a}e^{-i\omega_dt/2}+\hat{a}^\dagger e^{i\omega_dt/2}\right)^n\\
    + \left(G_{\rm 1,S} + \tilde{G}_{\rm 1,S}(e^{2i\omega_d t}+e^{-2i\omega_d t})\right)(\hat{a}^\dagger e^{i\omega_d t/2}+\hat{a} e^{-i\omega_d t/2})
    + G_{\rm 3,S}(\hat{a}^\dagger e^{i\omega_d t/2}+\hat{a} e^{-i\omega_d t/2})^3.
\end{multline}
The time-independent part of the STS Hamiltonian is given by
\begin{equation}
\hat{H}_{\rm st} = \delta \hat{a}^\dagger \hat{a}- K \hat{a}^{\dagger^2} \hat{a}^2 + \left(G_{\rm 2,S}+6G_{\rm 4,S}\right)(\hat{a}^2+\hat{a}^{\dagger^2}) + 4G_{\rm 4,S}\left(\hat{a}^{\dagger^3}\hat{a}+\hat{a}^\dagger \hat{a}^3\right).
\label{eq:zeroth_order}
\end{equation}
We will break the time-dependent part of the total Hamiltonian into different orders in $\varphi_{\rm zps}$ contributions: 
\begin{align}
&{\rm osc}\bigg(\hat{H}^{(1)}(t)\bigg) = \left(G_{1,S} + \tilde{G}_{1,S}\left(e^{2i\omega_d t}+e^{-2i\omega_d t}\right)\right)\hat{a} e^{-i\omega_d t/2} + {\rm h.c.},\nonumber \\
&{\rm osc}\bigg(\hat{H}^{(2)}(t)\bigg) = G_{2,S}\left(\left(e^{i\omega_dt}+e^{-i\omega_dt}\right)\hat{a}^\dagger \hat{a} +e^{-2i\omega_dt}\hat{a}^2\right) + {\rm h.c.},\nonumber \\
&{\rm osc}\bigg(\hat{H}^{(3)}(t)\bigg) = G_{3,S}\Big(3e^{-i\omega_d t/2 }\hat{a}+ e^{-3i\omega_d t/2}\hat{a}^3\nonumber+ 3e^{-i\omega_d t/2 }\hat{a}^\dagger \hat{a}^2\Big) + {\rm h.c.},\nonumber \\
&{\rm osc}\bigg(\hat{H}^{(4)}(t)\bigg) =G_{4,S}\Big(6\left(e^{i\omega_dt}+e^{-i\omega_dt}\right)\hat{a}^\dagger \hat{a} + 6 e^{-2i\omega_d t}\hat{a}^2+\left(e^{-i\omega_d t}+e^{-3i\omega_d t}\right)\hat{a}^4 + 3 \left(e^{i\omega_dt}+e^{-i\omega_dt}\right)\hat{a}^{\dagger^2} \hat{a}^2\Big)+ {\rm h.c.},
\label{eq:diff_ord_Ham}
\end{align}
where $\hat{H}^{(i)}$ represent the ${\cal O}(\varphi_{\rm zps}^i)$ contribution. ${\rm osc}~(g(t)) = g(t) - \int_0^T g(t)dt$ is the oscillating part of $g$; $T$ is the periodicity of $g(t)$. While writing Eqs.~(\ref{eq:diff_ord_Ham}), we neglected all the scalar terms. The first and third order terms originate from $\hat{H}_{\rm SQ}^{\rm asym}$ whereas the second and the fourth order terms originate from $\hat{H}_{\rm SQ}^{\rm sym}$. 

We follow Ref.~\cite{jaya2022} to obtain a time-independent Hamiltonian at different orders of $\varphi_{\rm zps}$ using a generalized Schieffer-Wolff transformation. The canonical transformation that leads to an effective time-independent Hamiltonian is given by
\begin{equation}
\hat{\cal H}=e^{\hat{L}_S}(\hat{H})+\int_0^1 d\epsilon e^{\epsilon \hat{L}_S }(\dot{\hat{S}}(t)),
\end{equation}
where $\hat{S}(t)$ is the generator of the transformation and $\hat{L}_S(\hat{O}(t)) = \frac{1}{i\hbar}[\hat{S}(t),\hat{O}(t)]$, where $\hat{O}(t)$ can be any operator. 
Expanding the above equation, we find following expressions for the total system Hamiltonian at different orders of $\varphi_{zps}$
\begin{align}
\hat{\cal H}^{(0)} &= 0~~~{\rm since~} \hat{H}^{(0)}=0,\nonumber \\
\hat{\cal H}^{(1)} &= \hat{H}^{(1)}(t) + {\partial_t} \hat{S}^{(1)}, \nonumber \\
\hat{\cal H}^{(2)} &= \hat{H}^{(2)}(t) + \frac{1}{i\hbar} \left[\hat{S}^{(1)},\hat{H}^{(1)}\right]+\frac{1}{2i\hbar}\left[\hat{S}^{(1)},{\partial_t}\hat{S}^{(1)} \right]+{\partial_t} \hat{S}^{(2)},\nonumber \\
\hat{\cal H}^{(3)} &=\hat{H}^{(3)}(t) + \frac{1}{i\hbar} \left[\hat{S}^{(1)},\hat{H}^{(2)}\right] + \frac{1}{i\hbar} \left[\hat{S}^{(2)},\hat{H}^{(1)}\right] + \frac{1}{2! (i\hbar)^2}\left[\hat{S}^{(1)},\left[\hat{S}^{(1)},\hat{H}^{(1)}\right]\right]\nonumber \\
&~~+\frac{1}{2i\hbar}\left(\left[\hat{S}^{(1)},{\partial_t}\hat{S}^{(2)} \right]+\left[\hat{S}^{(2)},{\partial_t}\hat{S}^{(1)} \right]\right)+\frac{1}{3!(i\hbar)^2}\left[\hat{S}^{(1)},\left[\hat{S}^{(1)},\partial_t \hat{S}^{(1)}\right]\right]+\partial_t \hat{S}^{(3)},
\label{eq:eff_H_diff_ord}
\end{align}
and 
\begin{multline}
\hat{\cal H}^{(4)} =  \hat{H}^{(4)} + [\hat{S}^{(1)},\hat{H}^{(3)}]
+[\hat{S}^{(2)},\hat{H}^{(2)}]+[\hat{S}^{(3)},\hat{H}^{(1)}]+\frac{1}{2!(i\hbar)^2}\Big([\hat{S}^{(1)},[\hat{S}^{(1)},\hat{H}^{(2)}]]+[\hat{S}^{(1)},[\hat{S}^{(2)},\hat{H}^{(1)}]]\\
+[\hat{S}^{(2)},[\hat{S}^{(1)},\hat{H}^{(1)}]]\Big)+\frac{1}{3!(i\hbar)^3}[\hat{S}^{(1)},[\hat{S}^{(1)},[\hat{S}^{(1)},\hat{H}^{(1)}]]]+\frac{1}{2i\hbar}\Big([\hat{S}^{(1)},\partial_t\hat{S}^{(3)}]+[\hat{S}^{(2)},\partial_t\hat{S}^{(2)}]+[\hat{S}^{(3)},\partial_t\hat{S}^{(1)}]\Big)\\
 + \frac{1}{3!(i\hbar)^2}\Big([\hat{S}^{(1)},[\hat{S}^{(1)},\partial_t\hat{S}^{(2)}]]+[\hat{S}^{(1)},[\hat{S}^{(2)},\partial_t\hat{S}^{(1)}]]+[\hat{S}^{(2)},[\hat{S}^{(1)},\partial_t\hat{S}^{(1)}]]\Big)+\partial_t \hat{S}^{(4)},
\end{multline}
where we expanded $\hat{S} (t)= \sum_n \hat{S}^{(n)}(t)$ at different orders of $\varphi_{\rm zps}$ and we choose
\begin{align}
\hat{S}^{(1)}(t) &= -\int dt\,  {\rm osc}\left(\hat{ H}^{(1)}(t) \right), \nonumber \\
\hat{S}^{(2)}(t) &= -\int dt\,  {\rm osc}\left(\hat{ H}^{(2)}(t) + \frac{1}{i\hbar} \left[\hat{S}^{(1)}(t),\hat{H}^{(1)}(t)\right] + \frac{1}{2i\hbar}\left[\hat{S}^{(1)}(t),\partial_t \hat{S}^{(1)}(t)\right] \right),
\label{eq:par_s1}
\end{align}
and so on. Note that  all the terms in ${ H}^{(n)}(t)$ are time-dependent. The time independent terms were all added to Eq.~(\ref{eq:zeroth_order}). 
We find,
\begin{equation}
\hat{S}^{(1)} = \frac{2i}{\omega_{\rm d}} e^{i\omega_{\rm d} t/2} \bigg[G_{\rm 1,S} \left(\hat{a}^\dagger - \hat{a}e^{-i\omega_d t}\right)+\tilde{G}_{\rm 1,S}\hat{a}^\dagger \left(\frac{1}{5}e^{2i\omega_{\rm d} t}-\frac{1}{3}e^{-2i\omega_{\rm d} t}\right)
+\tilde{G}_{\rm 1,S}\hat{a}\left(\frac{1}{3}e^{i\omega_{\rm d} t}-\frac{1}{5}e^{-3i\omega_{\rm d} t}\right)\bigg].
\end{equation}
For $\hat{S}^{(2)}$, we obtain
\begin{equation}
    \hat{S}^{(2)}(t)=\frac{iG_{\rm 2,S}}{2\omega_d}(\hat{a}^{\dagger 2}e^{2i\omega_d t}-\hat{a}^{2}e^{-2i\omega_d t})+\frac{2iG_{\rm 2,S}}{\omega_d}\hat{a}^\dagger \hat{a}(e^{i\omega_{\rm d} t}-e^{-i\omega_{\rm d} t})+{\rm cons}.
\end{equation}
Similarly, at ${\cal O}(\varphi_{\rm zps}^3)$ we have
\begin{equation}
\hat{S}^{(3)}(t) = -\int dt\,  {\rm osc}\bigg(\hat{H}^{(3)}(t) + \frac{1}{i\hbar} \left[\hat{S}^{(1)},\hat{H}^{(2)}\right] + \frac{1}{i\hbar} \left[\hat{S}^{(2)},\hat{H}^{(1)}\right]
+\frac{1}{2i\hbar}\left(\left[\hat{S}^{(1)},{\partial_t}\hat{S}^{(2)} \right]+\left[\hat{S}^{(2)},{\partial_t}\hat{S}^{(1)} \right]\right)
\bigg),
\end{equation}
where we used $\left[\hat{S}^{(1)},\left[\hat{S}^{(1)},\hat{H}^{(1)}\right]\right]=0$ and $\left[\hat{S}^{(1)},\left[\hat{S}^{(1)},\partial_t \hat{S}^{(1)}\right]\right]=0$.
Since, 
\begin{equation}
-\int dt {~\rm osc} \left(H^{(3)}(t)\right) =  \frac{6iG_{\rm 3,S}}{\omega_{\rm d}}\left(e^{i\omega_{\rm d} t/2}\left(\hat{a}^\dagger + \hat{a}^{\dagger^2}\hat{a}\right)-e^{-i\omega_{\rm d} t/2}\left(\hat{a} + \hat{a}^2\hat{a}^\dagger\right)\right)
+\frac{2iG_{\rm 3,S}}{3\omega_{\rm d}}\left(e^{3i\omega_{\rm d}t/2}\hat{a}^{\dagger^3}-e^{-3i\omega_{\rm d}t/2}\hat{a}^3\right),
\end{equation}
we find
\begin{multline}
\hat{S}^{(3)}(t)={\cal G}_{31}(t)\hat{a}^\dagger + {\cal G}_{31}^\dagger(t)\hat{a}+\frac{2i}{3\omega_{\rm d}}G_{\rm 3,S}e^{3i\omega_{\rm d}t/2}\hat{a}^\dagger\hat{a}^\dagger\hat{a}^\dagger + \frac{6i}{\omega_{\rm d}}G_{\rm 3,S}e^{i\omega_{\rm d}t/2}\hat{a}^\dagger \hat{a}^\dagger \hat{a}\\
-\frac{6i}{\omega_{\rm d}}G_{\rm 3,S}e^{-i\omega_{\rm d}t/2}\hat{a}^\dagger \hat{a}\hat{a}-\frac{2i}{3\omega_{\rm d}}G_{\rm 3,S}e^{-3i\omega_{\rm d}t/2}\hat{a}\hat{a}\hat{a},
\end{multline}
where
\begin{multline}
{\cal G}_{31}(t)=\frac{i}{\hbar\omega_{\rm d}^2}\bigg(\tilde{G}_{\rm 1,S}G_{\rm 2,S}\left(\frac{38}{25}e^{-5i\omega_{\rm d}t/2}+\frac{71}{105}e^{7i\omega_{\rm d}t/2}+\frac{46}{45}e^{3i\omega_dt/2}+\frac{19}{5}e^{-i\omega_{\rm d}t/2}\right)\\
+G_{\rm 1,S}G_{\rm 2,S}\left(3e^{3i\omega_{\rm d}t/2}-2e^{-i\omega_{\rm d}t/2}\right)+6\hbar \omega_{\rm d}G_{\rm 3,S}e^{i\omega_{\rm d}t/2}\bigg).
\end{multline}

Since all the oscillating terms in $\hat{H}^{(i)}$ gets cancelled with the corresponding terms in $\partial_t\hat{{S}}^{(i)}$ and its commutators, Eqs.~(\ref{eq:eff_H_diff_ord}) will be largely simplified. The resulting static Hamiltonian can be written as
\begin{align}
\hat{\cal H}_{\rm st} &= \sum_{i=1}^4 \hat{\cal H}^{(i)}={\rm non~osc} \bigg(\frac{1}{i\hbar} \left[\hat{S}^{(1)},\hat{H}^{(1)}\right]+\frac{1}{2i\hbar}\left[\hat{S}^{(1)},{\partial_t}\hat{S}^{(1)} \right]+ \frac{1}{i\hbar} \left[\hat{S}^{(1)},\hat{H}^{(2)}\right]  \nonumber \\
&~~+ \frac{1}{i\hbar} \left[\hat{S}^{(2)},\hat{H}^{(1)}\right]+ \frac{1}{2! (i\hbar)^2}\left[\hat{S}^{(1)},\left[\hat{S}^{(1)},\hat{H}^{(1)}\right]\right]+\frac{1}{2i\hbar}\Big(\left[S^{(1)},{\partial_t}S^{(2)} \right],\nonumber \\
&~~+\left[S^{(2)},{\partial_t}S^{(1)} \right]\Big)+\frac{1}{3!(i\hbar)^2}\left[S^{(1)},\left[S^{(1)},\partial_t S^{(1)}\right]\right] + {\cal O}(\varphi_{zps}^4)\bigg),
\end{align}
where ${\rm non~osc}~(g(t))$ extracts the time-independent terms in $g(t)$. We find
\begin{equation}
\hat{\cal H}_{\rm st} =\bigg(-\frac{2G^2_{\rm 2,S}}{\omega_d}- 2K\bigg) \hat{a}^\dagger \hat{a}
\end{equation}
Hence, the static effective Hamiltonian up to ${\cal O}(\varphi_{\rm zps}^4)$ is given by
\begin{equation}
\hat{\cal H}_{\rm S} = \hat{H}_{\rm st}+ \hat{\cal H}_{\rm st}=\Delta \hat{a}^\dagger \hat{a} + \epsilon_2\left(\hat{a}^{\dagger^2}+\hat{a}^2\right) - K\hat{a}^{\dagger^2}\hat{a}^2 + \Lambda \left(\hat{a}^{\dagger^3}\hat{a}+\hat{a}^3\hat{a}^\dagger\right),
\label{eq:kerr_cat_ham}
\end{equation}
where $\Delta = \delta -\frac{2G^2_{\rm 2,S}}{\omega_{\rm d}}-\frac{24G_{\rm 1,S}G_{\rm 3,S}}{\omega_{\rm d}}-2K$, $K = E_{\rm C}/2$, $\epsilon_2 = G_{\rm 2,S} + 6 G_{\rm 4,S}$ and $\Lambda = 4 G_{\rm 4,S}$.

\section{STS - Master Equation}
\label{app:master}
In this section, we will study the dynamics of the STS in the presence of a thermal environment. The Kerr-cat Hamiltonian in Eq.~(\ref{eq:kerr_cat_ham}) acts as the system Hamiltonian. The system bath Hamiltonian in the rotating frame is given by
\begin{equation}
\hat{H}_{\rm SB}(t)=i\left(\hat{a}e^{-i\omega_dt/2}-\hat{a}^\dagger e^{i\omega_dt/2}\right)\hat{B}(t),
\end{equation}
where $\hat{ B}(t)=\sum_j ih_j\left(\hat{b}_je^{-i\omega_j\tau}-\hat{b}_j^\dagger e^{i\omega_j\tau}\right)$, $\hat{b}_j(\hat{b}_j^\dagger)$ are the annihilation (creation) operators of mode $j$ of the bosonic environment and $h_j$ determines the system-bath coupling strength. The system-bath Hamiltonian, $\hat{H}_{\rm SB}(t)$ is independent of the zero point spread of the phase operator $({\cal O}(\varphi_{\rm zps}^0))$. Similar to the case of system Hamiltonian, we do the generalized Schrieffer-Wolff transformation to the system-bath coupling Hamiltonian at various orders of zero point spread of the phase operator. We find
\begin{align}
\hat{\cal H}_{\rm SB}^{(0)} &=  \hat{H}_{\rm SB}, \nonumber \\
\hat{\cal H}_{\rm SB}^{(1)} &=  \frac{1}{i\hbar} \left[\hat{S}^{(1)},\hat{H}_{SB}\right], \nonumber \\
\hat{\cal H}_{\rm SB}^{(2)}  &= \frac{1}{i\hbar} \left[\hat{S}^{(2)},{H}_{\rm SB}\right]+ \frac{1}{2! (i\hbar)^2}\left[\hat{S}^{(1)},\left[\hat{S}^{(1)},\hat{H}_{SB}\right]\right],\nonumber \\
\hat{\cal H}_{\rm SB}^{(3)}  &=\frac{1}{i\hbar} \left[\hat{S}^{(3)},\hat{H}_{SB}\right]+\frac{1}{2!(i\hbar)^2}\Big([\hat{S}^{(1)},[\hat{S}^{(2)},\hat{H}_{SB}]]+[\hat{S}^{(2)},[\hat{S}^{(1)},\hat{H}_{SB}]]\Big)+\frac{1}{3!(i\hbar)^3}[\hat{S}^{(1)},[\hat{S}^{(1)},[\hat{S}^{(1)},\hat{H}_{SB}]]].
\label{eq:eff_Hsb_diff_ord}
\end{align}
We dropped the time dependence of the generator and the Hamiltonian, for the sake of simplicity. Since, both $\hat{S}^{(1)}$ and $\hat{H}_{\rm SB}$ are linear in the annihilation and creation operators, $\hat{\cal H}_{\rm SB}^{(1)}$ will be a scalar and would not contribute towards the system-bath dynamics. By the same reasoning, $\left[\hat{S}^{(1)},\left[\hat{S}^{(1)},\hat{H}_{\rm SB}\right]\right] = 0$ and hence
\begin{equation}\label{eq_H_sb_2}
    \hat{\cal H}_{\rm SB}^{(2)}(t)=\frac{[\hat{S}^{(2)}(t),\hat{H}_{\rm SB}(t)]}{i\hbar},
\end{equation}
which simplifies to
\begin{equation} \hat{\cal H}_{\rm SB}^{(2)}(t)=\frac{iG_{\rm 2,S}}{\hbar\omega_d}\left\{3(-\hat{a}^\dagger e^{3i\omega_dt/2}+\hat{a} e^{-3i\omega_dt/2})+2(-\hat{a} e^{i\omega_dt/2}+\hat{a}^\dagger e^{-i\omega_dt/2})\right\}\hat{B}(t).
\end{equation}
The asymmetry in Josephson junction has no contribution on the bath induced dynamics of the system till ${\cal O}(\varphi_{zps}^2)$. Disregarding the scalars, the system-bath Hamiltonian at ${\cal O}(\varphi_{zps}^3)$ is given by
\begin{equation}
\hat{\cal H}_{\rm SB}^{(3)}(t)  =\frac{1}{i\hbar} \left[\hat{S}^{(3)}(t),\hat{H}_{SB}(t)\right],
\end{equation}
which evaluates to
\begin{equation}
{\cal H}_{\rm SB}^{(3)} (t) = \frac{8iG_{\rm 3,S}}{\hbar \omega_{\rm d}}\left(e^{-i\omega_{\rm d}t}\hat{a}^2-e^{i\omega_{\rm d}t}\hat{a}^{\dagger^2}\right).
\end{equation}
Similarly, the fourth order contribution $({\cal O}(\varphi_{\rm zps}^4))$ can be simplified to
\begin{equation}
\hat{\cal H}_{\rm SB}^{(4)} (t) =\frac{1}{i\hbar} \left[\hat{S}^{(4)}(t),\hat{H}_{\rm SB}(t)\right] +\frac{1}{2!(i\hbar)^2}\bigg\{\left[\hat{S}^{(2)}(t),\left[\hat{S}^{(2)}(t),\hat{H}_{\rm SB}(t)\right]\right]
+\left[\hat{S}^{(1)}(t),\left[\hat{S}^{(3)}(t),\hat{H}_{\rm SB}(t)\right]\right]\bigg\},
\end{equation}
where the fourth order transformation is given as
\begin{multline}
    \hat{S}^{(4)}(t)=-\int \textrm{osc}\bigg(\hat{H}^{(4)}+\frac{[\hat{S}^{(1)},\hat{H}^{(3)}]}{i}+\frac{[\hat{S}^{(2)},\hat{H}^{(2)}]}{i}+\frac{[\hat{S}^{(3)},\hat{H}^{(1)}]}{i}
+\frac{[\hat{S}^{(1)},\partial_t\hat{S}^{(3)}]}{2i}+\frac{[\hat{S}^{(2)},\partial_t\hat{S}^{(2)}]}{2i}+\frac{[\hat{S}^{(3)},\partial_t\hat{S}^{(1)}]}{2i}   \bigg)dt
\end{multline}
After some calculations, we obtain
\begin{multline}
\hat{\cal H}_{\rm SB}^{(4)} (t) =i\bigg\{\left(\frac{13G_{2,S}^2}{2\hbar^2\omega_{\rm d}^2}\hat{a}-\frac{24G_{1,S}G_{3,S}}{\hbar^2\omega_{\rm d}^2}\hat{a}+\frac{12G_{4,S}}{\hbar\omega_{\rm d}}\hat{a}^\dagger+\frac{4G_{4,S}}{\hbar\omega_{\rm d}}\hat{a}^{3}+\frac{12G_{4,S}}{\hbar\omega_{\rm d}}\hat{a}^{\dagger^2} \hat{a}\right)e^{-i\omega_{\rm d}t/2}\\
+\left(\frac{18G_{4,S}}{\hbar\omega_{\rm d}}\hat{a}-\frac{11G_{2,S}^2}{\hbar^2\omega_{\rm d}^2}\hat{a}^\dagger+\frac{12\Tilde{G}_{1,S}G_{3,S}}{5\hbar^2\omega_{\rm d}^2}\hat{a}^\dagger+\frac{12G_{4,S}}{\hbar\omega_{\rm d}}\hat{a}^{\dagger}\hat{a}^2\right)e^{-i3\omega_{\rm d}t/2}\\
+\left(\frac{13G_{2,S}^2}{3\hbar^2\omega_{\rm d}^2}\hat{a}+\frac{4\Tilde{G}_{1,S}G_{3,S}}{3\hbar^2\omega_{\rm d}^2}\hat{a}+\frac{4G_{4,S}}{3\hbar\omega_{\rm d}}\hat{a}^{3}\right)e^{-i5\omega_{\rm d}t/2}\bigg\}\hat{B}(t)+{\rm h.c.}
\end{multline}
The total system-bath Hamiltonian
\begin{equation}
\hat{\cal H}_{\rm SB}(t) = \sum_{k=0}^4\hat{\cal H}_{\rm SB}^{(k)}(t)
\end{equation}
We start by writing the system-bath coupling Hamiltonian as 
\begin{equation}
\hat{\cal H}_{\rm SB}(t)=\sum_{\alpha=1}^n \hat{A}_\alpha (t)\otimes \hat{B}_\alpha(t),
\end{equation}
where $\hat{B}_\alpha(t) = \hat{B}(t)$ for all $\alpha$. In addition, we group the system operators such that $\hat{A}_\alpha(\tau) = \hat{A}_\alpha^\dagger(\tau)$ for the sake of simplification. We also define $\hat{A}_\alpha=\hat{A}_\alpha(\tau=0)$. In the interaction picture, the system bath coupling Hamiltonian will read as
\begin{equation}
\hat{\tilde{\mathcal{H}}}_{\rm SB}(t)=e^{\frac{i}{\hbar}\hat{\cal H}_{\rm S}t}\hat{\cal H}_{\rm SB}(t)e^{-\frac{i}{\hbar}\hat{\cal H}_{\rm S}t}=\sum_{\alpha} e^{\frac{i}{\hbar}{\cal H}_st} A_\alpha e^{-\frac{i}{\hbar}{\cal H}_st}\otimes B_\alpha(t)=\sum_{\alpha} \tilde{ A}_\alpha(t) \otimes \tilde{B}_\alpha(t),
\label{eq:hsb_til}
\end{equation}
where $\tilde{B}_\alpha(t)={B}_\alpha(t)$. In the interaction picture, the Liouville equation for the density matrix can be written as
\begin{equation}
\frac{d\hat{\tilde{\rho}}(t)}{dt}=-\frac{i}{\hbar}\left[\hat{\tilde{\mathcal{H}}}_{\rm SB}(t), \hat{\tilde{\rho}}(t)\right].
\label{eq:liou_int}
\end{equation}
Integrating above equation on both sides with respect to time, we obtain
\begin{equation}
\hat{\tilde{\rho}}(t)=-\frac{i}{\hbar}\int_0^t\left[\hat{\tilde{\mathcal{H}}}_{\rm SB}(t'), \hat{\tilde{\rho}}(t')\right]dt' + \hat{\tilde{\rho}}(0)
\label{eq:int_rho}
\end{equation}
Substituting Eq.~(\ref{eq:int_rho}) in Eq.~(\ref{eq:liou_int}), we obtain
\begin{equation}
\frac{d\hat{\tilde{\rho}}(t)}{dt}=-\frac{i}{\hbar}\left[\hat{\tilde{\mathcal{H}}}_{\rm SB}(t),\hat{\tilde{\rho}}(0)\right]-\frac{1}{\hbar^2}\int_0^t \left[\hat{\tilde{\mathcal{H}}}_{\rm SB}(t),\left[\hat{\tilde{\mathcal{H}}}_{\rm SB}(t'),\hat{\tilde{\rho}}(t')\right]\right]dt'.
\end{equation}
Tracing out the bath degrees of freedom from both sides of above equation, we get
\begin{equation}
\frac{d\tilde{\rho}_{\rm S}(t)}{dt}=-\frac{i}{\hbar}{\rm Tr}_{\rm B}\left[\hat{\tilde{\mathcal{H}}}_{\rm SB}(t),\hat{\tilde{\rho}}(0)\right]-\frac{1}{\hbar^2}{\rm Tr}_{\rm B}\int_0^t \left[\hat{\tilde{\mathcal{H}}}_{\rm SB},\left[\hat{\tilde{\mathcal{H}}}_{\rm SB}(t'),\hat{\tilde{\rho}}(t')\right]\right]dt'.
\label{eq:bef_born}
\end{equation}

{\bf Born Approximation:}\\
Considering macroscopic and bulky environment, we can assume that the bath dynamics is fast and there is no change to bath statistics due to system-bath coupling. Factorising the system and bath degrees of freedom in the density matrix, we have
\begin{equation}
\hat{\tilde{\rho}}(t)=\hat{\tilde{\rho}}_{\rm S}(t)\otimes \hat{\tilde{\rho}}_{\rm B}.
\label{eq:born}
\end{equation}
Eq.~(\ref{eq:bef_born}) reduces to
\begin{equation}
\frac{d\hat{\tilde{\rho}}_{\rm S}(t)}{dt}=-\frac{i}{\hbar}{\rm Tr}_{\rm B}\left[\hat{\tilde{\mathcal{H}}}_{\rm SB}(t),\hat{\tilde{\rho}}_{\rm S}(0)\otimes \hat{\tilde{\rho}}_{\rm B}\right]-\frac{1}{\hbar^2}\int_0^t \left[\hat{\tilde{\mathcal{H}}}_{\rm SB}(t),\left[\hat{\tilde{\mathcal{H}}}_{\rm SB}(t'),\hat{\tilde{\rho}}_{\rm S}(t')\otimes \hat{\tilde{\rho}}_{\rm B}\right]\right]dt'.
\label{eq:b18}
\end{equation}
Above equation is second order in $\hat{\tilde{\mathcal{H}}}_{\rm SB}$ and all the higher order terms are neglected. This leads from the approximation in Eq.~(\ref{eq:born}) and is known as the Born approximation. The first term on the right hand side of above equation vanishes since 
\begin{equation}
{\rm Tr}_{\rm B}\left[\hat{B}(t)\hat{\tilde{\rho}}_{\rm B}(0)\right]=0.
\end{equation}
Using Eqs.~(\ref{eq:hsb_til}) and (\ref{eq:b18}), we obtain
\begin{equation}
\frac{d\hat{\tilde{\rho}}_{\rm S}(t)}{dt}=-\frac{1}{\hbar^2}\sum_{\alpha,\beta}\int_0^t dt'{\rm Tr}_{\rm B}\left[\hat{\tilde{A}}_\alpha(t)\otimes \hat{\tilde{B}}_{\alpha}(t),\left[\hat{\tilde{A}}_\beta(t')\otimes\hat{\tilde{B}}_{\beta}(t'),\hat{\tilde{\rho}}_{\rm S}(t')\otimes \hat{\tilde{\rho}}_{\rm B}\right]\right]
\end{equation}
Expanding the commutators, we obtain
\begin{align}
\frac{d\hat{\tilde{\rho}}_{\rm S}(t)}{dt}=-\frac{1}{\hbar^2}\sum_{\alpha\beta}\int_0^tdt'&\bigg[\hat{\tilde{A}}_\alpha(t)\hat{\tilde{A}}_\beta(t')\hat{\tilde{\rho}}_{\rm S}(t'){\rm Tr}_{\rm B}\left[\hat{\tilde{B}}_\alpha(t)\hat{\tilde{B}}_\beta(t')\hat{\tilde{\rho}}_{\rm B}\right]\nonumber \\
&-\hat{\tilde{A}}_\alpha(t)\hat{\tilde{\rho}}_{\rm S}(t')\hat{\tilde{A}}_\beta(t'){\rm Tr}_{\rm B}\left[\hat{\tilde{B}}_\alpha(t)\hat{\tilde{\rho}}_{\rm B}\hat{\tilde{B}}_\beta(t')\right]\nonumber \\
&-\hat{\tilde{A}}_\beta(t')\hat{\tilde{\rho}}_{\rm S}(t')\hat{\tilde{A}}_\alpha(t){\rm Tr}_{\rm B}\left[\hat{\tilde{B}}_\beta(t')\hat{\tilde{\rho}}_{\rm B}\hat{\tilde{B}}_\alpha(t)\right]\nonumber \\
&+\hat{\tilde{\rho}}_{\rm S}(t')\hat{\tilde{A}}_\beta(t')\hat{\tilde{A}}_\alpha(t){\rm Tr}_{\rm B}\left[\hat{\tilde{\rho}}_{\rm B}\hat{\tilde{B}}_\beta(t')\hat{\tilde{B}}_\alpha(t)\right]\bigg].
\end{align}
Using the cyclic property of trace and organizing the bath operators, we obtain
\begin{equation}
\frac{d\hat{\tilde{\rho}}_{\rm S}(t)}{dt}=-\frac{1}{\hbar^2}\sum_{\alpha\beta}\int_0^tdt'\bigg[C_{\alpha\beta}(t,t')\left[\hat{\tilde{A}}_\alpha(t),\hat{\tilde{A}}_\beta(t')\hat{\tilde{\rho}}_{\rm S}(t')\right]+C_{\beta\alpha}(t',t)\left[\hat{\tilde{\rho}}_{\rm S}(t')\hat{\tilde{A}}_\beta(t'),\hat{\tilde{A}}_\alpha(t)\right]\bigg],
\end{equation}
where
\begin{equation}
C_{\alpha\beta}(t_1,t_2)={\rm Tr}_{\rm B}\left[\hat{B}_\alpha(t_1)\hat{B}_\beta(t_2)\hat{\tilde{\rho}}_{\rm B}\right].
\end{equation}

{\bf Markov Approximation:}\\
Since, $\left[\hat{\tilde{\mathcal{H}}}_{\rm B},\hat{\tilde{\rho}}_{\rm B}\right]=0$, we have $C_{\alpha\beta}(t_1,t_2)=C_{\alpha\beta}(t_1-t_2)=C_{\alpha\beta}(\tau)$, where $\tau=t_1-t_2$. Further, assuming the bath operators in the contact Hamiltonian to be hermitian, i.e. $B_\alpha=B_\alpha^\dagger$, we have $C_{\alpha\beta}(\tau)=C_{\beta\alpha}^\dagger (-\tau)$. We will assume that the system density matrix varies very slowly compared to the decay time of the bath correlations. This assumptions leads us to the first Markov approximation,
\begin{equation}
\frac{d\tilde{\rho}_{\rm S}(t)}{dt}=-\frac{1}{\hbar^2}\sum_{\alpha\beta}\int_0^tdt'\bigg[C_{\alpha\beta}(\tau)\left[\hat{\tilde{A}}_\alpha(t),\hat{\tilde{A}}_\beta(t')\hat{\tilde{\rho}}_{\rm S}(t)\right]+C_{\beta\alpha}(-\tau)\left[\hat{\tilde{\rho}}_{\rm S}(t)\hat{\tilde{A}}_\beta(t'),\hat{\tilde{A}}_\alpha(t)\right]\bigg],
\end{equation}
where we replaced $\hat{\tilde{\rho}}_{\rm S}(t')$ with $\hat{\tilde{\rho}}_{\rm S}(t)$. Since, the bath correlation function decays rapidly, the integration limit can be extended to $t\rightarrow \infty$ without affecting the dynamics. We are only interested in the time far longer than the characteristic decay time of bath correlations. This is the second Markov approximation. Using the two Markov approximations the master equation is given by
\begin{equation}
\frac{d\hat{\tilde{\rho}}_{\rm S}(t)}{dt}=-\frac{1}{\hbar^2}\sum_{\alpha\beta}\int_0^\infty d\tau \bigg[C_{\alpha\beta}(\tau)\left[\hat{\tilde{A}}_\alpha(t),\hat{\tilde{A}}_\beta(t-\tau)\hat{\tilde{\rho}}_{\rm S}(t)\right]+C_{\beta\alpha}(-\tau)\left[\hat{\tilde{\rho}}_{\rm S}(t)\hat{\tilde{A}}_\beta(t-\tau),\hat{\tilde{A}}_\alpha(t)\right]\bigg].
\end{equation}
Going back to the Schr\"odinger picture, we obtain
\begin{equation}
\frac{d\hat{\rho}_{\rm S}(t)}{dt}=-\frac{i}{\hbar^2}\left[\hat{\cal H}_{\rm S},\hat{\rho}_{\rm S}(t)\right]-\frac{1}{\hbar}\sum_{\alpha\beta}\int_0^\infty d\tau \bigg[C_{\alpha\beta}(\tau)\left[\hat{A}_\alpha,\hat{A}_\beta(-\tau)\hat{\rho}_{\rm S}(t)\right]
+C_{\beta\alpha}(-\tau)\left[\hat{\rho}_{\rm S}(t)\hat{A}_\beta(-\tau),\hat{A}_\alpha\right]\bigg].
\label{eq:lind_red}
\end{equation}
Since $\hat{B}_\alpha(\tau)=\sum_j ih_j\left(\hat{b}_je^{-i\omega_j\tau}-\hat{b}_j^\dagger e^{i\omega_j\tau}\right)$, $C_{\alpha\beta}(\tau)$ reduces to
\begin{equation}
C_{\alpha\beta}(\tau)=\text{Tr}_{\rm B}\left[\hat{B}_\alpha(\tau)\hat{B}_\beta \rho_{\rm B} \right]=\sum_j|h_j|^2\left[e^{-i\omega_j\tau}\left(1+n(\omega_j)\right)+e^{i\omega_j\tau}n(\omega_j)\right],
\end{equation}
where $n(\omega_j)=\text{Tr}_{\rm B}\left[\hat{b}_j^\dagger \hat{b}_j \rho_{\rm B}\right]$ is the Bose-Einstein distribution function. Next, we will demonstrate the calculation of up to $({\cal O}(\varphi_{zps}^2))$ master equation, i.e. for $\hat{\cal H}_{\rm SB} = \hat{H}_{\rm SB} + \hat{\cal H}_{\rm SB}^{(2)}$. Note that, $\hat{\cal H}_{\rm SB}^{(1)}$, which is the first order in $\varphi_{\rm zps}$ contribution, is a scalar and does not induce any dynamics. In this case, we have
\begin{align}
\hat{A}_1(\tau)&=i\left(\hat{a}e^{-i\omega_d\tau/2}-\hat{a}^\dagger e^{i\omega_d\tau/2}\right),\nonumber \\
\hat{A}_2(\tau)&=\frac{3iG_{\rm 2,S}}{\hbar\omega_d}\left(\hat{a} e^{-3i\omega_d\tau/2}-\hat{a}^\dagger e^{3i\omega_d\tau/2}\right),\nonumber \\
\hat{A}_3(\tau)&=\frac{2iG_{\rm 2,S}}{\hbar\omega_d}\left(-\hat{a} e^{i\omega_d\tau/2}+\hat{a}^\dagger e^{-i\omega_d\tau/2}\right).
\end{align}
The general form of $\hat{A}_{\beta}$ can be written as
\begin{equation}
    \hat{A}_{\beta}(\tau)=i{A}_{\beta}^{(0)}(\hat{X}_\beta e^{i\omega_\beta \tau}-\hat{X}_{\beta}^{\dagger}e^{-i\omega_{\beta}\tau}),
\end{equation}
where $A_\beta^{(0)}$ is a constant. This rewriting of the system operator will hugely simplify the calculation later when we encounter multi-photon effects while considering $\hat{\cal H}_{\rm SB}^{(3)}$ and $\hat{\cal H}_{\rm SB}^{(4)}$.

The general expression for the integrals in Eq.~(\ref{eq:lind_red}) can then be written as
\begin{multline}
    I_{+\beta}=\frac{1}{\hbar^2}\int_{0}^{\infty}C_{\alpha\beta}(\tau)\hat{A}_{\beta}(-\tau)d\tau=\frac{iA^{(0)}_{\beta}}{2\hbar^2}\Bigg\{n'(\omega_j)\hat{X}_{\beta}\left[2\pi\sum_{j}|h_j|^2 \delta(-\omega_{\beta}-\omega_j)\right]
    -n'(\omega_j)\hat{X}_{\beta}^{\dagger}\left[2\pi\sum_{j}|h_j|^2 \delta(\omega_{\beta}-\omega_j)\right]\\
    +n(\omega_j)\hat{X}_{\beta}\left[2\pi\sum_{j}|h_j|^2 \delta(\omega_j-\omega_{\beta})\right]
    -n(\omega_j)\hat{X}^{\dagger}_{\beta}\left[2\pi\sum_{j}|h_j|^2 \delta(\omega_{\beta}+\omega_j)\right]\Bigg\},
    \label{eq:red_int_1}
\end{multline}
where $n^\prime(\omega_j) = (1 + n (\omega_j))$. Similarly,
\begin{multline}
    I_{-\beta}=\frac{1}{\hbar^2}\int_{0}^{\infty}C_{\alpha\beta}(-\tau)\hat{A}_{\beta}(-\tau)d\tau=\frac{iA_{\beta}^{(0)}}{2\hbar^2}\Bigg\{n^\prime(\omega_j)X_{\beta}\left[2\pi\sum_{j}|h_j|^2 \delta(\omega_j-\omega_{\beta})\right]
    -n^\prime(\omega_j)X^{\dagger}_{\beta}\left[2\pi\sum_{j}|h_j|^2 \delta(\omega_{\beta}+\omega_j)\right]\\
    +n(\omega_j)X_{\beta}\left[2\pi\sum_{j}|h_j|^2 \delta(-\omega_j-\omega_{\beta})\right]
    -n(\omega_j)X^{\dagger}_{\beta}\left[2\pi\sum_{j}|h_j|^2 \delta(\omega_{\beta}-\omega_j)\right]\Bigg\}
    \label{eq:red_int_2}
\end{multline}
Defining the spectral density of the bath as
\begin{equation}
\kappa(\omega)=\frac{2\pi}{\hbar} \sum_j |h_j|^2 \delta(\omega_j-\omega),
\end{equation}
we obtain
\begin{align}
&\hat{I}_{+\beta} = \frac{iA_{\beta}^{(0)}\hat{X}_\beta}{2\hbar}\left(n(\omega_\beta)\kappa(\omega_\beta)+n^\prime(-\omega_\beta)\kappa(-\omega_\beta)\right)-\frac{iA_{\beta}^{(0)}\hat{X}_\beta^\dagger}{2\hbar}\left(n^\prime(\omega_\beta)\kappa(\omega_\beta)+n(-\omega_\beta)\kappa(-\omega_\beta)\right),\nonumber \\
&\hat{I}_{-\beta} = \frac{iA_{\beta}^{(0)}\hat{X}_\beta}{2\hbar}\left(n^\prime(\omega_\beta)\kappa(\omega_\beta)+n(-\omega_\beta)\kappa(-\omega_\beta)\right)-\frac{iA_{\beta}^{(0)}\hat{X}_\beta^\dagger}{2\hbar}\left(n^\prime(-\omega_\beta)\kappa(-\omega_\beta)+n(\omega_\beta)\kappa(\omega_\beta)\right).
\end{align}
Using above equations, Eq.~(\ref{eq:lind_red}) reduces to
\begin{equation}
\frac{d\hat{\rho}_{\rm S}(t)}{dt}=-\frac{i}{\hbar}\left[\hat{\cal H}_{\rm S},\hat{\rho}_{\rm S}(t)\right]-\sum_{\alpha\beta}\left(\left[\hat{A}_\alpha,\hat{I}_{+\beta}\hat{\rho}_{\rm S}(t)\right]+\left[\hat{\rho}_{\rm S}(t)\hat{I}_{-\beta},\hat{A}_\alpha\right]\right).
\label{eq:lind_red_2}
\end{equation}
Let us consider one of the possible cases, $\alpha=\beta =1$, such that $A_\alpha = A_1 = i\left(\hat{a}-\hat{a}^\dagger\right)$, $A_{\beta}^{(0)}=1$, $X_{\beta}=-\hat{a}^\dagger$, $X^{\dagger}_{\beta}=-\hat{a}$, and $\omega_{\beta}=\frac{\omega_d}{2}$. We find
\begin{equation}
\hat{I}_{\pm 1} = \frac{-i\hat{a}^\dagger}{2}\gamma(\pm \omega_d/2)+\frac{i\hat{a}}{2}\gamma(\mp\omega_d/2),
\end{equation}
where we defined the transition rate as $\gamma(\omega) = \kappa(\omega)n(\omega)/\hbar$. Similarly, we can calculate the integrals $\hat{I}_{\pm \beta}$ for all $\beta$ and substitute it in Eq.~(\ref{eq:lind_red_2}). After some calculations, we obtain
\begin{multline}
\frac{d\hat{\rho}_{\rm S}}{dt} =\frac{-i}{\hbar}[\mathcal{H}_s,\hat{\rho}_{\rm S}(t)]+\frac{1}{\hbar}\kappa(\omega_d/2)\bigg\{n(\omega_d/2)\mathcal{D}\bigg[a^\dagger+ \frac{2G_{2,{\rm S}}}{\hbar\omega_d}a\bigg]\hat{\rho}_{\rm S}(t)+[n(\omega_d/2)+1] \mathcal{D}\bigg[a+\frac{2G_{2,{\rm S}}}{\hbar\omega_d}a^\dagger \bigg]\hat{\rho}_{\rm S}(t)\bigg\}\\
    +\frac{1}{\hbar}\bigg(\frac{3G_{2,{\rm S}}}{\hbar\omega_d}\bigg)^2\kappa(3\omega_d/2)\bigg\{ n(3\omega_d/2)\mathcal{D}[a^\dagger]\hat{\rho}_{\rm S}(t)+[n(3\omega_d/2)+1]\mathcal{D}[a]\hat{\rho}_{\rm S}(t)\bigg\}. 
\end{multline}
The above master equation is ${\cal O}(\varphi_{\rm zps}^2)$. In order to go beyond second order in $\varphi_{\rm zps}$, we will consider the higher order contribution to the system-bath coupling Hamiltonian. Taking $\hat{\cal H}_{\rm SB}^{(3)}$ and $\hat{\cal H}_{\rm SB}^{(4)}$, we obtain the effective Lindblad master equation given by
\begin{multline}
\frac{d\hat{\rho}_{\rm S}^{3,4}}{dt} =\frac{1}{\hbar}\kappa(\omega_d)\bigg\{n(\omega_d)\mathcal{D}\bigg[\frac{8G_{\rm 3,S}}{\hbar\omega_{\rm d}}\hat{a}^{\dagger^2}\bigg]\hat{\rho}_{\rm S}(t)+[n(\omega_d)+1] \mathcal{D}\bigg[\frac{8G_{\rm 3,S}}{\hbar\omega_{\rm d}}\hat{a}^{2}\bigg]\hat{\rho}_{\rm S}(t)\bigg\}\\
    +\frac{1}{\hbar}\kappa(\omega_d/2)\bigg\{n(\omega_d/2)\mathcal{D}\bigg[\frac{3G_{\rm 2,S}^2}{2\hbar^2\omega_{\rm d}^2}\hat{a}^\dagger-\frac{24G_{\rm 1,S}G_{\rm 3,S}}{\hbar^2\omega_{\rm d}^2}\hat{a}^\dagger+\frac{12G_{\rm 4,S}}{\hbar\omega_{\rm d}}\hat{a}+\frac{4G_{\rm 4,S}}{\hbar\omega_{\rm d}}\hat{a}^{\dagger^3}+\frac{12G_{\rm 4,S}}{\hbar\omega_{\rm d}}\hat{a}^\dagger \hat{a}^2\bigg]\hat{\rho}_{\rm S}(t)\\+[n(\omega_d/2)+1] \mathcal{D}\bigg[\frac{3G_{\rm 2,S}^2}{2\hbar^2\omega_{\rm d}^2}\hat{a}-\frac{24G_{1,S}G_{\rm 3,S}}{\hbar^2\omega_{\rm d}^2}\hat{a}+\frac{12G_{\rm 4,S}}{\hbar\omega_{\rm d}}\hat{a}^\dagger+\frac{4G_{\rm 4,S}}{\hbar\omega_{\rm d}}\hat{a}^{3}+\frac{12G_{\rm 4,S}}{\hbar\omega_{\rm d}}\hat{a}^{\dagger^2} \hat{a}\bigg]\hat{\rho}_{\rm S}(t)\bigg\}\\
    +\frac{1}{\hbar}\kappa(3\omega_d/2)\bigg\{ n(3\omega_d/2)\mathcal{D}\bigg[\frac{18G_{\rm 4,S}}{\hbar\omega_{\rm d}}\hat{a}^\dagger-\frac{11G_{\rm 2,S}^2}{\hbar^2\omega_{\rm d}^2}\hat{a}+\frac{12\Tilde{G}_{\rm 1,S}G_{\rm 3,S}}{5\hbar^2\omega_{\rm d}^2}\hat{a}+\frac{12G_{\rm 4,S}}{\hbar\omega_{\rm d}}\hat{a}^{\dagger^2}\hat{a}\bigg]\hat{\rho}_{\rm S}(t)\\
    +[n(3\omega_d/2)+1]\mathcal{D}\bigg[\frac{18G_{\rm 4,S}}{\hbar\omega_{\rm d}}\hat{a}-\frac{11G_{\rm 2,S}^2}{\hbar^2\omega_{\rm d}^2}\hat{a}^\dagger+\frac{12\Tilde{G}_{\rm 1,S}G_{\rm 3,S}}{5\hbar^2\omega_{\rm d}^2}\hat{a}^\dagger+\frac{12G_{\rm 4,S}}{\hbar\omega_{\rm d}}\hat{a}^{\dagger}\hat{a}^2\bigg]\hat{\rho}_{\rm S}(t)\bigg\}\\
    +\frac{1}{\hbar}\kappa(5\omega_d/2)\bigg\{ n(5\omega_d/2)\mathcal{D}\bigg[\frac{5G_{\rm 2,S}^2}{3\hbar^2\omega_{\rm d}^2}\hat{a}^\dagger+\frac{4\Tilde{G}_{\rm 1,S}G_{\rm 3,S}}{3\hbar^2\omega_{\rm d}^2}\hat{a}^\dagger+\frac{4G_{\rm 4,S}}{3\hbar\omega_{\rm d}}\hat{a}^{\dagger^3}\bigg]\hat{\rho}_{\rm S}(t)\\
    +[n(5\omega_d/2)+1]\mathcal{D}\bigg[\frac{5G_{\rm 2,S}^2}{3\hbar^2\omega_{\rm d}^2}\hat{a}+\frac{4\Tilde{G}_{\rm 1,S}G_{\rm 3,S}}{3\hbar^2\omega_{\rm d}^2}\hat{a}+\frac{4G_{\rm 4,S}}{3\hbar\omega_{\rm d}}\hat{a}^{3}\bigg]\hat{\rho}_{\rm S}(t)\bigg\},
    \label{eq:mas_eq_deph}
\end{multline}
where the terms in the first line are ${\cal O}(\varphi_{\rm zps}^3)$ whereas the latter terms are ${\cal O}(\varphi_{\rm zps}^4)$.

\subsection{Single photon processes and RWA}
\label{app:RWA_lifetime}
The master equation in the RWA can be written as
\begin{equation}
\partial_t\hat{\rho}_{\rm S}=\hat{\cal L}\hat{\rho}_{\rm S}=-\frac{i}{\hbar}\left[\hat{\cal H}_{\rm S},\hat{\rho}_{\rm S}\right]+\frac{1}{\hbar}\kappa {n}(\omega_d/2){\cal D}[\hat{a}^\dagger]\hat{\rho}_{\rm S}+\frac{1}{\hbar}\kappa (1+{n}(\omega_d/2)){\cal D}[\hat{a}]\hat{\rho}_{\rm S},
\end{equation}
where $\hat{\cal L}$ is the Liouvillian. We consider the spectral function to be energy independent such that $\kappa(\omega_d/2)=\kappa$. 
First we will calculate the eigenstates and eigenventors of $\hat{\cal H}_{\rm S}$,
\begin{equation}
\hat{\cal H}_{\rm S} \ket{\psi_{m}^{\pm}}= \omega_m^{\pm} \ket{\psi_{m}^{\pm}}.
\end{equation}
The lowest energy eigenstates of the Hamiltonian are given by the degenerate cat state manifold $\psi_0^\pm = {\cal C}_\alpha^{\pm}$. The lowest eigenvalue of the Liouvillian ($\lambda$) will give the coherence life time of the qubit, $T_X = [-{\rm Re}\lambda]^{-1}$. The Liouvillian can be calculated from the master equation,
\begin{equation}
\hat{\cal L}\cdot=-\frac{i}{\hbar}\left[\hat{\cal H}_{\rm S},\cdot\right] + \frac{1}{\hbar}\kappa {n}(\omega_d/2) {\cal D}[\hat{a}^\dagger]\cdot +\frac{1}{\hbar} \kappa (1+{n}(\omega_d/2)) {\cal D}[\hat{a}]\cdot
\label{eq:liou_rwa}
\end{equation}
Following Ref.~\cite{frattini2022}, we separate out the population and coherence subspace, we have
\begin{align}
&\hat{\cal B}_{\rm pop} =\left\{\ket{\psi_m^+}\bra{\psi_m^+},\ket{\psi_m^+}\bra{\psi_m^+},\cdots\right\}~~~~{\rm for~} m\geq 0,\\ \nonumber
&\hat{\cal B}_{\rm coh} =\left\{\ket{\psi_m^+}\bra{\psi_m^-},\ket{\psi_m^-}\bra{\psi_m^+},\cdots\right\}~~~~{\rm for~} m\geq 0.
\end{align}
We do not need to consider all $m$, since the higher lying states have smaller probabilities of getting populated. For example, for $\epsilon_2=0$, all the states with $\omega_n^+ -\omega_n^- >\kappa$ are beyond the reach of being excited. 

Furthermore, we observe that under single photon loss and gain, the master equation corresponding to population and coherence are decoupled. Hence, we can study the decoherence mechanism separately from the population. In addition, the change in $\hat{\rho}_{\rm S}$ is induced by the change in coherent state, the population states describe the steady state of the Lindbladian. The density matrix vector takes the following form
\begin{equation}
\begin{bmatrix}
{\rho}_{\rm S,0}^{\pm} &
{\rho}_{\rm S,1}^{\pm}&
{\rho}_{\rm S,2}^{\pm}&
\hdots &
{\rho}_{\rm S,0}^{\mp}&
{\rho}_{\rm S,1}^{\mp}&
{\rho}_{\rm S,2}^{\mp}&
\hdots
\end{bmatrix}^T,
\end{equation}
where ${\rho}_{{\rm S},m}^{\pm} = \bra{\psi_m^+}\hat{\rho}_{\rm S}\ket{\psi_m^-}$. The corresponding Liouvillian superoperator is given by
\begin{equation}
\hat{\bf L} = \hat{\bf L}_{\rm H} + \hat{\bf L}_{\rm D},
\end{equation}
where $H(D)$ determined the unitary (dissipative) part of the Liouvillian. First of all, let's calculate the unitary part
\begin{equation}
-i\bra{\psi_m^+}\left[\hat{\cal H}_{\rm S},\hat{\rho}_{\rm S}\right] \ket{\psi_m^-}=-i\left(\omega_m^+-\omega_m^-\right)\rho_{{\rm S},m}^{\pm}=-i\delta_m \rho_{{\rm S},m}^{\pm},
\end{equation}
where $\delta_m = \omega_m^+ - \omega_m^-$. Note that, $\omega_0^+-\omega_0^-=\delta_0=0$. Hence, in the block diagonal form,
\begin{equation}
\hat{\bf L}_{\rm H}=\begin{bmatrix}
-\Delta & 0 \\
0 & \Delta 
\end{bmatrix},
\end{equation}
where 
\begin{equation}
\Delta =\begin{bmatrix}
0 & ~ & ~ & ~\\
~ & i\delta_1 & ~ &~\\
~&~&i\delta_2 & ~\\
~&~&~ & \ddots
\end{bmatrix}.
\end{equation}
To calculate $\hat{\mathbf{L}}_{\rm D}$, we first calculate $\bra{\psi^+_m}\mathcal{D}[\hat{O}]\hat{\rho}_{\rm S}\ket{\psi^-_m}$ for a general operator $\hat{O}$. We can write $\hat{\rho}_{\rm S}$ in the $\hat{\mathcal{B}}_{\mathrm{coh}}$ basis and rewrite above expression as
\begin{multline}
        \bra{\psi^+_m}\mathcal{D}[\hat{O}]\hat{\rho}_{\rm S}\ket{\psi^-_m}=\bra{\psi^+_m}\hat{O}\bigg(\sum_{\alpha\beta, i}\rho^{\alpha\beta}_{{\rm S},i}\ket{\psi^\alpha_i}\bra{\psi^\beta_i}\bigg)\hat{O}^\dagger\ket{\psi^-_m}\\
        -\frac{1}{2}\bra{\psi^+_m}\bigg[ \hat{O}^\dagger \hat{O}\bigg(\sum_{\alpha\beta ,i}\rho^{\alpha\beta}_{{\rm S},i}\ket{\psi^\alpha_i}\bra{\psi^\beta_i}\bigg)+\bigg(\sum_{\alpha\beta, i}\rho^{\alpha\beta}_{{\rm S},i}\ket{\psi^\alpha_i}\bra{\psi^\beta_i}\bigg)\hat{O}^\dagger \hat{O} \bigg]\ket{\psi^-_m},
\end{multline}
where $\alpha,\beta=+,-$. With further simplification, the above equation reduces to
\begin{equation}
\bra{\psi^+_m}\mathcal{D}[\hat{O}]\hat{\rho}_{\rm S}\ket{\psi^-_m}=
    \sum_{\alpha\beta, i}\bra{\psi^{+}_m}\hat{O}\ket{\psi^\alpha_i}\bra{\psi^\beta_i}\hat{O}^\dagger\ket{\psi^-_m}\rho^{\alpha\beta}_{{\rm S},i}-\frac{1}{2}\bigg[ \sum_{\alpha}\bra{\psi^+_m}\hat{O}^\dagger \hat{O}\ket{\psi^\alpha_m}\rho^{\alpha -}_{{\rm S}
    ,m}+\sum_{\beta}\bra{\psi^\beta_m}\hat{O}^\dagger \hat{O}\ket{\psi^-_m}\rho^{+ \beta}_{{\rm S},m} \bigg].
\end{equation}
The $\alpha$ in the second summation can only be $+$ and the $\beta$ in the third summation can only be $-$, so the second and third summations can be dropped and we obtain
\begin{equation}
\bra{\psi^+_m}\mathcal{D}[\hat{O}]\hat{\rho}_{\rm S}\ket{\psi^-_m}=
    \sum_{\alpha\beta, i}\bra{\psi^{+}_m}\hat{O}\ket{\psi^\alpha_i}\bra{\psi^\beta_i}\hat{O}^\dagger\ket{\psi^-_m}\rho^{\alpha\beta}_{{\rm S},i}-\frac{1}{2}\bigg[\bra{\psi^+_m}\hat{O}^\dagger \hat{O}\ket{\psi^+_m}+\bra{\psi^-_m}\hat{O}^\dagger \hat{O}\ket{\psi^-_m} \bigg]\rho^{\pm}_{{\rm S},m}.
    \label{eq:simpl_liou_rwa}
\end{equation}
Substituting Eq.~(\ref{eq:simpl_liou_rwa}) into Eq.~(\ref{eq:liou_rwa}), we can rewrite the dissipative part of the Liouvillian superoperator in the matrix form as

\begin{equation}
    \hat{\mathbf{L}}_{\rm D}=\kappa(1+n(\omega_d/2))\begin{bmatrix}
    -\mathbf{A} & \mathbf{B}\\
    \mathbf{B} & -\mathbf{A}\\
    \end{bmatrix}+\kappa n (\omega_d)\begin{bmatrix}
     -\mathbf{C} & \mathbf{D}\\
    \mathbf{D} & -\mathbf{C}\\   
    \end{bmatrix},
\end{equation}
where the block matrices $\mathbf{A,B,C,D}$ are given by
\begin{equation}
   \mathbf{A}=\begin{bmatrix}
 A_0 & ~ & ~ & ~\\
~ & A_1 & ~ &~\\
~&~& A_2 & ~\\
~&~&~ & \ddots      
\end{bmatrix},
\mathbf{B}=\begin{bmatrix}
 B_{00} & B_{01} & B_{02} & ~\\
B_{10} & B_{11} & B_{12} & ~\\
B_{20} & B_{21} & B_{22} & ~\\
~&~&~ & \ddots      
\end{bmatrix},
       \mathbf{C}=\begin{bmatrix}
 C_0 & ~ & ~ & ~\\
~ & C_1 & ~ &~\\
~&~& C_2 & ~\\
~&~&~ & \ddots      
\end{bmatrix},
\mathbf{D}=\begin{bmatrix}
 D_{00} & D_{01} & D_{02} & ~\\
D_{10} & D_{11} & D_{12} & ~\\
D_{20} & D_{21} & D_{22} & ~\\
~&~&~ & \ddots      
\end{bmatrix},
\end{equation}
where the entries for each matrices are given by
\begin{equation}
    A_{m}=\frac{1}{2}\bigg[\bra{\psi^+_{m}}\hat{O}^\dagger \hat{O}\ket{\psi^+_m}+\bra{\psi^-_{m}}\hat{O}^\dagger \hat{O}\ket{\psi^-_m}\bigg],~~
    B_{mp}=\bra{\psi^-_{m}}\hat{O}\ket{\psi^+_{p}}\bra{\psi^-_{p}}\hat{O}^\dagger\ket{\psi^+_{m}},
    \label{eq:mat_el_AB}
\end{equation}
and
\begin{equation}
    C_{m}=\frac{1}{2}\bigg[\bra{\psi^+_{m}}\hat{O}\hat{O}^\dagger\ket{\psi^+_m}+\bra{\psi^-_{m}}\hat{O}\hat{O}^\dagger\ket{\psi^-_m}\bigg],~~
    D_{mp}=\bra{\psi^-_{m}}\hat{O}^\dagger\ket{\psi^+_{p}}\bra{\psi^-_{p}}\hat{O}\ket{\psi^+_{m}}.
    \label{eq:mat_el_CD}
\end{equation}

\subsection{Beyond RWA and multi-photon processes}
\label{eq:nonrwa_lifetime}
The Liouvillian that we want to calculate is for the master equation up to $\varphi_{\rm zps}^3$ which includes multi-photon effects
\begin{multline}
\frac{\partial \hat{\rho}_{\rm S}}{\partial t}=-\frac{i}{\hbar}[\mathcal{H}_{\rm S},\hat{\rho}_{\rm S}(t)]+\frac{1}{\hbar}\kappa(\omega_d/2)\bigg\{n(\omega_d/2)\mathcal{D}\bigg[\hat{a}^\dagger+\frac{2G_{\rm 2,S}}{\hbar\omega_{\rm d}}\hat{a}\bigg]\hat{\rho}_{\rm S}+[1+n(\omega_{\rm d}/2)]\mathcal{D}\bigg[\hat{a}+\frac{2G_{\rm 2,S}}{\hbar\omega_{\rm d}}\hat{a}^\dagger\bigg]\hat{\rho}_{\rm S}\bigg\}\\
+\frac{1}{\hbar}\bigg(\frac{3G_{\rm 2,S}}{\hbar\omega_{{\rm d}}}\bigg)^2\kappa(3\omega_{\rm d}/2)\bigg\{n(3\omega_{\rm d}/2)\mathcal{D}[a^{\dagger}]\hat{\rho}_{\rm S}+[1+n(3\omega_{\rm d}/2)]\mathcal{D}[\hat{a}]\hat{\rho}_{\rm S}\bigg\}\\
+\frac{1}{\hbar}\bigg(\frac{8 G_{\rm 3,S}}{\hbar\omega_{{\rm d}}}\bigg)^2\kappa(\omega_{\rm d})\bigg\{n(\omega_{\rm d})\mathcal{D}[\hat{a}^{\dagger^2}]\hat{\rho}_{\rm S}+[1+n(\omega_{\rm d})]\mathcal{D}[\hat{a}^{2}]\hat{\rho}_{\rm S}\bigg\}.
\end{multline}
The Liouvillian for the unitary part is the same as in the single photon case. The effective Liouvillian for the decoherence part is
\begin{equation}
\hat{\mathbf{L}}_{\rm eff}=\hat{\mathbf{L}}_{\rm H}+\hat{\mathbf{L}}_{\rm D1}+\hat{\mathbf{L}}_{\rm D2}+\hat{\mathbf{L}}_{\rm D3}
\end{equation}
where
\begin{equation}
    \hat{\mathbf{L}}_{\rm Di}=\frac{1}{\hbar}\kappa(\omega_i)(1+n(\omega_i))\begin{bmatrix}
    -\mathbf{A}^{{\rm D}i} & \mathbf{B}^{{\rm D}i}\\
    \mathbf{B}^{{\rm D}i} & -\mathbf{A}^{{\rm D}i}\\
    \end{bmatrix}+\frac{1}{\hbar}\kappa(\omega_i) n (\omega_i)\begin{bmatrix}
     -\mathbf{C}^{{\rm D}i} & \mathbf{D}^{{\rm D}i}\\
    \mathbf{D}^{{\rm D}i} & -\mathbf{C}^{{\rm D}i}\\ 
    \end{bmatrix}.
\end{equation}
In above equation, we defined $\omega_1 = \omega_{\rm d}/2$, $\omega_2 = 3\omega_{\rm d}/2$ and $\omega_3 = \omega_{\rm d}$. The matrix elements of the matrices $\mathbf{A},\mathbf{B},\mathbf{C}$ and $\mathbf{D}$ are given by  Eqs.~(\ref{eq:mat_el_AB}) and (\ref{eq:mat_el_CD}) by suitably replacing the operator $\hat{O}$ by the corresponding single photon or multi-photon annihilation and creation operator. For instance, in the case of $\hat{\mathbf{L}}_{\rm D3}$, $\hat{O}=\left(\frac{G_{\rm 3,S}}{8\hbar\omega_{\rm d}}\right)^2\hat{a}^2$ for the matrices $\mathbf{A}$ and $\mathbf{B}$, and $\hat{O}=\left(\frac{G_{\rm 3,S}}{8\hbar\omega_{\rm d}}\right)^2\hat{a}^{\dagger^2}$ for the matrices $\mathbf{C}$ and $\mathbf{D}$.

\subsection{Effect of Dephasing}
\label{app:dephasing}
In order to understand the effect of dephasing on $T_\alpha$, we add an extra term to the master equation, $ \gamma_\phi \mathcal{D}[\hat{a}^\dagger \hat{a}]\hat{\rho}_{\rm S}$, where $\gamma_\phi$ determines the dephasing strength. In Fig.~\ref{fig:lifetime_deph}, we plot the lifetime as a function of the two-photon drive strength for different values of the dephasing strength. We consider 10 STS connected in series with a single junction in each branch. The dot-dashed purple curve done for no dephasing matches with the plot in Fig.~\ref{fig:lifetime_phizps}. However, we observe a reduction in lifetime when the dephasing is introduced. Although the lifetime follows a staircase pattern, the lifetime where the plateaus occur are reduced. Further, the reduction is not linear as a function of dephasing strength. We find that for $\gamma_\phi/K = 10^{-6}$, the lifetime plot (dotted red curve) runs very close to the plot for $\gamma_\phi=0$. However, the lifetime gets drastically reduced for ten fold increase in the dephasing strength (black dashed curve). The reduction is even more significant for $\gamma_\phi/K = 10^{-4}$ (solid blue curve) to the point that the lifetime for the two-photon drive strength $\epsilon_2/K\approx 2$ is almost the same as for the two-photon drive strength $\epsilon_2/K\approx 10$.
\begin{figure}[H]
\centering
\includegraphics[width=0.6\columnwidth]{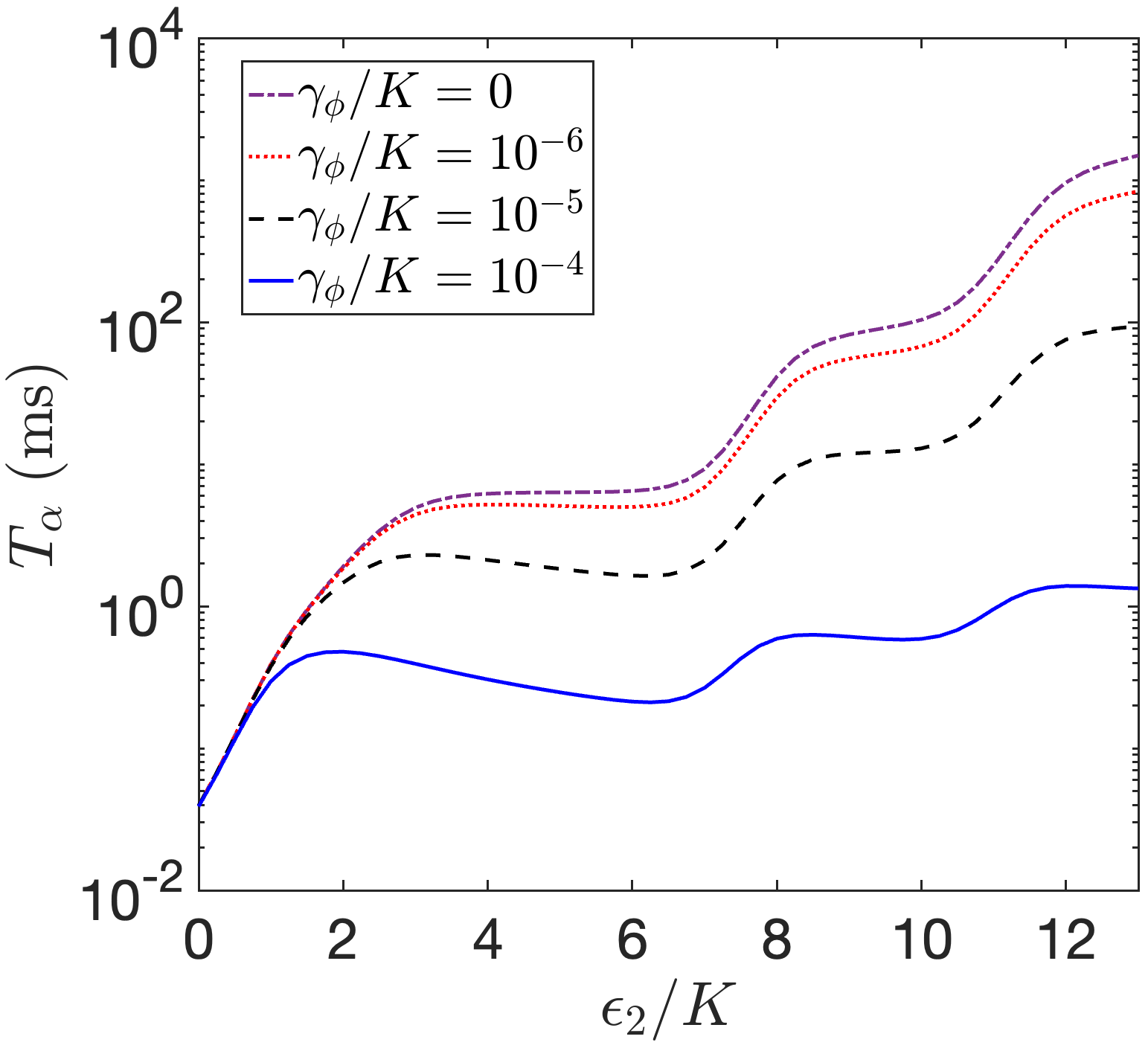}
 \caption{$T_\alpha$ as a function of the two-photon drive strength $\epsilon_2/K$ in the presence and absence of dephasing for different values of $\gamma_\phi$. We use the same parameters as in Fig.~\ref{fig:lifetime_phizps}.} 
\label{fig:lifetime_deph}
\end{figure}

\subsection{Effect of stronger modulation depth}
\label{app:strong}
In this section, we will study the effect of higher order terms in modulation depth to the master equation of STS taking symmetric junctions. If we have a stronger modulation depth, we must further expand $\sin{(\delta\varphi\cos(\omega_d t))}$ to third order, giving us a modulation depth of 
\begin{equation}
    \sin{(\delta\varphi\cos(\omega_d t))}\approx\delta\varphi\cos(\omega_d t)-\frac{1}{3!}[\delta\varphi\cos(\omega_d t)]^3.
\end{equation}
This modifies the Hamiltonian in Eq.~(\ref{eq:main_H_non_rot}) to
\begin{equation}
    \hat{H}_{\textrm{lab}}=4E_{\rm C}\hat{n}^2-E_{\rm J2}\cos{\hat{\varphi}}-2E_{\rm J\Sigma}\bigg[\delta\varphi\cos(\omega_d t)-\frac{1}{3!}[\delta\varphi\cos(\omega_d t)]^3\bigg]\cos\hat{\varphi},
\end{equation}
where we assumed $E_{\rm J\Delta}=0$. We write $\cos(\omega_d t)$  in terms of exponential and Taylor expand $\cos{\hat{\varphi}}$ to fourth order to obtain
\begin{multline}
    \hat{H}_{\textrm{lab}}=4E_{\rm C}\hat{n}^2-E_{\rm J2}\cos{\hat{\varphi}}-2E_{\rm J\Sigma}\bigg[\frac{\delta\varphi}{2}(e^{i\omega_d t}+e^{-i\omega_d t})
    -\frac{\delta\varphi^3}{48}(e^{3i\omega_d t}+3e^{i\omega_d t}+3e^{-i\omega_d t}+e^{-3i\omega_d t})\bigg]\bigg(1-\frac{\hat{\varphi}^2}{2!}+\frac{\hat{\varphi}^4}{4!}\bigg).
\end{multline}
If we write $\hat{\varphi}$ in terms of annihilation and creation operators as well as drop all scalars, we will get the Hamiltonian
\begin{equation}
\hat{H}_{\rm lab}=\epsilon_c \hat{a}^\dagger \hat{a} - K \hat{a}^{\dagger^2} \hat{a}^2+2E_{\rm J\Sigma}\bigg[\bigg(\frac{\delta\varphi}{2}-\frac{\delta\varphi^3}{16}\bigg)(e^{i\omega_d t}+e^{-i\omega_d t})
-\frac{\delta\varphi^3}{48}(e^{3i\omega_d t}+e^{-3i\omega_d t})\bigg]
\bigg[\frac{\varphi_{zps}^2}{2!}(\hat{a}^{\dagger}+\hat{a})^2-\frac{\varphi_{zps}^4}{4!}(\hat{a}^\dagger+\hat{a})^4\bigg].
\end{equation}
Now if we move to the rotating frame using $\hat{a}\longrightarrow \hat{a}e^{-i\omega_dt/2}$ and rearrange some terms, we obtain
\begin{equation}
\hat{H}_{\rm T} =  \delta \hat{a}^\dagger \hat{a} - K \hat{a}^{\dagger^2} \hat{a}^2,
\end{equation}
and
\begin{multline}
    \hat{H}_{\textrm{SQ}}^{\rm sym}=2E_{J\Sigma}\bigg[\bigg(\frac{\delta\varphi}{2}-\frac{\delta\varphi^3}{16}\bigg)(e^{i\omega_d t}+e^{-i\omega_d t})-\frac{\delta\varphi^3}{48}(e^{3i\omega_d t}+e^{-3i\omega_d t})\bigg]\\
    \bigg[\frac{\varphi_{zps}^2}{2!}(a^\dagger e^{i\omega_dt/2}+ae^{-i\omega_dt/2})^2-\frac{\varphi_{zps}^4}{4!}(a^\dagger e^{i\omega_dt/2}+ae^{-i\omega_dt/2})^4\bigg]
\end{multline}
for the transmon and symmetric SQUID Hamiltonian, respectively. Following Appendix~\ref{app:static_eff}, we perform the generalized Schrieffer-Wolff transformation and obtain the following static effective Hamiltonian
\begin{equation}
    \hat{\mathcal{H}}_{\rm S}=\Delta \hat{a}^\dagger \hat{a}+(G_{2,S}-G_{2,S}'+6G_{4,S}-6G_{4,S}')(\hat{a}^{\dagger^2}+\hat{a}^2)
    -K\hat{a}^{\dagger^2}\hat{a}^2
    +4(G_{4,S}-G_{4,S}')(\hat{a}^\dagger \hat{a}^3+\hat{a}^{\dagger^3}\hat{a})
\end{equation}
where $G_{\rm 2,S}=\frac{\delta\varphi E_{\rm J\Sigma}\varphi_{\rm zps}^2}{2}$ , $G_{\rm 2,S}'=\frac{\delta\varphi^3E_{\rm J\Sigma}\varphi_{\rm zps}^2}{16}$, $G_{\rm 4,S}=-\frac{\varphi_{\rm zps}^2}{12}G_{\rm 2,S}$, $G_{\rm 4,S}'=\frac{\varphi_{\rm zps}^2}{12}G_{\rm 2,S}'$, and the detuning, $\Delta=\delta-\frac{2(G_{\rm 2,S}-G'_{\rm 2,S})^2}{\omega_{\rm d}}+\frac{G_{\rm 2,S}'^2}{9\omega_{\rm d}}$. Similarly, the static effective interaction Hamiltonian is given by
\begin{multline}
    \hat{\mathcal{H}}_{\rm SB}=\frac{E_{\rm J\Sigma}\varphi^2_{\rm zps}}{i\omega_{\rm d}}\bigg[\frac{3}{2}\bigg(\delta\varphi-\frac{\delta\varphi^3}{8}\bigg)(\hat{a}^\dagger e^{i3\omega_{\rm d}t/2}-\hat{a}e^{-i3\omega_{\rm d}t/2})+\frac{5\delta\varphi^3}{144}(\hat{a}^\dagger e^{-i5\omega_{\rm d}t/2}-\hat{a}e^{i5\omega_{\rm d} t/2})
    \\
    +\frac{11\delta\varphi^3}{576}(\hat{a}e^{-i7\omega_{\rm d} t/2}-\hat{a}^\dagger e^{i7\omega_{\rm d} t/2})+\bigg(\delta\varphi-\frac{\delta\varphi^3}{8}\bigg)(\hat{a}e^{i\omega_{\rm d} t/2}-\hat{a}^\dagger e^{-i\omega_{\rm d} t/2})\bigg]\hat{B}(t).
\end{multline}
Finally, the master equation up to order $\varphi_{zps}^2$ is given by
\begin{multline}
    \partial_t\hat{\rho}_{\rm S}(t)=\frac{1}{\hbar}\kappa(\omega_{\rm d}/2)\bigg\{n(\omega_{\rm d}/2)\mathcal{D}\bigg[\hat{a}^\dagger+ \frac{2(G_{\rm 2,S}-G_{\rm 2,S}')}{\hbar\omega_{\rm d}}\hat{a}\bigg]\hat{\rho}_{\rm S}(t)
    +[n(\omega_{\rm d}/2)+1] \mathcal{D}\bigg[\hat{a}+\frac{2(G_{\rm 2,S}-G_{\rm 2,S}')}{\hbar\omega_{\rm d}}\hat{a}^\dagger \bigg]\hat{\rho}_{\rm S}(t)\bigg\}\\
    +\frac{1}{\hbar}\bigg[\frac{3(G_{\rm 2,S}-G_{\rm 2,S}')}{\hbar\omega_{\rm d}}\bigg]^2\kappa(3\omega_{\rm d}/2)
   \bigg\{ n(3\omega_{\rm d}/2)\mathcal{D}[\hat{a}^\dagger]\hat{\rho}_{\rm S}(t) +[n(3\omega_{\rm d}/2)+1]\mathcal{D}[\hat{a}]\hat{\rho}_{\rm S}(t)\bigg\}\\
    +\frac{1}{\hbar}\bigg(\frac{5G_{\rm 2,S}'}{8\hbar\omega_{\rm d}}\bigg)^2\kappa(5\omega_{\rm d}/2)\bigg\{ n(5\omega_{\rm d}/2)\mathcal{D}[\hat{a}]\hat{\rho}_{\rm S}(t)
    +[n(5\omega_{\rm d}/2)+1]\mathcal{D}[\hat{a}^\dagger]\hat{\rho}_{\rm S}(t)\bigg\}\\
    +\frac{1}{\hbar}\bigg(\frac{11G_{\rm 2,S}'}{36\hbar\omega_{\rm d}}\bigg)^2\kappa(7\omega_{\rm d}/2)\bigg\{ n(7\omega_{\rm d}/2)\mathcal{D}[\hat{a}^\dagger]\hat{\rho}_{\rm S}(t)+[n(7\omega_{\rm d}/2)+1]\mathcal{D}[\hat{a}]\hat{\rho}_{\rm S}(t)\bigg\}-\frac{i}{\hbar}[\hat{\mathcal{H}}_{\rm S},\hat{\rho}_{\rm S}(t)].
\end{multline}

In order to understand the effect of stronger modulation depth and higher Kerr-nonlinearity on the the lifetime of the coherent states, we plot $T_\alpha$ for $N=1$, i.e. $K/h \approx 50 {\rm ~MHz}$ (i.e., $E_{\rm C}/h = 100 ~\textrm{MHz}$) under different considerations in Fig.~\ref{fig:lifetime_kerr_mul}. This allows us to investigate the strong modulation depth regime (up to third order in $\delta\phi$) with $M=N=1$. The solid red curve gives the $T_\alpha$ plot taking into account higher-order modulation depth corrections to single-photon effects, finite $\Lambda$ and stronger modulation depth. The solid blue curve in Fig.~\ref{fig:lifetime_kerr2} was obtained for the same set of parameters but taking only leading order term in the modulation depth $\delta \phi$. The difference is a longer first plateau when stronger modulation depth is considered (compare the plateau for dashed blue curve in Fig.~\ref{fig:lifetime_kerr_mul} and the green dotted curve in the inset of Fig.~\ref{fig:lifetime_kerr2}). We find that once we set $\Lambda = 0$, the dip in $T_\alpha$ vanishes (dashed blue curve). However, $T_\alpha$ decreases as a function of the two-photon drive strength for $\epsilon_2/K\gtrsim 2$ and finally aligns with the solid red curve before increasing again. Next, we consider only the RWA terms neglecting all the higher-order corrections taking $\Lambda = 0$ and find that the staircase type behavior of $T_\alpha$ has been restored (dotted green curve). The first plateau aligns with the case of small Kerr coefficient obtained with $M=N=10$ (dot dashed purple curve). Note that even when the single-photon heating and cooling effects as well as finite detuning due to $\Lambda$ are neglected, $T_\alpha$ plateau gets longer for higher Kerr coefficient demanding stronger two-photon drive for a similar enhancement in $T_\alpha$.
\begin{figure}[H]
\centering
\includegraphics[width=0.6\columnwidth]{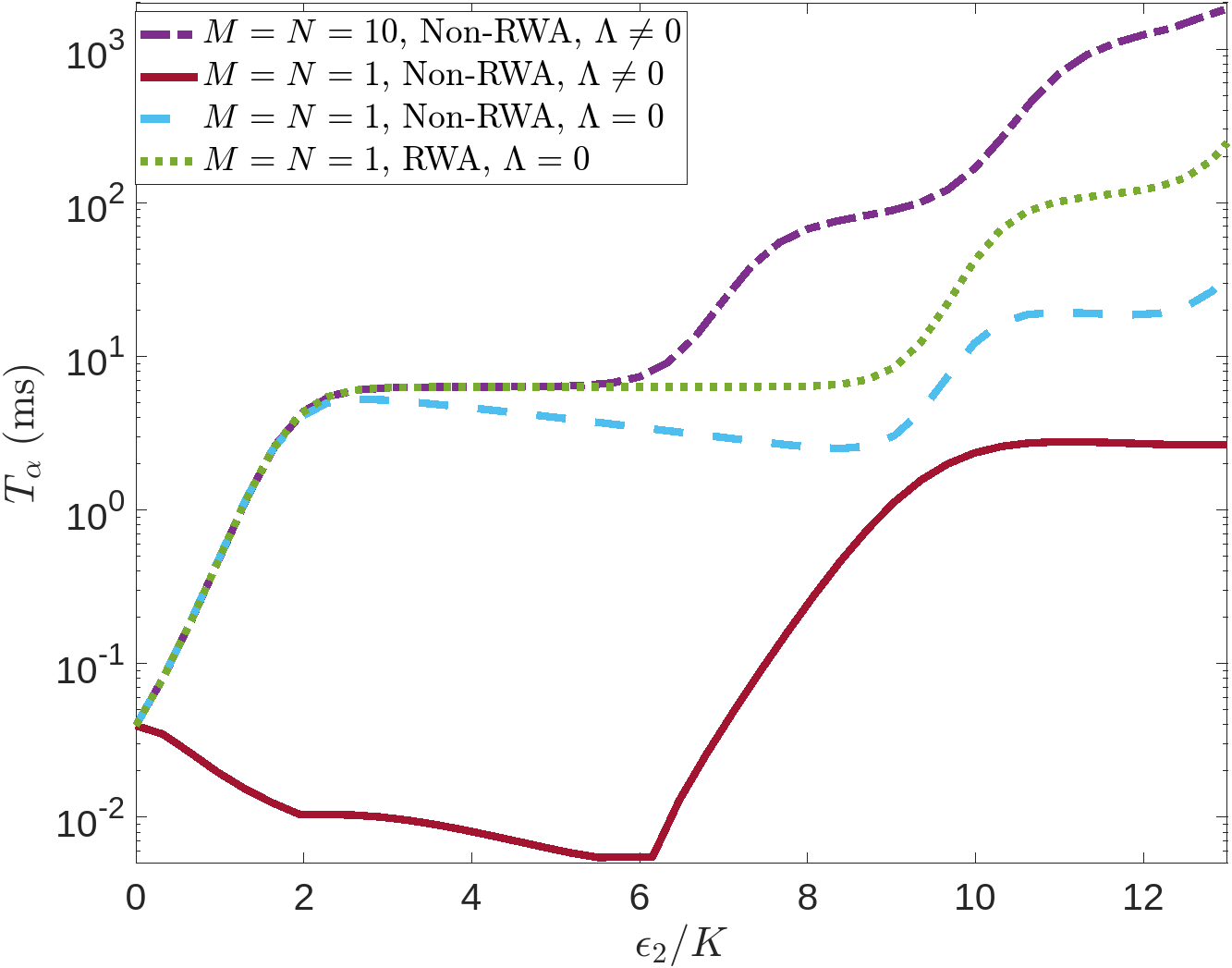}
 \caption{$T_\alpha$ of the Kerr-cat qubit as a function of the two-photon drive strength $\epsilon_2/K$ under strong modulation depth. The purple dot-dashed curve is obtained for 10 STS connected in series with $\tilde{N}=1$ number of junction in the transmon branch which results in a small Kerr coefficient, $K/h=~1{\rm MHz}$. For the rest of the curves, we consider $M=N=1$. The solid red curve is obtained by going up to ${\cal O}(\delta\phi^3)$ in the modulation depth and includes both non-RWA effects up to ${\cal O}(\varphi_{\rm zps}^2)$ and effects due to finite $\Lambda$ up to ${\cal O}(\varphi_{\rm zps}^4)$. For $\Lambda= 0$, we have the dotted green and dashed blue curves obtained with RWA and taking into account the non RWA effects, respectively. We chose $E_{\rm C}/h=100\textrm{MHz}$ for this calculation only.}
\label{fig:lifetime_kerr_mul}
\end{figure}
\section{SQUID Kerr-cat qubit - Hamiltonian and master equation}
\label{app:squid}
In this section, we will study the effective Hamiltonian and effective master equation for a driven SQUID and compare it to the case of STS. The Hamiltonian for a SQUID with symmetric junctions is given by (see Eq.~(\ref{eq:ham_double_squid}))
\begin{align}
\hat{H}_{\rm lab, SQ} = 4 E_{\rm C} \hat{n}^2 -  2E_{\rm J\Sigma}\cos\hat{\varphi} \cos\tilde{\phi}_e,
\label{eq:single_squid_ham}
\end{align}
where $\tilde{\phi}_e = \phi_e/2$ with $2\pi \phi_e/\phi_0$ being the external flux threading the SQUID. The charging energy is $E_{\rm C}=e^2/2C$, where $C$ is the total self-capacitance of the constituent Josephson junctions. We observe that for $\tilde{\phi}_e = \pi/2$, $\hat{H}_{\rm lab, SQ} = 4 E_{\rm C} \hat{n}^2$. This implies that the circuit becomes purely capacitive, and hence can't be operated as a oscillator around $\tilde{\phi}_e = \pi/2$. However, a STS acts as a non-linear oscillator for $\varphi_\Sigma=\pi/2$ (see Eq.~(\ref{eq:ham_double_squid})) and hence can be operated around $\varphi_\Sigma = \pi/2$. The SQUID can however be operated around $\tilde{\phi}_e = \pi/4$. Let us consider
\begin{equation}
\tilde{\phi}_e = \pi/4 + \delta \phi \cos(\omega_d t).
\label{eq:ext_flux_sq}
\end{equation}
Inserting Eq.~(\ref{eq:ext_flux_sq}) in Eq.~(\ref{eq:single_squid_ham}), we obtain
\begin{equation}
\hat{H}_{\rm lab, SQ} = 4E_{\rm C}\hat{n}^2 -\sqrt{2}E_{\rm J\Sigma}\left(1-\delta\phi^2/4\right)\cos{\hat{\phi}}+\sqrt{\frac{1}{8}}\delta\phi^2 E_{\rm J\Sigma} \cos\hat{\phi}\cos(2\omega_{\rm d}t)
+\sqrt{2}E_{\rm J \Sigma}\cos\hat{\phi}\delta\phi \cos(\omega_{\rm d} t).
\end{equation}

Following Appendix~\ref{app:static_eff}, we perform the generalized Schrieffer-Wolff transformation and obtain the following static effective Hamiltonian in the rotating frame
\begin{multline}
\label{eq:SQUID_static_H}
\hat{\cal H}_{\rm SQ} = \Delta \hat{a}^\dagger \hat{a}-\frac{\sqrt{2}E_{\rm J\Sigma}\varphi_{\rm zps}^4}{4}\bigg(1-\frac{\delta\phi^2}{4}\bigg)\hat{a}^{\dagger ^2}\hat{a}^2-\frac{\sqrt{2}\delta\phi E_{\rm J\Sigma}}{4}\bigg[
    \bigg(\varphi_{\rm zps}^2-\frac{\varphi_{\rm zps}^4}{2}\bigg)(\hat{a}^{\dagger ^2}+\hat{a}^2)\\
-\frac{\varphi_{\rm zps}^4}{3}(\hat{a}^\dagger \hat{a}^3+\hat{a}^{\dagger ^3}\hat{a})\bigg]+\frac{\sqrt{2}\delta\phi^2E_{\rm J\Sigma}\varphi_{\rm zps}^4}{192}(\hat{a}^{\dagger ^4}+\hat{a}^4)
+\frac{\delta\phi^2E_{\rm J\Sigma}\varphi_{\rm zps}^4}{2\omega_d}\bigg[\bigg(-1+\frac{\delta\phi^2}{12}\bigg)\hat{a}^\dagger \hat{a}+\frac{\delta\phi}{4}(\hat{a}^{\dagger ^2}+\hat{a}^2)\bigg].
\end{multline}
Similarly, the static effective master equation is given by
\begin{multline}
\label{eq:SQUID_master_eqn}
\frac{\partial \hat{\rho}_{\rm S}}{\partial t} = -\frac{i}{\hbar}[\hat{\cal H}_{\rm SQ},\hat{\rho}_{\rm S}]+\frac{1}{\hbar}\kappa(\omega_{\rm d}/2)\bigg\{n(\omega_{\rm d}/2)\mathcal{D}\bigg\{\hat{a}^\dagger-\frac{\sqrt{2}G_{2,S}}{\hbar\omega_{\rm d}}\hat{a}\bigg\}\hat{\rho}_{\rm S}+[1+n(\omega_d/2)]\mathcal{D}\bigg\{\hat{a}-\frac{\sqrt{2}G_{2,S}}{\hbar\omega_{\rm d}}\hat{a}^\dagger\bigg\}\hat{\rho}_{\rm S}\bigg\}\\
+\frac{1}{\hbar}\bigg(\frac{\sqrt{2}\delta\phi E_J^S\varphi_{zps}^2}{4\hbar\omega_d}\bigg)^2\kappa(3\omega_d/2)\bigg\{n(3\omega_d/2)\mathcal{D}\{3\hat{a}^\dagger+\delta\phi \hat{a}\}\hat{\rho}_{\rm S}+[1+n(3\omega_d/2)]\mathcal{D}\{3\hat{a}+\delta\phi \hat{a}^\dagger\}\hat{\rho}_{\rm S}\bigg\}\\
+\frac{1}{\hbar}\bigg(\frac{\delta\phi\sqrt{2}G_{2,S}}{3\hbar\omega_d}\bigg)^2\kappa(5\omega_d/2)\bigg\{
n(5\omega_d/2)\mathcal{D}\{\hat{a}^\dagger\}\hat{\rho}_{\rm S}+[1+n(5\omega_d/2)]\mathcal{D}\{\hat{a}\}\hat{\rho}_{\rm S}\bigg\}.
\end{multline}

We can see from Eq.~\ref{eq:SQUID_static_H} that the static effective Hamiltonian parameters also depend on the modulation depth and has an additional four-photon drive term. The effects of the modulation depth on the Hamiltonian parameters are negligible due to the modulation depth being a small quantity, so the parameters are effectively constant. From Eq.~\ref{eq:SQUID_master_eqn} we find that, similar to the STS, there are only single photon heating and cooling contributions at $\mathcal{O}(\varphi_{\rm zps}^2)$. However, the SQUID Kerr cat also has an effective higher frequency single-photon decay as well as some additional heating terms that do not appear for the STS. We find that the rise in lifetime for the STS begins for a smaller modulation depth compared to the SQUID Kerr cat qubit, as shown in Fig. \ref{fig:single_vs_double}. The SQUID Kerr cat qubit has both a larger dip in lifetime and larger plateau than the STS Kerr cat qubit when plotted directly as a function of the modulation depth.

\begin{figure}[H]
\centering
\includegraphics[width=0.6\columnwidth]{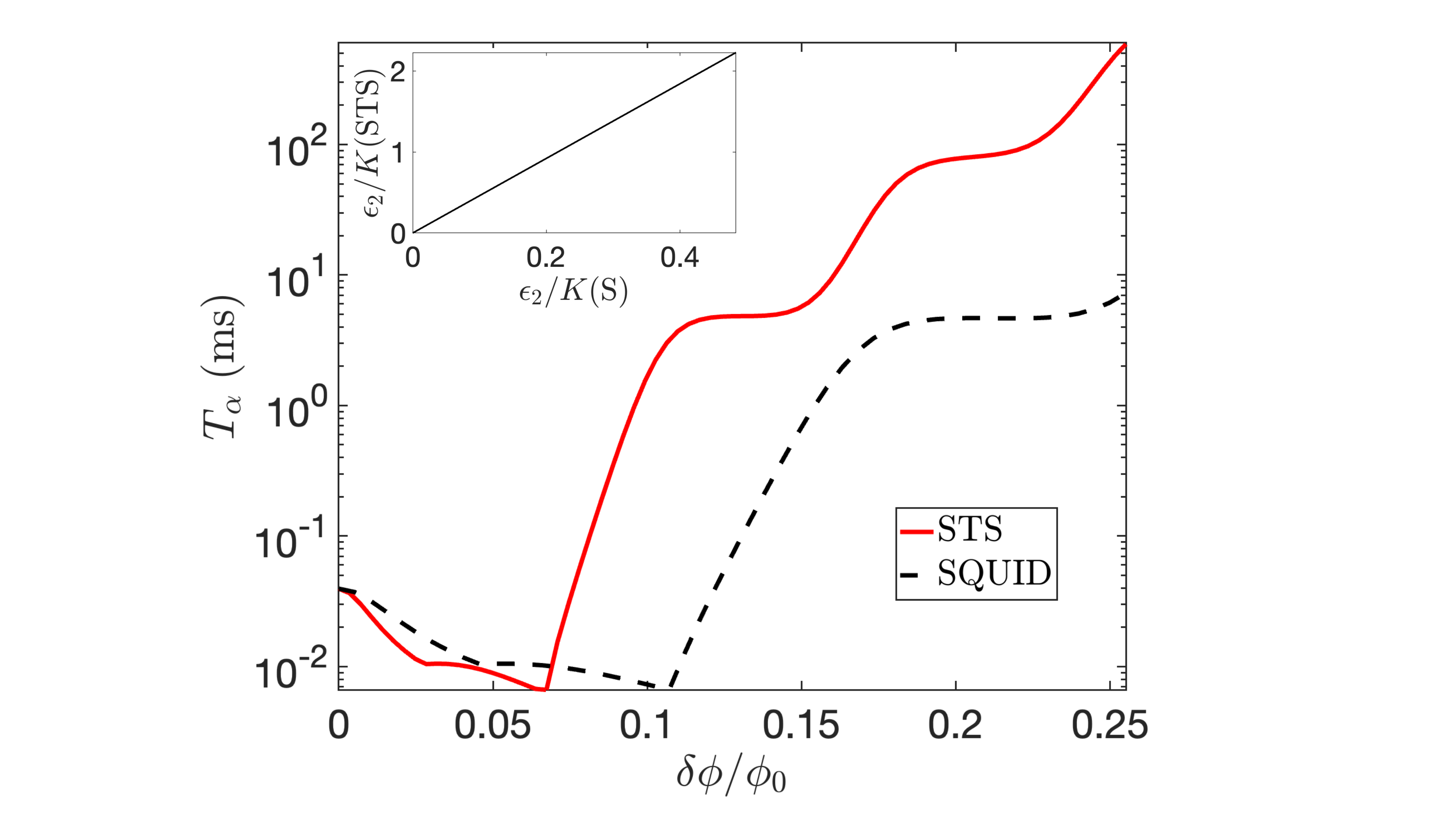}
 \caption{$T_\alpha$ as a function of the modulation depth for a single SQUID (black dashed curve) compared to the case of STS (solid red curve) for $K/h = 31.3~{\rm MHz}$, $\phi_0$ is the flux quantum. In the inset, we plot the two-photon drive strength obtained for the STS compared to a SQUID (S) by varying the modulation depth.} 
\label{fig:single_vs_double}
\end{figure}
\section{Leakage and Energy Spectrum in Detuned Kerr-cat Qubit}
\label{app:detuned_leak_energy}
We evaluate the energy eigenvalue spectra for degenerate energy levels as a function of the drive $\epsilon_2$ in Fig.~\ref{fig:detunedenergyspec} for the Hamiltonian of the Kerr-cat qubit, given in Eq.~\ref{eq:Kerrcat}, with detuning $\Delta/K=2m$, where $m=1$ for the left plot and $m=2$ for the right plot. For detuning $\Delta/K=2m$, we get $m+1$ separate pairs of degenerate energy levels. Moreover, each degenerate pair that is not the ground energy shows non-monotonous behavior as a function of $\epsilon_2/K$, which is demonstrated in the insets of the plots in Fig.~\ref{fig:detunedenergyspec} (for $\Delta/K =2,4$, we observe two and four non-monotonous eigenenergies, respectively). 
\begin{figure}[H]
\includegraphics[width=0.95\columnwidth]{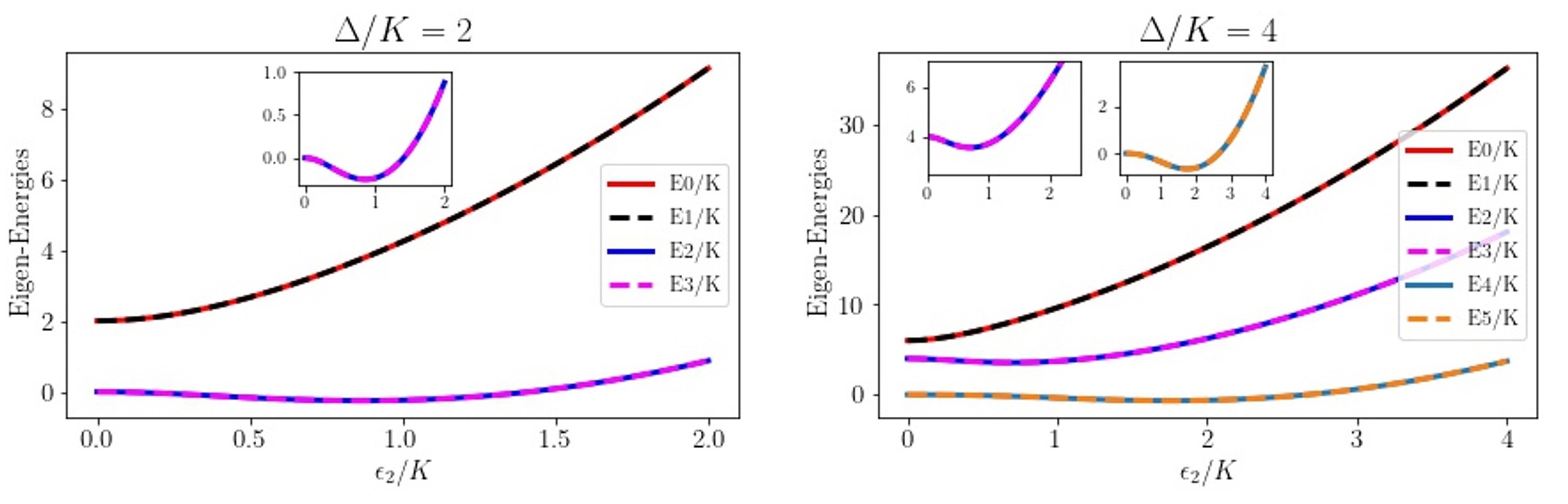}
 \caption{Energy eigenvalue spectra as a function of the two-photon drive strength $\epsilon_2/K$ for detuned Kerr-cat qubit with (a) $\Delta/K=2$ with the inset showing the energy spectra for $\rm E_2$ and $\rm E_3$, and (b) $\Delta/K=4$ with the leftmost inset showing the energy spectra for $\rm E_2$ and $\rm E_3$, and the rightmost inset showing the energy spectra for $\rm E_4$ and $\rm E_5$.} 
\label{fig:detunedenergyspec}
\end{figure}

\section{SNAIL and STS Kerr-cat qubit}
\label{app:snail_vs_squid}
\begin{figure}[hbt!]
\includegraphics[width=0.95\columnwidth]{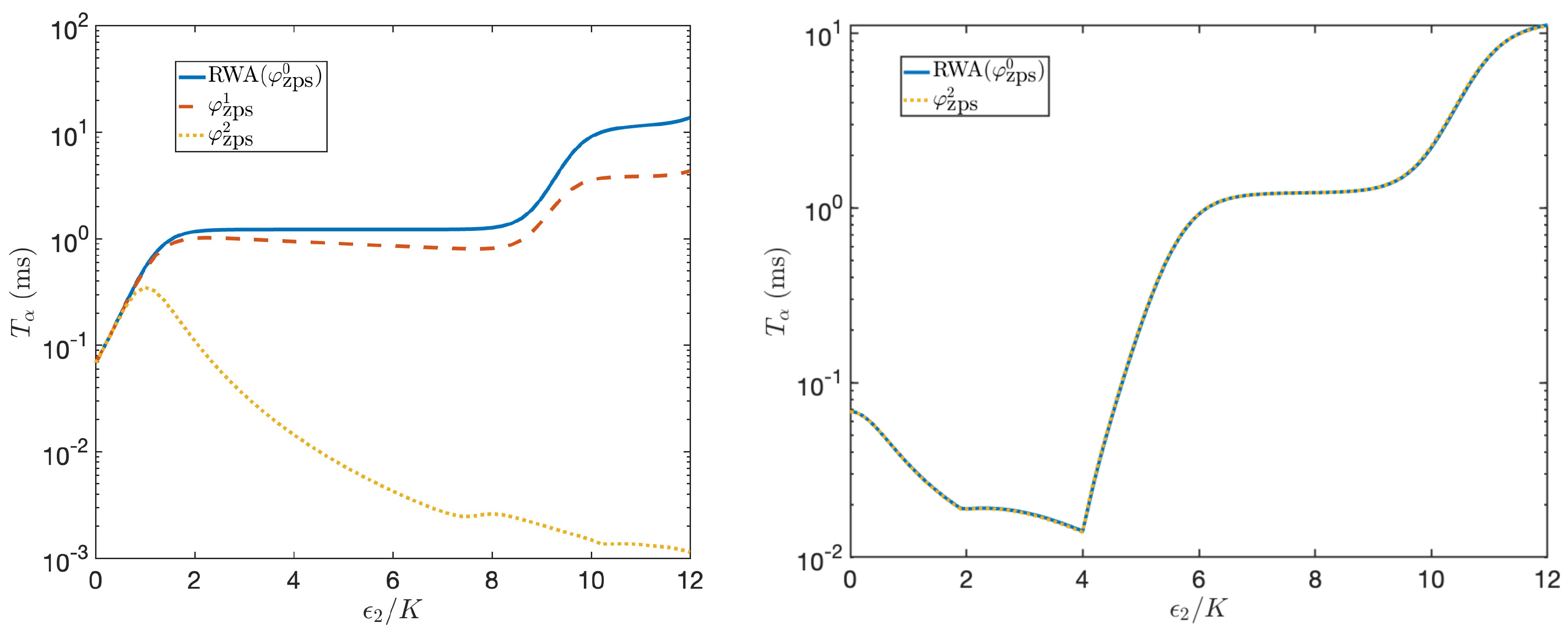}
 \caption{$T_\alpha$ of the Kerr-cat qubit as a function of the two-photon drive strength $\epsilon_2/K$ taking $K=14.4 {\rm MHz}$ (a) for a SNAIL design and (b) for a STS  design.} 
\label{fig:snailvssquid}
\end{figure}
We compare the SNAIL and STS Kerr-cat qubit with realistic parameters from experiments. Using the quantum master equations calculated in Ref.~\cite{venkatraman2022}, we plot $T_\alpha$ of a SNAIL Kerr-cat qubit in Fig.~\ref{fig:snailvssquid} (a) for a Kerr coefficient of $14.4 {\rm~MHz}$, temperature of $T_{\omega_{\rm d}/2}=82.5\textrm{~mK}$, and decay rate of $\kappa(\omega_{\rm d}/2)/h=26\textrm{~kHz}$. All the other parameters were taken from Ref.~\cite{venkatraman2022}. Keeping the coupling and temperature of the environment the same as above, and fixing the parameters in STS to obtain $K/h = 14.4 {\rm ~MHz}$, we plot $T_\alpha$ for the STS Kerr-cat qubit in Fig.~\ref{fig:snailvssquid}(b). We take up to ${\cal O}(\varphi_{\rm zps}^2)$ terms in both cases in the master equation. Unlike the case of STS where the system bath coupling at ${\cal O}(\varphi_{\rm zps}^1)$ is a scalar and does not induce any dynamics, in the case of SNAIL it leads to a small reduction in lifetime (see the dashed orange curve in Fig.~\ref{fig:snailvssquid}(a)). Further, the  ${\cal O}(\varphi_{\rm zps}^2)$ which leads to two-photon dissipative effects in the SNAILs leading to a strong reduction in lifetime (see the yellow dashed curve in Fig.~\ref{fig:snailvssquid}(a)), has no effect in the lifetime in the case of STS. We observed in the Appendix~\ref{app:master} that the two-photon heating and cooling effects enter the STS dynamics only at ${\cal O}(\varphi_{\rm zps}^3)$ and is proportional to the asymmetry of the SQUID junctions. Hence, the two-photon dissipative effects can be mitigated by making the junctions as symmetric as possible. One feature that leads to the reduction in $T_\alpha$ of the STS Kerr-cat qubit is the presence of the extra term $\Lambda$ in the STS Kerr-cat Hamiltonian in Eq.~(\ref{eq:ham_main}) compared to the standard Kerr-cat Hamiltonian in Eq.~(\ref{eq:Kerrcat}) (leading to a dip in lifetime for $\epsilon_2/K\leq 4$ in Fig.~\ref{fig:snailvssquid}(b)). In the inset of Fig.~\ref{fig:lifetime_kerr2}, we showed that the effect of $\Lambda$ can be canceled by adding a drive-dependent detuning to the Kerr-cat Hamiltonian leading to the restoration of the staircase type lifetime plot.

\end{widetext}
\bibliography{refs_sup}

\end{document}